\documentclass[letterpaper]{article}
\usepackage{graphicx}
\usepackage{enumerate}
\usepackage[colorlinks=true, allcolors=blue]{hyperref}
\usepackage{nameref, zref-xr}
\zxrsetup{toltxlabel}
\zexternaldocument*{Supp_au}

\usepackage[font=small]{caption}
\usepackage{natbib}
\usepackage[top=3cm, bottom=3cm, left=2cm, right=2cm]{geometry}
\usepackage[english]{babel}

\usepackage{tikz}
\usetikzlibrary{fit,positioning}
\usepackage{amsthm,amsmath,amssymb,mathrsfs,mathtools,bm,dutchcal,color}
\usepackage{multirow,dsfont}
\usepackage{subcaption}
\usepackage{booktabs}
\usepackage[capitalise]{cleveref}
\usepackage{verbatim,listings}
\usepackage[section]{placeins}

\usepackage{setspace}
\usepackage[ruled,norelsize,vlined]{algorithm2e}

\usepackage{todonotes}

\usepackage{enumitem}
\newcommand{\RNum}[1]{\uppercase\expandafter{\romannumeral #1\relax}}

\newcommand{\iidsim}{\overset{\text{iid}}{\sim}} 
\newcommand{\indsim}{\overset{\text{ind}}{\sim}} 
\newcommand{\ed}{\overset{\text{d}}{=}} 
\newcommand{\Norm}{\textsc{N}} 
\newcommand{\Dir}{\textsc{Dir}} 
\newcommand{\DP}{\textsc{DP}} 

\newcommand{\Gem}{\textsc{GEM}}

\newcommand{\HyDP}{\textsc{HHDP}}
\newcommand{\NDP}{\textsc{NDP}}
\newcommand{\HDP}{\textsc{HDP}}
\newcommand{\NIG}{\textsc{NIG}}

\newcommand{\Var}{\mathrm{Var}} 

\newcommand{\indic}{\mathbb{1}}

\newcommand{\E}{\mathbb{E}}

\renewcommand{\P}{\mathbb{P}}

\newcommand{\X}{\mathbb{X}}

\newcommand \dd[1]  { \,\textrm d{#1}}   

\def\simiid{\stackrel{\mbox{\scriptsize{iid}}}{\sim}}

\newtheorem{theorem}{Theorem}

\newtheorem{proposition}{Proposition}

\theoremstyle{definition}
\newtheorem{definition}{Definition}

\usepackage{xpatch}

\makeatletter
\xpatchcmd{\pprintMaketitle}{%
	\hrule\vskip12pt%
}{}{\typeout{Success}}{}

\xpatchcmd{\pprintMaketitle}{%
	\hrule\vskip12pt%
}{\@date}{\typeout{Success}}{}
\makeatother

\begin{document}
	
	\begin{center}
		{\Large
			\textbf{\newline{Flexible clustering via hidden hierarchical Dirichlet priors}}
		}
		\newline
		\\
		Antonio Lijoi \textsuperscript{1},
		Igor Pr\"unster \textsuperscript{1},
		Giovanni Rebaudo\textsuperscript{2}
		\\
		\bigskip
		{\mbox{\bf{1}} \small Department of Decision Sciences and BIDSA, Bocconi University, via R\"ontgen 1, 20136 Milan, Italy}
		\\
		{\mbox{\bf{2}} \small Department of Statistics and Data Sciences, University of Texas at Austin, 
		TX 78712-1823, USA}
	\end{center}
\date{}

	\renewcommand{\refname}{{REFERENCES}}

	\begin{abstract}
		\baselineskip=12pt
		The Bayesian approach to inference stands out for naturally allowing borrowing information across heterogeneous populations, with different samples possibly sharing the same distribution. A popular Bayesian nonparametric model for clustering probability distributions is the nested Dirichlet process, which however has the drawback of grouping distributions in a single cluster when ties are observed across samples. 
		With the goal of achieving a flexible and effective clustering method for both samples and observations, we investigate a nonparametric prior that arises as the composition of two different discrete random structures and derive a closed-form expression for the induced distribution of the random partition, the fundamental tool regulating the clustering behavior of the model. On the one hand, this allows to gain a deeper insight into the theoretical properties of the model and, on the other hand, it yields an MCMC algorithm for evaluating Bayesian inferences of interest. Moreover, we single out limitations of this algorithm when working with more than two populations and, consequently, devise an alternative more efficient sampling scheme, which as a by-product, allows testing homogeneity between different populations. Finally, we perform a comparison with the nested Dirichlet process and provide illustrative examples of both synthetic and real data.
	\end{abstract}
	
	\section{Introduction}\label{sec:Intro}
	Dirichlet process (DP) mixtures are well-established and highly successful Bayesian nonparametric models for density estimation and clustering, which also enjoy appealing frequentist asymptotic properties \citep{Lo1984a,Escobar1994,Escobar1995,Ghosal2017a}. However, they are not suitable to model data $\{(X_{j,1},\ldots, X_{j,I_j}): j=1,\ldots,J\}$ 
	that are recorded under $J$ different, though related, experimental conditions. This is due to exchangeability implying a common underlying distribution across populations, a homogeneity assumption which is clearly too restrictive. To make things concrete we consider the Collaborative Perinatal Project, which is a large prospective epidemiologic study conducted from 1959 to 1974 (analyzed in Section \ref{subsec:real_data_appl}), where pregnant women were enrolled in 12 hospitals and followed over time. Using a standard DP mixture on the patients enrolled across all 12 hospitals would correspond to ignoring the information on the specific center $j$ where the data are collected and, thus, the heterogeneity across samples. The opposite, also unrealistic, extreme case corresponds to modeling data from each hospital independently, thus ignoring possible similarities among them. 
	
	A natural compromise between the aforementioned extreme cases 
	is \emph{partial exchangeability} \citep{DeFinetti1938}, which entails exchangeability within each experimental condition (but not across) and \textit{dependent} population--specific distributions (thus allowing borrowing of information). See \cite{Kallenberg2006} for a detailed account of the topic. In this framework the proposal of dependent versions of the DP date back to the seminal papers of \cite{Cifarelli1978c} and \cite{MacEachern1999,MacEachern2000}. Dependent DPs can be readily used within mixtures leading to several success stories in topic modeling, biostatistics, speaker diarization, genetics, fMRI analysis, and so forth. See \cite{Dunson2010d, Teh2010b, Foti2015, ddp2021_Quintana} and references therein.
	
	Two hugely popular dependent nonparametric priors, which will also represent the key ingredients of the present contribution, 
	are the hierarchical Dirichlet process (HDP) \citep{Teh2006} and the nested Dirichlet process (NDP) \citep{Rodriguez2008a}.
	The HDP clusters observations within and across populations. The NDP aims to cluster both population distributions and observations, but as shown in \citet{Camerlenghi2019c}, does not achieve this goal. In fact, if there is a cluster of observations shared by different samples, the model degenerates to  exchangeability across samples. This issue is successfully overcome in \citet{Camerlenghi2019c} by introducing \textit{latent nested nonparametric priors}. However, while this proposal has the merit of being the first to solve the degeneracy problem, it suffers from other limitations in terms of implementation and modeling: (a) with data from more than two populations the analytical and computational burden implied by the additive structure becomes overwhelming; (b) the model lacks the flexibility needed to capture different weights that common clusters may feature across different populations. More details can be found in the discussion to \citet{Camerlenghi2019c}.
	
	The goal of this paper is thus to devise a principled Bayesian nonparametric approach, which allows to cluster simultaneously distributions and observations (within and across populations). We achieve this by blending peculiar features of both the NDP and the HDP into a model, which we term \textit{Hidden Hierarchical Dirichlet Process} (HHDP). 
	Importantly, the HHDP overcomes the above-mentioned theoretical, modeling, and computational limitations since it, respectively, does not suffer from the degeneracy flaw, is able to effectively capture different weights of shared clusters and allows to handle several populations as showcased in the real data application. 
	Note that the idea of the model was first hinted at in \cite{james2008discussion} and, later, considered in \cite{Agrawal2013} from a mere computational point of view without providing results on distributional properties that are relevant for Bayesian inference. Hence, as a by-product, our theoretical results shed also some light on the topic modeling applications of \cite{Agrawal2013}.
	Additionally, the same model was independently applied in \cite{balocchi2021clustering} to successfully cluster urban areal units at different levels of resolution simultaneously.
	
	\cref{sec:BNPClust} concisely reviews the HDP and the NDP with a focus on the random partitions they induce. In \cref{sec:HyDP} we define the HHDP 
	and investigate its properties, foremost  its clustering structure (induced by a partially exchangeable array of observations). These findings lead to the development of marginal and conditional Gibbs sampling schemes in \cref{sec:Inference}.
	In \cref{sec:Illustration} we draw a comparison between HHDP and NDP on synthetic data and present a real data application for our model.
	Finally, \cref{sec:Conclusion} is devoted to some concluding remarks and possible future research.
	
	\section{Bayesian nonparametric priors for clustering}\label{sec:BNPClust}
	The assumption of exchangeability that characterizes widely used Bayesian inferential procedures is equivalent to assuming data homogeneity. This is not realistic in many applied contexts, 
	for instance, for data recorded under $J$ different experimental conditions inducing heterogeneity. A natural assumption that relaxes exchangeability and is suited for arrays of random variables $\{(X_{j,i})_{i \ge 1}: j=1,\ldots,J \}$ is \textit{partial exchangeability}, which amounts to assuming homogeneity within each population, though not across different populations. This is characterized by 
	\[
	\{(X_{j,i})_{i \ge 1}: j=1,\ldots,J \} \ed \{(X_{j,\sigma_j(i)})_{i \ge 1}: j=1,\ldots,J \},
	\]
	for every finitary permutation $\{\sigma_j : j=1,\ldots,J\}$ with $\ed$ henceforth denoting equality in distribution. 
	Thanks to de Finetti's representation theorem for partially exchangeable arrays, the dependence structure is effectively 
	represented through the following hierarchical formulation
	\begin{equation}
		\label{eq:hierarchic_part_exchange}
		\begin{split}
			X_{j,i} \mid (G_1,\ldots,G_J)
			&\indsim G_{j}, \quad \quad (i=1,\ldots,I_j, j=1,\ldots,J)\\
			(G_1,\ldots,G_J) 
			&\sim \mathcal{L}.
		\end{split}
	\end{equation}
	Here we focus on priors $\mathcal{L}$ defined as compositions of discrete random structures and including, as special cases, both the HDP and the NDP. More specifically, we consider $\mathcal{L}$ in \eqref{eq:hierarchic_part_exchange}  that is defined as follows 
	\begin{equation}
		\label{eq:composition_prior}
			G_j\,|\, Q\,
			\iidsim \,\mathcal{L}(G_j|\,Q)\quad (j=1,\ldots,J);
			\qquad 
			Q \,|\,G_0\,
			\sim \, \mathcal{L}(Q|\, G_0); 
			\qquad G_0\,
			\sim\, \mathcal{L}(G_0),
	\end{equation}
	with discrete random probability measures $G_j$ ($j=1,\ldots,J$), $Q$ and $G_0$.
	The data are denoted by $\bm{X} = \{\bm{X}_1, \ldots, \bm{X}_J \}$ with $\bm{X}_j= ( X_{j,1}, \ldots, X_{j,I_j} ) $ and $I_j$ the size of the $j$th sample. 
	Discreteness of these random structures entails that with positive probability there are ties within each sample $\bm{X}_j$ and also across samples $j=1,\ldots,J$, i.e.\ $\P(X_{j,i}=X_{j,\ell})>0$ for any $i\ne \ell$, and  $\P(X_{j,i}=X_{\kappa,\ell})>0$ for any $j\ne \kappa$. 
	Hence, $\bm{X}$ induces a random partition of the integers $\{1,2,\ldots,n\}$ with $n=I_1+\,\cdots\,+I_J$, whose distribution encapsulates the whole probabilistic clustering of the model and is, therefore, the key quantity to study.  
	Importantly, the random partition can be characterized in terms of the 
	partially exchangeable partition probability function (pEPPF) as defined in \citet{Camerlenghi2019}.
	The pEPPF is the natural generalization 
	of the concept of exchangeable partition probability function (EPPF) for the exchangeable case \citep[see e.g.][]{Pitman2006}. 
	More precisely, $D$ is the number of distinct values among the $n= \sum_{j=1}^J I_j$ observations in the overall sample $\bm{X}$. 
	The vector of frequency counts is denoted by $\bm{n}_j=(n_{j,1},\ldots,n_{j,D})$ with $n_{j,d}$ indicating the number of elements in the $j$th sample that coincide with the $d$th distinct value in order of arrival.
	Clearly, $n_{j,d} \ge 0$ and $\sum_{i=1}^{J} n_{i,d} \ge 1$. One may well have $n_{j,d}=0$, which implies that the $d$th distinct value is 
	not recorded in the $j$th sample, though by virtue of $\sum_{i=1}^{J} n_{i,d} \ge 1$ it must be recorded 
	at least in one of the samples. 
	The $d$th distinct value is shared by any two samples $j$ and $j^\prime$ if and only if $n_{j,d} \, n_{j^{\prime},d} \ge 1$.
	The probability law of the random partition is characterized by the pEPPF defined as
	\begin{equation}\label{eq:pEPPF}
		\Pi_D^{(n)}(\bm{n}_1,\ldots,\bm{n}_J) = \E \int_{\mathbb{X}^{D}_{\ast}} \prod_{d=1}^D \{G_1(\mathrm{d}x_d)\}^{n_{1,d}} \ldots \{G_J(\mathrm{d}x_d)\}^{n_{J,d}},
	\end{equation}
	with the constraint $\sum_{d=1}^D n_{j,d} = I_j$, for each $j=1,\ldots,J$ and where $\mathbb{X}$ is the space in which the $X_{j,i}$'s take values and $\mathbb{X}^{D}_{\ast}$ is the collection of vectors in $\X^D$ whose entries are all distinct. 
	We stress 
	that the expected value in \eqref{eq:pEPPF} is computed with respect to the joint law of the vector of random probabilities $(G_1,\ldots,G_J)$, that is the de Finetti measure $\mathcal{L}$ in \eqref{eq:hierarchic_part_exchange}. Hence, the pEPPF may also be interpreted as a marginal likelihood when $(G_1,\ldots,G_J)$ directly model the observations according to \eqref{eq:hierarchic_part_exchange}.
	Obviously, for a single population, that is $J=1$, the standard EPPF is recovered  and 
	\eqref{eq:pEPPF} is further interpretable as an extension of a product partition model to a multiple samples framework. As such, 
	it provides an alternative approach to popular covariate--dependent product partition models. See, e.g., \cite{MQR}, \cite{Page_Quintana_2016} and \cite{Page_Quintana}. 
	
	If we 
	specify $\mathcal{L}(\,\cdot\,|Q)$ and $Q$ such that they give rise to an NDP, then one may have ties also among the population probability distributions $G_1,\ldots,G_J$, i.e.\ $\P(G_j=G_\kappa)>0$ for any $j\ne \kappa$. Therefore, in the framework of \eqref{eq:hierarchic_part_exchange} and \eqref{eq:composition_prior}, one may investigate two types of clustering: (i) \textit{distributional 
		clustering}, 
	which is related to $G_1,\ldots,G_J$ and (ii) \textit{observational clustering}, which refers to $\bm{X}$. The composition of these two clustering structures is the main tool we rely on to devise a simple, yet effective, model that considerably improves over existing alternatives.%
	
	\subsection{Hierarchical Dirichlet process}\label{subsec:HDP}
	Probably the most popular nonparametric prior for the partially exchangeable case is the HDP of \cite{Teh2006}, which can be nicely framed in the composition scheme \eqref{eq:composition_prior} as 
	\begin{equation}
		\label{eq:HDP}
		\mathcal{L}(G_j|Q)=\DP(G_j|\beta,Q),\quad \mathcal{L}(Q|G_0)=\delta_{G_0}(Q),\quad
		\mathcal{L}(G_0)=\DP(G_0|\beta_0;H),
	\end{equation}
	where $\DP(\,\cdot\,|\alpha,P)$ denotes the law of a DP with concentration parameter $\alpha>0$ and baseline probability measure $P$. Here we assume that $H$ is a non--atomic probability measure on $\X$ and we refer to such prior as the $J$-dimensional HDP denoted by $(G_1,\ldots,G_J) \sim \HDP(\beta,\beta_0;H)$. Hence, the $G_j$'s share the atoms through $G_0$ and this leads to the creation of shared clusters of observations (or latent features) across the $J$ groups. The pEPPF 
	induced by a partially exchangeable array 
	in \eqref{eq:hierarchic_part_exchange} with  $\mathcal{L} = \HDP(\beta,\beta_0;H)$ 
	has been determined in \citet{Camerlenghi2019}. It is important to stress that the model is not suited for comparing populations' distributions since $\P(G_j=G_\kappa)=0$ for any $j\ne \kappa$ (unless the $G_j$'s are degenerate at $G_0$, in which case all distributions are equal). Similar compositions have been considered in \citet{Camerlenghi2019} and, later, in \cite{argiento} and \cite{bassetti}. 
	Hierarchically dependent mixture hazards have been introduced in \cite{camerlenghi2021survival}.
	Anyhow, the HDP and its variations cannot be used to cluster both populations and observations. To achieve this, one has to rely on priors induced by nested structures, the most popular being the NDP.
	\subsection{Nested Dirichlet process}\label{subsec:NDP}
	The NDP, introduced by \cite{Rodriguez2008a}, is the most widely used nonparametric prior allowing to cluster both observations and populations. However, as proved in \citet{Camerlenghi2019c}, it suffers from a \emph{degeneracy issue}, because even a single tie shared across samples is enough to group the $J$ population distributions into a single cluster.
	
	Like the HDP, also the NDP can be framed in the composition structure \eqref{eq:composition_prior} as
	\begin{equation}
		\label{eq:NDP}
		\mathcal{L}(G_j|Q)=Q(G_j),\quad \mathcal{L}(Q|G_0)=\DP (Q|\alpha;G_0),\quad \mathcal{L}(G_0)=\delta_{\DP(\beta;H)}(G_0),
	\end{equation}
	where $Q$ is a random probability measure on the space $\mathscr{P}_{\X}$ of probability measures on $\X$ and $G_0$ is degenerate at the atom $\DP(\beta;H)$, which is the law of a DP on the sample space $\X$. 
	As in \eqref{eq:HDP}, $H$ is assumed to be a non-atomic probability measure on $\X$. Henceforth, we write $(G_1,\ldots,G_J) \sim \NDP(\alpha, \beta; H)$. By virtue of the well--known stick--breaking representation of the DP \citep{Sethuraman1994a} one has
	\begin{equation}
		\label{eq:ndp_gemstructure}
		Q = \sum_{k \ge 1} \pi^\ast_k \delta_{G^\ast_k},\quad (\pi^\ast_k)_{k \ge 1} \sim \Gem(\alpha),
		\quad G^\ast_k \iidsim \DP(\beta; H),
	\end{equation}
	where the weights $(\pi^\ast_k)_{k \ge 1}$ and the 
	random distributions $(G^\ast_k)_{k\ge 1}$ are independent. Recall that $\Gem$ stands for the distribution of probability weights after Griffiths, Engen, and McCloskey, according to the well-established terminology of \cite{Ewens1990}. Given a sequence $(V_i)_{i\ge 1}$ such that $V_i\simiid \mbox{Beta}(1,\alpha)$, this means that $\pi_1^*=V_1$ and $\pi_k^*=V_k\,\prod_{i=1}^{k-1}(1-V_i)$, for any $k\ge 2$.
	Since $\P(G_j=G_\kappa)=1/(\alpha+1)$ for any $j\ne \kappa$, $Q$ generates ties among the random distributions $G_j$'s with positive probability and, thus, clusters populations.
	Furthermore, a structure similar to the one displayed in \eqref{eq:ndp_gemstructure} holds for each $G^\ast_k$, i.e.
	\begin{equation*}
		G^\ast_k = \sum_{l \ge 1} \omega_{k,l} \delta_{X^\ast_{k,l}},\quad
		(\omega_{k,l})_{l \ge1} \iidsim \Gem(\beta),\quad
		X^{\ast}_{k,l} \iidsim H,
	\end{equation*}
	and, due to the non--atomicity of $H$, the $X_{k,l}^\ast$ are all distinct values. 
	
	The discrete structure of the $G_k^\ast$'s generates ties across the samples $\{\bm{X}_j: j=1,\ldots,J\}$ with positive probability. For example, $\P(X_{j,i}=X_{j^\prime,i^\prime})=1/\{(\alpha+1)(\beta+1)\}$ for any $j \ne j^\prime$.  Hence, the $G_k^\ast$'s induce the clustering of the observations $\bm{X}$.
	
	If the data $\bm{X}$ are modeled as in \eqref{eq:hierarchic_part_exchange}, with $(G_1,\ldots,G_J)\sim\textsc{NDP}(\alpha,\beta;H)$, conditional on a partition of the $G_j$'s the observations from populations allocated to the same cluster are exchangeable and those from populations allocated to distinct clusters are independent. This potentially appealing feature of the NDP is however the one responsible for the above-mentioned \textit{degeneracy issue}. For exposition clarity, consider the case of $J=2$ populations. If the two populations belong to different clusters, i.e.\ $G_1 \ne G_2$, they cannot share even a single atom $X_{k,l}^{\ast}$ due to the non--atomicity of $H$. Hence, $\P(X_{1,l}=X_{2,l'}|G_1\ne G_2)=0$ for any $l$ and $l'$. 
	Therefore 
	there is neither clustering of observations nor borrowing of information across different populations. 
	On the contrary, 
	$\P(X_{1,i}=X_{2,i'}|G_1= G_2)=1/(\beta+1)>0$. These two findings are quite intuitive. Indeed, $G_1\ne G_2$ means they are independent realizations of a DP 
	with atoms iid from the same non-atomic probability distribution $H$ and, thus, they are almost surely different. 
	Instead, $G_1=G_2$ corresponds to all observations coming from the same population distribution, more precisely from the same DP, and ties occur with positive probability. 
	A less intuitive fact is that when a single atom, say $X_{k,l}^{\ast}$, is shared between $G_1$ and $G_2$ the model degenerates to the exchangeable case, namely $\P(G_1=G_2|X_{1,i}=X_{2,i'})=1$ and the two populations have (almost surely) equal distributions. Hence, the NDP is not an appropriate specification when aiming at clustering both populations and observations across different populations. 
	This was shown in \citet{Camerlenghi2019c} where, spurred by this anomaly of the NDP, a novel class of priors named \textit{latent nested processes} (LNP) 
	designed to ensure that $\P(G_1\ne G_2|X_{1,i}=X_{2,i'})>0$ is proposed. However, while this formally solves the problem, it has computational and modeling limitations. On the one hand, the implementation of LNPs with more than two samples is not feasible due to severe computational hurdles. On the other hand, LNPs have limited flexibility since the weights of the common clusters of observations across different populations are the same. This feature is not suited to several applications and the discussion to \citet{Camerlenghi2019c} provides interesting examples. 
	See also \cite{soriano_ma2019,christ_ma2020,denti2021common,beraha2021semi} for further stimulating contributions to this literature.
	
	Hence, within the composition structure framework   \eqref{eq:composition_prior}, our goal is to obtain a prior distribution able to infer the clustering structure of both populations and observations, which is  highly flexible and implementable for a large number of populations and associated samples. 
	
	\section{Hidden hierarchical Dirichlet process}\label{sec:HyDP}
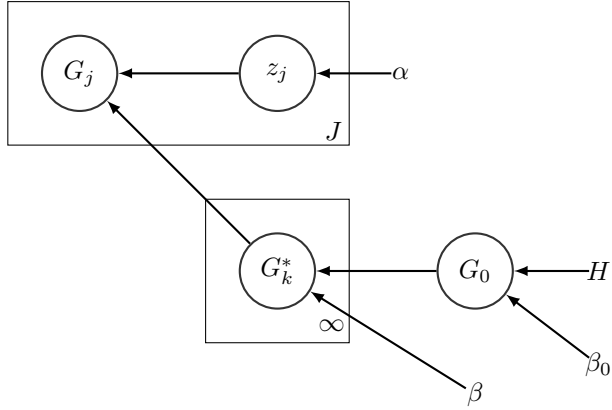
\begin{figure}[!h]
				\centering
				\begin{tikzpicture}
					\tikzstyle{main}=[circle, minimum size = 10mm, thick, draw =black!80, node distance = 16mm]
					\tikzstyle{connect}=[-latex, thick]
					\tikzstyle{box}=[rectangle, draw=black!100]
					\node[main] (Gj) 		[label=center:$G_j$] { };
					\node[main] (zj) 		 [right=of Gj,label=center:$z_j$] { };
					\node[]        (alpha)   [right= of zj, label=center:$\alpha$] { };
					\node[main] (Gstark) [below=of zj,label=center:$G^\ast_k$] { };
					\node[main] (Gstar0) [right=of Gstark,label=center:$G_0$] { };
					\node[]        (beta) 	 [below=of Gstar0,label=center:$\beta$] { };
					\node[]        (H) 	 	  [right=of Gstar0,label=center:$H$] { };
					\node[]        (beta0)  [below=of H,label=center:$\beta_0$] { };
					\path (beta0) edge [connect] (Gstar0)
					(H) edge [connect] (Gstar0)
					(Gstar0) edge [connect] (Gstark)
					(beta) edge [connect] (Gstark)
					(Gstark) edge [connect] (Gj)
					(zj) edge [connect] (Gj)
					(alpha) edge [connect] (zj);
					\node[rectangle, inner sep=0mm, fit= (Gj) (zj),label=below right:$J$, xshift=13mm] {};
					\node[rectangle, inner sep=4.4mm,draw=black!100, fit= (Gj) (zj)] {};
					\node[rectangle, inner sep=0mm, fit= (Gstark) ,label=below right:$\infty$, xshift=-1mm] {};
					\node[rectangle, inner sep=4.4mm, draw=black!100, fit = (Gstark)] {};
				\end{tikzpicture}
				\caption{Graphical model representing the dependencies for a $\HyDP(\alpha,\beta,\beta_0;H)$. Here the $z_j$'s are auxiliary integer--valued random variables that assign each $G_j$ to a specific atom $G_k^\ast$ of $Q$.}
				\label{fig:graphical_model}
			\end{figure}

	Our proposal consists in blending the HDP and the NDP in a way to leverage on their strengths, namely clustering data across multiple heterogeneous samples for the HDP and clustering different populations (or probability distributions) for the NDP. More precisely we combine these two models in a structure \eqref{eq:composition_prior} as
	\[
	\mathcal{L}(G_j|Q)=Q(G_j),\quad \mathcal{L}(Q|G_0)=\DP(Q|\alpha;\DP(\beta;G_0)),\quad \mathcal{L}(G_0)=\DP(G_0|\beta_0;H).
	\] 
	This leads to the following definition.
	\begin{definition}\label{def:HyDP}
		The vector of random probability measures $(G_1,\ldots,G_J)$ is a \emph{hidden hierarchical Dirichlet process} (HHDP) if
				\[
				G_j \mid Q \iidsim Q, 
				\quad
				Q = \sum_{k \ge 1} \pi^\ast_k \delta_{G^\ast_k},\quad 
				(\pi^\ast_k)_{k \ge 1} \sim \Gem(\alpha),\quad 
				(G^\ast_k)_{k \ge 1} \sim \HDP(\beta,\beta_0; H),
				\]
				with $(\pi^\ast_k)_{k \ge 1}$ and $(G^\ast_k)_{k \ge 1}$ independent. In the sequel we write $(G_1,\ldots,G_J) \sim \HyDP(\alpha,\beta, \beta_0;H)$.
			\end{definition}
			In terms of a graphical model, the HHDP can be represented as in Figure~\ref{fig:graphical_model}. 
			
			The sequence $(G^\ast_k)_{k \ge 1}$ acts as a hidden, or latent, component that is crucial to avoid the bug of the NDP, namely clustering of populations when they share some observations. Moreover, by extending \eqref{eq:HDP} to $J=\infty$, it can be more conveniently represented as
			\begin{equation}
				\label{eq:hidden_hdp}
				\begin{split}
					G_k^\ast =\sum_{l\ge 1}\omega_{k,l}\,\delta_{Z_{k,l}},\quad Z_{k,l}|G_0
					&\iidsim 
					G_0 ,\quad 
					G_0 =\sum_{l\ge 1}\omega_{0,l}\,\delta_{X_l^\ast},\quad
					X_\ell^\ast\iidsim H,\\
					(\omega_{k,l})_{l\ge 1}\iidsim \Gem(\beta),
					&\quad (\omega_{0,l})_{l\ge 1}\sim\Gem(\beta_0),
				\end{split}
			\end{equation}
			where independence holds true between the sequences $(\omega_{k,l})_{l\ge 1}$ and $(Z_{k,l})_{l\ge 1}$ and between $(\omega_{0,l})_{l\ge 1}$ and $(X_{l}^\ast)_{l\ge 1}$.  Combining the stick-breaking representation and a closure property of the DP with respect to grouping, one further has
					\begin{equation*}
						G^\ast_k = \sum_{l \ge 1} \omega^\ast_{k,l} \delta_{X_l^\ast}, 
						G_0 = \sum_{l \ge 1} \omega_{0,l} \delta_{X_l^\ast},
					\end{equation*}
			where $((\omega^\ast_{k,l})_{l \ge 1} \mid \bm{\omega}_0) \iidsim \DP(\beta; \bm{\omega}_0)$, 
			$\bm{\omega}_{0}=(\omega_{0,l})_{l \ge 1} \sim \Gem(\beta_0)$ and $X_l^\ast \iidsim H$, for $l\ge 1$.
			
			In this scheme, the clustering of populations is governed, \textit{a priori}, by the NDP layer $Q$ 
			through $(\pi^\ast_k)_{k \ge 1} \sim \Gem(\alpha)$. However, the aforementioned degeneracy issue of the NDP, \textit{a posteriori}, is successfully avoided. The intuition is quite straightforward: unlike for the NDP, the distinct distributions $G^\ast_k$ in the HHDP 
			are dependent and have a common random discrete base measure $G_0$, which leads to shared atoms across the $G_k^\ast$'s and thus borrowing of information, similarly to the HDP case.

			\subsection{Some distributional properties}\label{subsec:PropertiesHyDP}
			Given the discreteness of $(G_1,\ldots,G_J)\sim \HyDP(\alpha,\beta, \beta_0;H)$, the key quantity to derive is the induced random partition, which controls the clustering mechanism of the model. 
			However, it is useful to start with a description of pairwise dependence of the elements of the vector $(G_1,\ldots,G_J)$, which allows a better understanding of the model and intuitive parameter elicitation. To this end, as customary, we evaluate the correlation between $G_j(A)$ and $G_{j^\prime}(A)$: whenever it does not depend on the specific measurable set $A\subset \X$, it is used as a measure of overall dependence between $G_j$ and $G_{j^\prime}$.
			\begin{proposition}\label{eq:MeanVarCovHyDP}
				If $(G_1,\ldots,G_J) \sim \HyDP(\alpha, \beta,\beta_0;H)$ and $A$ is a measurable subset of $\X$, then
				\begin{align*}
					\Var[G_j(A)] &= \dfrac{H(A) [1-H(A)](\beta_0+ \beta +1)}{(\beta+1) (\beta_0+1)} &(j=1,\ldots,J),\\
					\mathrm{Corr}[G_j(A),G_{j^\prime}(A)] &= 1 - \dfrac{\alpha \beta_0}{(\alpha+1) (\beta+\beta_0+1)} 	&(j\ne j').
				\end{align*}
			\end{proposition}
			
			Arguments similar to those in the proof of Proposition~\ref{eq:MeanVarCovHyDP} lead to determine the correlation between observations, either from the same or from different samples.
			\begin{proposition}\label{eq:CovXHyDP}
				If  $\{\bm{X}_j: j=1,\ldots,J \}$ are 
				from $(G_1, \ldots, G_J) \sim \HyDP(\alpha, \beta, \beta_0; H)$ according to \eqref{eq:hierarchic_part_exchange}, then
				\begin{align*}
					\mathrm{Corr}(X_{j,i},X_{j^\prime,i^\prime}) = \P(X_{j,i}=X_{j,i^\prime}) = \begin{cases}
						\dfrac{1}{\beta_0+1} +\dfrac{\beta_0}{(1+\alpha)(1+\beta)(1+\beta_0)} & (j \ne   j^\prime)\\[7pt]
						\dfrac{\beta +\beta_0 +1}{(\beta+1)(\beta_0+1)} 					  
						& (j = j^\prime).
					\end{cases}
				\end{align*}
			\end{proposition}
			\begin{figure}
				\centering
				\begin{subfigure}{0.3\linewidth}
					\includegraphics[width=\linewidth]{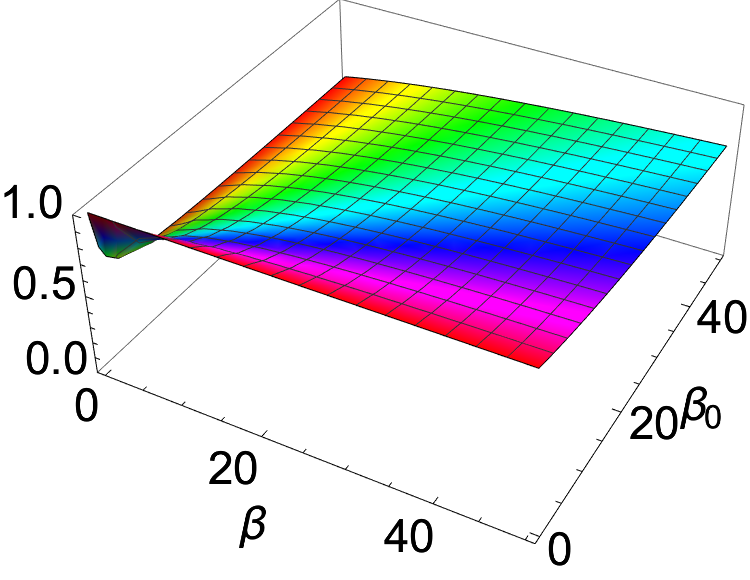}
				\end{subfigure}
				\begin{subfigure}{0.3\linewidth}
					\includegraphics[width=\linewidth]{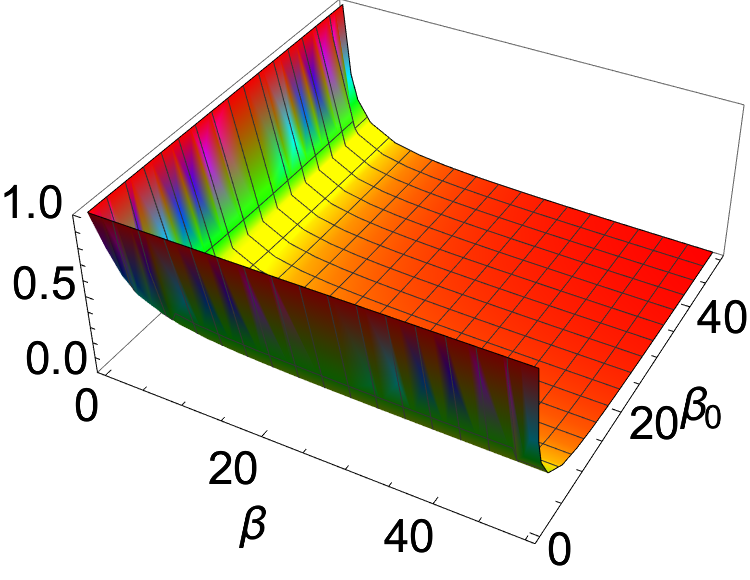}
				\end{subfigure}
				\begin{subfigure}{0.38\linewidth}
					\includegraphics[width=\linewidth]{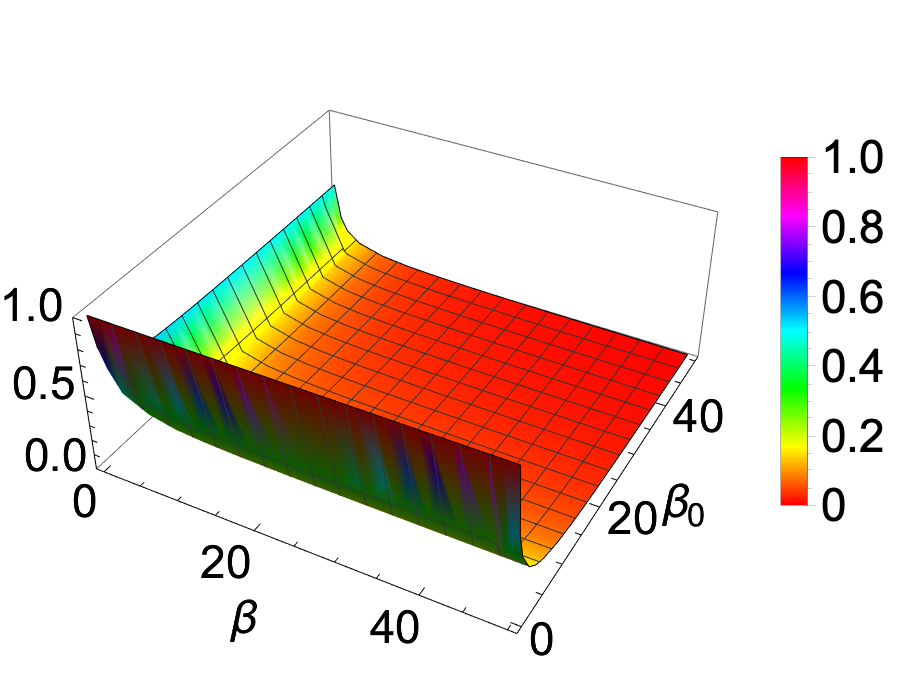}
				\end{subfigure}
				\caption{Correlations as functions of the hyperparameters $\beta$ and $\beta_0$ with $\alpha=1$. The left plot represents the correlation between random probabilities $G_j(A)$, the middle one between observations collected in the same population and the right one between observations from different populations. }
				\label{fig:hyperparameterscorrrandprobs}
			\end{figure}
			The correlation between observations of the same sample depends only on the parameters of the underlying HDP$(\beta,\beta_0;H)$ that governs the atoms $G_k^\ast$: this is not surprising since, whatever the value of the parameter $\alpha$ at the NDP layer, observations from the same sample are exchangeable. Moreover, an appealing feature is that such a correlation is higher than for the case of observations from different samples, i.e.\ $j\ne j'$. As for the dependence on the hyperparameters $(\alpha,\beta_0,\beta)$, when $\alpha\to\infty$ the $G_j$'s ar forced to equal different unique distributions $G_k^\ast$, similarly to the NDP case. However, unlike the NDP, this does not imply that the distributions are independent, and the correlation is controlled by the hyperparameters $\beta$ and $\beta_0$ (increasing in $\beta$ and decreasing in $\beta_0$). In \cref{fig:hyperparameterscorrrandprobs} we report the aforementioned correlations as functions of $\beta$ and $\beta_0$ with $\alpha$ set equal $1$. Finally, if $\alpha \rightarrow 0$ the a priori probability to degenerate to the exchangeable case, i.e. all  $G_j$'s coincide a.s., tends to $1$ and so does also $\mathrm{Cor}[G_j(A),G_{j^\prime}(A)]$.
			
		We now  investigate the random partition structure associated with a HHDP, namely the partition of $\{1,\ldots,n\}$, with $n=\sum_{j=1}^J I_j$, induced by a partially exchangeable sample $\bm{X}$ modeled as in \eqref{eq:hierarchic_part_exchange}. Since a $\HyDP(\alpha, \beta,\beta_0;H)$ arises from the composition of two discrete random structures, it is clear that the partition induced by $\bm{X}$ will depend on the partition, say $\Psi^{(J)}$, of the random probability measures $G_1,\ldots,G_J$. As for the latter, the $G_i$'s are drawn from a discrete random
		probability measure on $\mathscr{P}_{\X}$ whose weights
		have a $\Gem(\alpha)$ distribution and whose atoms are almost surely different since they are sampled from an $\textsc{HDP}(\beta,\beta_0;H)$. Then the probability distribution of $\Psi^{(J)}$ is the well--known Ewens sampling formula, namely
		\begin{equation*}
			\P[\Psi^{(J)}=\{B_1,\ldots,B_R\}]=\phi_{R}^{(J)}(m_1,\ldots,m_{R} ) = \frac{\alpha^{R}}{\alpha_{(J)}} \prod_{r=1}^{R} (m_r -1)!,
		\end{equation*}
		where $\{B_1,\ldots,B_R\}$ is a partition of $\{1,\ldots,J\}$, with $1\le R\le J$, the frequencies $m_r=\mbox{card}(B_r)$ are such that $\sum_{r=1}^R m_r=J$ and $\alpha_{(J)}=\Gamma(\alpha+J)/\Gamma(\alpha)$. This structure \textit{a priori} implies, as in the NDP case, that $\P(G_j=G_\kappa)\in(0,1)$ for any $j\ne \kappa$. However, unlike the NDP, \textit{a posteriori} the HHDP yields $\P(G_j=G_\kappa \mid \bm{X})<1$, regardless of the shared clusters across the samples $\bm{X}$.  
		Moreover, let $\Phi_{D,R}^{(n)}(\,\cdots\,;\beta,\beta_0)$ denote the pEPPF of a $\HDP (\beta,\beta_0;H)$, namely 
		\begin{equation*}
			\Phi_{D,R}^{(n)}(\bm{n}_1^\ast,\ldots,\bm{n}_{R}^\ast;\beta,\beta_0)=\E \int_{\X_*^D} 
			\prod_{d=1}^{D} \hat{G}_1(\dd x_d)^{n_{1,d}^\ast}\:\cdots\:\hat{G}_{R}(\dd x_d)^{n_{R,d}^\ast }
			,
		\end{equation*}
		where $(\hat{G}_1,\ldots,\hat{G}_{R})\sim \HDP(\beta,\beta_0;H)$, $D\in\{1,\ldots,n\}$ and $\sum_{r=1}^R\sum_{d=1}^D n_{r,d}^*=n$. An explicit  expression of $\Phi_{D,R}^{(n)}$ has been established in \citet{Camerlenghi2019}, even beyond the DP case. Now we can state the 
		pEPPF induced by 
		$\{\bm{X}_{j}: j=1,\ldots,J \}$ in \eqref{eq:hierarchic_part_exchange}, where  $\mathcal{L}$ is the law of a $\HyDP(\alpha, \beta,\beta_0;H)$. 
		\begin{theorem}\label{th:pEPPFHyDP}
			The random partition induced by the partially exchangeable array $\{\bm{X}_j: j=1,\ldots,J \}$ drawn from $(G_1,\ldots,G_J) \sim \HyDP(\alpha, \beta,\beta_0;H)$, according to \eqref{eq:hierarchic_part_exchange}, is characterized by the following pEPPF
			\begin{equation}
				\label{eq:peppf_hhdp}
				\Pi_D^{(n)}(\bm{n}_1, \ldots, \bm{n}_J) = \sum \phi_{R}^{(J)}(m_1,\ldots,m_{R}; \alpha )  \Phi_{D,R}^{(n)}(\bm{n}_1^\ast,\ldots,\bm{n}^\ast_{R}; \beta, \beta_0 ),
			\end{equation}
			where the sum runs over all partitions $\{B_1,\ldots,B_{R}\}$ of $\{1,\ldots,J\}$ and $n^{\ast}_{r,d} = \sum_{j \in B_r} n_{j,d}$ for each $r\in\{1,\ldots,R\}$, $d\in\{1,\ldots,D\}$. 
		\end{theorem}
		Given the composition structure underlying the $\HyDP(\alpha,\beta,\beta_0;H)$, the pEPPF \eqref{eq:peppf_hhdp} unsurprisingly is a mixture of pEPPF's induced by different HDPs. For ease of interpretation consider the case of $J=2$ populations and note that the pEPPF boils down to
		\begin{equation}\label{eq:pEPPFHyDPJ2}
			\Pi_D^{(n)}(\bm{n}_1,\bm{n}_2)= \dfrac{1}{\alpha+1} \Phi_{D,1}(\bm{n}_1+\bm{n}_2) + \dfrac{\alpha}{\alpha+1} \Phi_{D,2}(\bm{n}_1,\bm{n}_2),
		\end{equation}
		where $\Phi^{(n)}_{D,1}$ is the EPPF of a single $\HDP(\beta, \beta_0;H)$, namely $J=1$, while $\Phi^{(n)}_{D,2}$ is the pEPPF of a 
		$\HDP(\beta, \beta_0;H)$ with two samples, namely $J=2$. Clearly \eqref{eq:pEPPFHyDPJ2} arises from mixing with respect to partitions of $\{G_1,G_2\}$ in either $R=1$ and $R=2$ groups, where the former corresponds to exchangeability across the two populations. Still for the case $J=2$, a straightforward application of the pEPPF leads to the posterior probability of gathering the two probability curves, $G_1$ and $G_2$, in the same cluster thus making the two samples exchangeable, or homogeneous. 
		\begin{proposition}\label{prop:probDegHyDP}
			If the sample $\{\bm{X}_j: j=1,2\}$ is 
			from $(G_1,G_2) \sim \HyDP(\alpha, \beta,\beta_0;H)$, according to \eqref{eq:hierarchic_part_exchange}, the posterior probability of degeneracy is
			\begin{equation}
				\P(G_1=G_2 \mid \bm{X}) = \dfrac{\Phi^{(n)}_{D,1}(\bm{n}_1+\bm{n}_2)}{ \Phi^{(n)}_{D,1}(\bm{n}_1+\bm{n}_2)+\alpha \, \Phi^{(n)}_{D,2}(\bm{n}_1, \bm{n}_2)},
			\end{equation}
			where $\Phi^{(n)}_{D,1}$ and $\Phi^{(n)}_{D,2}$ are the EPPF and the pEPPF induced by 
			the $\HDP(\beta, \beta_0;H)$ for a single exchangeable sample and for two partially exchangeable samples,  respectively.
		\end{proposition}
		
		The pEPPF is a fundamental tool in Bayesian calculus 
		and   it plays, in the partially exchangeable framework, the same role of the EPPF in the exchangeable case. Indeed, the pEPPF 
		governs the learning mechanism, \textit{e.g.} the strength of the borrowing information, clustering, and, in view of \cref{prop:probDegHyDP}, it allows to perform hypothesis testing for distributional homogeneity between populations. Finally, one can obtain a 
		P\'{o}lya urn scheme that is essential for 
		inference and prediction,  See \cref{subsec:MargGibbs} in the Supplementary Material. In the next section, we provide a characterization of the $\HyDP(\alpha,\beta,\beta_0;H)$ that is reminiscent of the popular Chinese restaurant franchise metaphor for the HDP and allows us to devise a suitable sampling algorithm and further understand the model behavior.
		
		\subsection{The hidden Chinese restaurant franchise}\label{subsec:LCRF}
		The marginalization of the underlying random probability measures, as displayed in Theorem~\ref{th:pEPPFHyDP}, can be characterized in terms of a
		\textit{hidden Chinese restaurant franchise} (HCRF) metaphor. This representation sheds further light on the HHDP and clarifies the sense in which it generalizes  the well-known Chinese restaurant (CRP) and franchise (CRF) processes induced by the DP and the HDP, respectively. For simplicity we consider the case $J=2$.
		
		As with simpler sampling schemes, all restaurants of the franchise share the same menu, which has an infinite number of dishes generated by the non--atomic base measure $H$. However, unlike the standard CRF, the restaurants of the franchise are merged into a single one if $G_1=G_2$, while they 
		differ if $G_1\ne G_2$. Moreover, each $X_{j,i}$ identifies the label of the dish that customer $i$ from the $j$--th population chooses from the shared menu $(X_{d}^\ast)_{d\ge 1}$, with the unique dishes $X^\ast_d \iidsim H$. If $G_1\ne G_2$, customers may be assigned to different restaurants and when $G_1=G_2$, they are all seated in the same restaurant. Given such a grouping of the restaurants, the customers are, then, seated according to the CRF applied either to a single restaurant or to two distinct restaurants \citep{Teh2006,Camerlenghi2018e}. Furthermore, each restaurant has infinitely many tables. The first customer $i$ who arrives at a previously unoccupied table chooses a dish that is shared by all the customers who will join the table afterward. It is to be noted that distinct tables within each restaurant and across restaurants may share the same dish. 
		An additional distinctive feature, compared to the CRF, is that tables can be shared across populations when they are assigned to the same restaurant, i.e.\ when $G_1=G_2$. 
		Accordingly, the allocation of each customer $X_{j,i}$ 
		to a specific restaurant clearly depends on having either $G_1=G_2$ or $G_1\ne G_2$. 

		The sampling scheme simplifies 
		if latent variables $T_{j,i}$'s, denoting the tables' labels for customer $i$ from population $j$, are introduced. 
		We stress that, if $G_1\ne G_2$, the number of shared tables across the two populations is zero, given the populations $j=1,2$ are assigned to different restaurants, labeled $r=1,2$, respectively.
		Conversely, if $G_1=G_2$, one may have shared tables across populations, since they are assigned to the same restaurant $r=1$.

		Now define $q_{r,t,d}$ as the frequencies of observations sitting at table $t$ eating the $d$th dish, for a table specific to restaurant $r$. 
		Moreover, $D_t$ is the dish label corresponding to table $t$ and $\ell_{r,d}$ the frequency of tables serving dish $d$ in restaurant $r$.
		Marginal frequencies are represented with dots, e.g.\ $\ell_{r,\cdot}$ is the number of tables in restaurant $r$.
		Throughout the symbol $\bm{x}^{-i}$ identifies either a set or a frequency obtained upon removing the element $i$ from $\bm{x}$.
		Finally, $\Delta$ stands for an indicator function such that $\Delta=1$ if $G_1=G_2$, while $\Delta=0$ if $G_1\ne G_2$.

		The stepwise structure of the sampling procedure reflects the composition of the three layers $\mathcal{L}(G_j|Q)$, $\mathcal{L}(Q|G_0)$ and $\mathcal{L}(G_0)$ in \eqref{eq:hidden_hdp} relying on a conditional CRF.
		First, one sample the populations' clustering $\Delta$ and, given the allocations of the populations to the restaurants, one has a CRF.
		Hence, the algorithm becomes
		\begin{enumerate}
			\item[\texttt{(1)}] Sample the population assignments to the restaurants from 
			$\P(\Delta=1) = 1/(\alpha+1)$.
			\item[\texttt{(2)}] Sequentially sample the table assignments $T_{j,i}$ and corresponding dishes $D_{T_{j,i}}$ from\\ 
			\begin{align*}
					p(T_{j,i}, D_{T_{j,i}} \mid\bm{T}^{-(ji+)}, \bm{X}^{-(ji+)},\Delta  ) \propto \begin{cases} 
						T_{j,i} =t                                                                     & \frac{q_{r,t,\cdot}^{-(ji+)}}{q_{r,\cdot,\cdot}^{-(ji+)}+\beta}  \\ 
						T_{j,i} = t^\text{new}, D_{t^\text{new}} =d                   & \frac{\beta}{q_{r,\cdot,\cdot}^{-(ji+)}+\beta}  \frac{\ell_{\cdot,d}^{-(ji+)}}{\ell_{\cdot,\cdot}^{-(ji+)}+\beta_0} \\
						T_{j,i} = t^\text{new}, D_{t^\text{new}} =d^\text{new}  & \frac{\beta}{q_{r,\cdot,\cdot}^{-(ji+)}+\beta}  \frac{\beta_0}{\ell_{\cdot,\cdot}^{-(ji+)}+\beta_0},
					\end{cases}
				\end{align*}
			\end{enumerate}
			where $(ji+)=\{(ji'): i' \ge i \} \cup \{(j'i'):j' \ge j \}$ is the index set associated to the future random variables not yet sampled.
			
			\section{Posterior Inference  for HHDP mixture models}\label{sec:Inference}
			Thanks to the results of Section~\ref{sec:HyDP}, we now devise MCMC algorithms for drawing posterior inferences with mixture models driven by a HHDP. 
			Though the samplers are tailored to mixture models,	they are easily adapted to other inferential problems such as e.g. survival analysis and species sampling. 
			Henceforth, $\mathcal{K}$ is a density kernel and we consider 
			\begin{equation}\label{eq:MM}
				\begin{aligned}
					X_{j,i}\mid\theta_{j,i} &\indsim \mathcal{K}(\cdot| \theta_{j,i}), \quad \quad &(i=1,\ldots,I_j \quad j=1,\ldots,J),\\
					\theta_{j,i} \mid G_j &\indsim G_j, \quad \quad \qquad \:\: &(i=1,\ldots,I_j, \quad j=1,\ldots,J),\\
					(G_1,\ldots,G_J) &\sim \HyDP(\alpha, \beta, \beta_0; H).
				\end{aligned}
			\end{equation}
			We develop two samplers: (i) a marginal algorithm that relies on the posterior degeneracy probability (\cref{prop:probDegHyDP}) in \cref{subsec:MargGibbs} of the Supplementary Material; (ii) a conditional blocked Gibbs sampler, in the same spirit of the sampler proposed for the NDP by \cite{Rodriguez2008a}, in \cref{subsec:BlockGibbs}. 
			As for (i), the underlying random probability measures $G_0$ and $G_k^\ast$'s are integrated out leading to urn schemes that extend the class of Blackwell-MacQueen P\'olya urn processes.
			In such a way we generalize the \textit{a posteriori} sampling scheme of the Chinese restaurant process for the DP mixture \cite{Neal2000} and the one of the Chinese restaurant franchise for the HDP mixture \citep{Teh2006}.
			In the Supplementary Material, we describe the marginal sampler for the case of $J=2$ populations. 
			Even if in principle it can be  generalized in a straightforward way, it is computationally intractable for a larger number of populations. Similarly to the hidden Chinese restaurant franchise situation, one has to evaluate the posterior probability of all possible groupings of $G_1,\ldots,G_J$, which boils down to $\P(G_1=G_2|\bm{X})$ when $J=2$ but becomes involved for $J>2$. 
			
			This shortcoming is overcome by the conditional algorithm we derive in \cref{subsec:BlockGibbs}, which relies on finite--dimensional approximations of the trajectories of the underlying random probability measure.
			Its effectiveness in dealing with $J>2$ populations is further illustrated in the synthetic data example \ref{subsec:syn_data_more} and in the application of \cref{subsec:real_data_appl}.

			\subsection{A conditional blocked Gibbs sampler}\label{subsec:BlockGibbs}
			A more effective algorithm is based on a simple blocked conditional procedure. 
			To this end, we use a finite approximation of the DP in the spirit of \cite{Muliere1998} and \cite{Ishwaran2001}.
			However, instead of truncating the stick--breaking representation of the DP, we use a finite Dirichlet approximation. See \cite{Ishwaran2002a}. 
			Therefore, we approximate $\bm{\pi}^\ast, \bm{\omega}_0^\ast$, with a $K$-- and an $L$--dimensional Dirichlet distribution, respectively.
			More precisely, we consider the following approximation
				\begin{equation}
					\label{eq:FinApprox}
						\bm{\pi}^{\ast} \sim \Dir(\alpha/K, \ldots, \alpha/K),\qquad 
						\bm{\omega}_0^{\ast} \sim \Dir\big(\beta_0/L, \ldots, \beta_0/L\big)\\
			\end{equation}
			implying that $(\bm{\omega}^{\ast}_k \mid \bm{\omega}^{\ast}_0) \iidsim \Dir(\beta\,\bm{\omega}_0^{\ast})$, for $k \ge 1$.
			
			Introduce the auxiliary variables $z_j$ and $\zeta_{j,i}$ which represent the distributional and observational cluster memberships, respectively, such that $z_j=k$ and $\zeta_{j,i}=l$ if and only if $G_j=G^\ast_k$ and $\theta_{j,i}=\theta^\ast_{l}$.
			Henceforth, $\bm{S} = \{(\theta^\ast_{l})_{l=1}^{L}, \bm{\pi}^{\ast}, \bm{\omega}^{\ast}_0, (\bm{\omega}^{\ast}_k)_{k=1}^K,(z_j)_{j=1}^J, (\zeta_{j,i})_{j,i}, (X_{j,i})_{j,i} \}$ 
			and, in order to identify the full conditionals of the Gibbs sampler, we note that under the finite Dirichlet approximation \eqref{eq:FinApprox}
			\begin{multline*}
				p(\bm{S}) = p( \bm{\pi}^{\ast} ) p(\bm{\omega}^{\ast}_0) \bigg[\prod_{l=1}^{L} p(\theta^\ast_{l}) \bigg] \bigg[\prod_{k=1}^{K} p(\bm{\omega}^{\ast}_k \mid \bm{\omega}^{\ast}_0) \bigg]
				\bigg\{ \prod_{j=1}^{J} p(z_j \mid \bm{\pi}^{\ast} ) \bigg[ \prod_{i=1}^{I_j} p(X_{j,i}\mid \theta^\ast_{\zeta_{j,i}}) p(\zeta_{j,i} \mid \bm{\omega}^{\ast}_{z_j}) \bigg] \bigg\}.
			\end{multline*}
			This leads to the following
			
			\begin{enumerate}
				\item[\texttt{(1)}] Sample the unique $\theta^\ast_{l}$ from 
				\[
				p(\theta^{\ast}_l \mid \bm{S}^{-\theta^\ast_{l}}) \propto H(\theta^\ast_{l}) \prod_{\{j, i : \zeta_{j,i}=l\}} \mathcal{K}(X_{j,i}\mid \theta^\ast_{l}).
				\]
				\item[\texttt{(2)}] Sample distributional cluster probabilities from 
				\[
				p(\bm{\pi}^{\ast} \mid \bm{S}^{-\bm{\pi}^{\ast}}) = \Dir(\bm{\pi}^{\ast} \mid \alpha/K+m_1, \ldots,\alpha/K+m_K), 
				\]
				with $m_k= \sum_{j=1}^{J} \indic\{z_j=k\}$.
				
				\item[\texttt{(3)}] Sample probability weights of the base DP from \begin{align}
					\begin{split}
						p(\bm{\omega}^{\ast}_0 \mid \bm{S}^{-\bm{\omega}^{\ast}_0})
						\propto \prod_{l=1}^{L} \bigg[ \dfrac{(\omega_{0,l}^{\ast})^{\beta_0/L-1} \xi_{l}^{\beta \omega_{0,l}^{\ast} }}{\Gamma(\beta_0 \omega_{0,l}^{\ast})^K}\bigg],
					\end{split}
				\end{align}
				with $\xi_l= \prod_{k=1}^{K} \omega_{k,l}^{\ast}$.
				\item[\texttt{(4)}] Sample the observational cluster probabilities independently 
				from 
				\[
				p(\bm{\omega}^{\ast}_k \mid \bm{S}^{-\bm{\omega}^{\ast}_k}) 
				= \Dir(\bm{\omega}^{\ast}_k \mid \beta \bm{\omega}^{\ast}_0 + \bm{n}_k ),
				\]
				with $n_{k,l}=\sum_{\{j : z_j =k\}} \sum_{i=1}^{I_j} \indic\{\zeta_{j,i}=l\}$.
				
				\item[\texttt{(5)}] Sample distributional and observational cluster membership from
						\begin{align*}
							p(z_j=k \mid \bm{S}^{-\{z_j, \bm{\zeta}_{j}\}})
							&\propto 
							\pi^{\ast}_k \prod_{i=1}^{I_j} \sum_{l=1}^{L} \omega^{\ast}_{k,l} \mathcal{K}(X_{j,i}\mid \theta^\ast_{l})  &(k=1,\ldots,K),\\
							p(\zeta_{j,i}=l \mid \bm{S}^{-\zeta_{j,i}}) 
							&\propto 
							\omega^{\ast}_{z_j l} \, \mathcal{K}(X_{j,i}\mid \theta^\ast_{l}) &(l=1,\ldots,L).
						\end{align*}
				\end{enumerate}
				
				Importantly, 
				all the full conditional distributions are available in simple closed forms, with the exception of the distributions of $\bm{\omega}_0^{\ast}$ and, possibly, of $\theta^{\ast}_l$. To update $\bm{\omega}_0^{\ast}$ we perform a Metropolis-Hastings step, where we work on the unconstrained space $\mathbb{R}^{L-1}$ after the transformation $[\log(\omega_{0,1}/\omega_{0,L}),\ldots,\log(\omega_{0,L-1}/\omega_{0,L})]$ and we adopt a component--wise adaptive random walk proposal following \cite{Roberts2009}. The update of the unique atoms $\theta^{\ast}_l$ is standard, as with the DP mixture model in the exchangeable case. 
				
				In \cref{sec:Illustration} we assume a Gaussian kernel $\mathcal{K}(\cdot| \theta) = \Norm(\cdot| \mu, \sigma^{2})$ and a conjugate Normal-inverse-Gamma base measure $H(\cdot)=\NIG(\cdot \mid \mu_0,\lambda_0,s_0,S_0)$ and obtain\begin{equation*}
					p(\theta^{\ast}_l \mid \bm{S}^{-\theta^\ast_{l}}) = \NIG(\theta^{\ast}_l  \mid \mu_l,\lambda_l,s_l,S_l),
				\end{equation*}
				with $\mu_l = \dfrac{n_l \bar{y}_l + \lambda_0 \mu_0}{\lambda_0 +n_l}$, $S_l= S_0 + \dfrac{1}{2} \bigg(e^2_l + \dfrac{n_l \lambda_0 (\bar{y}_l - \mu_0)^2}{\lambda_0+n_l} \bigg)$, $\lambda_l =\lambda_0 +n_l$, and $s_l=n_l/2 +s_0$, where $n_l = \sum_{j=1}^{J} \sum_{i=1}^{I_j} \indic\{\zeta_{j,i}=l\}$, $\bar{y}_l = \sum_{\{j,i : \zeta_{j,i}=l\}} X_{j,i}/n_l$, and $e^2_l = \sum_{\{j,i : \zeta_{j,i}=l\}} (X_{j,i}- \bar{y}_l)^2$ are the observational cluster sizes, means and deviances, respectively.
				
				\section{Illustration}\label{sec:Illustration}
				In this section, we compare the performance of our proposal \eqref{eq:MM} with the same model where the HHDP is replaced by a NDP as in \eqref{eq:NDP}, 
				on synthetic data involving $J=2$ and $J=4$ populations. Note that for the latter, the implementation of the latent nested prior process mixture of \cite{Camerlenghi2019c} is not feasible, while the proposed HHDP mixture model can easily handle that level of complexity. The inferential results that we display are obtained by relying on the blocked Gibbs sampler of Section \ref{sec:Inference}.

				\subsection{Inference with two populations}\label{sec:two_populations}
				The data are simulated from the same scenarios considered in \citet{Camerlenghi2019c}. More precisely, we consider two populations and the data in each population are iid from a mixture of two normals:
				\begin{itemize}
					\item[\textbf{Scen 1.}] We simulate the data from the two populations independently from the same density 
					\begin{equation*}
						X_{1,i} \ed X_{2,i^\prime} \simiid 0.5 \Norm(0,1) + 0.5 \Norm(0,1).
					\end{equation*}
					\item[\textbf{Scen 2.}] We simulate the data in the two populations independently from mixtures of two normals with one shared component
					\begin{equation*}
						X_{1,i} \simiid 0.9 \Norm(5,0.6) + 0.1 \Norm(10,0.6) \quad \quad X_{2,i^\prime} \simiid 0.1 \Norm(5,0.6) + 0.9 \Norm(0,0.6).
					\end{equation*}
					\item[\textbf{Scen 3.}] We simulate the data in the two populations independently from mixtures of two normals having the same components, though with different weights
					\begin{equation*}
						X_{1,i} \simiid 0.8 \Norm(5,1) + 0.2 \Norm(0,1) \quad \quad X_{2,i^\prime} \simiid 0.2 \Norm(5,1) + 0.8 \Norm(0,1).
					\end{equation*}
				\end{itemize}
				In all these scenarios we consider balanced sample sizes $I_1=I_2=100$ and 
				an HHDP mixture model 
				\eqref{eq:MM}, with $\alpha=1$, $\beta=1$, $\beta_0=1$ and  
				$H(\cdot)=\NIG(\cdot \mid \mu_0, \lambda_0, s_0, S_0).$
				We set standard values of the hyperparameters in terms of the mean $\bar{y}$ and variance $\Var(y)$ of the data, i.e.\ $\mu_0=\bar{y}$, $\lambda_0=1/(3 \, \Var(y))$, $s_0=1$ and $S_0=4$. 
				In drawing the comparison between \eqref{eq:MM} and the $\NDP(\alpha, \beta; H)$, we further set $\alpha=\beta=1$. 
				Furthermore, we set the concentration parameters all equal to 1. In \cref{app:hyperpar} we perform a sensitivity analysis with respect to hyperparameters' specifications as done, for instance, by \cite{Zuanetti2018} for the NDP.
				The mean measure of the marginal underlying random distributions $\E[G_j(A)]=H(A)$ is the same for all populations. Also variances are comparable (see \cref{eq:MeanVarCovHyDP}) since $\Var[G_j(A)]$ equals $H(A)[1-H(A)]/2$ for the NDP and $3 H(A)[1-H(A)]/4$ for the HHDP. The sensitivity analysis leads, for  all the considered settings, to the same conclusions in terms of comparison of the two models.
				Moreover, we fix the dimensions of the finite approximations $L=K=50$ in \eqref{eq:FinApprox} and we do the same for the truncation levels in the algorithm of \cite{Rodriguez2008a}.
				In the Supplementary Material, we perform an empirical analysis trying different levels of $L$ and $K$ which corroborates the fact that the approximation error is negligible in terms of inferential results.
				
\begin{figure}[!h]
					\makebox[\linewidth]{\hspace*{2.5cm} True \hspace*{2cm} HHDP \hspace*{1.9cm} NDP \hspace*{1.5cm}}
					\makebox[\linewidth]{\hspace*{1.7cm} Pop 1 \hspace*{0.35cm} Pop 2\hspace*{0.55cm} Pop 1 \hspace*{0.35cm} Pop 2 \hspace*{0.45cm} Pop 1 \hspace*{0.35cm} Pop 2 \hspace*{0.8cm}}
					\rotatebox[origin=c]{90}{\makebox[3cm]{Scenario \RNum{1}}}
					\rotatebox[origin=c]{90}{\makebox[3cm]{ Pop 1 \hspace*{0.15cm} Pop 2}}
					\centering
					\begin{subfigure}{3cm}
						\includegraphics[width=\textwidth,height=3cm]{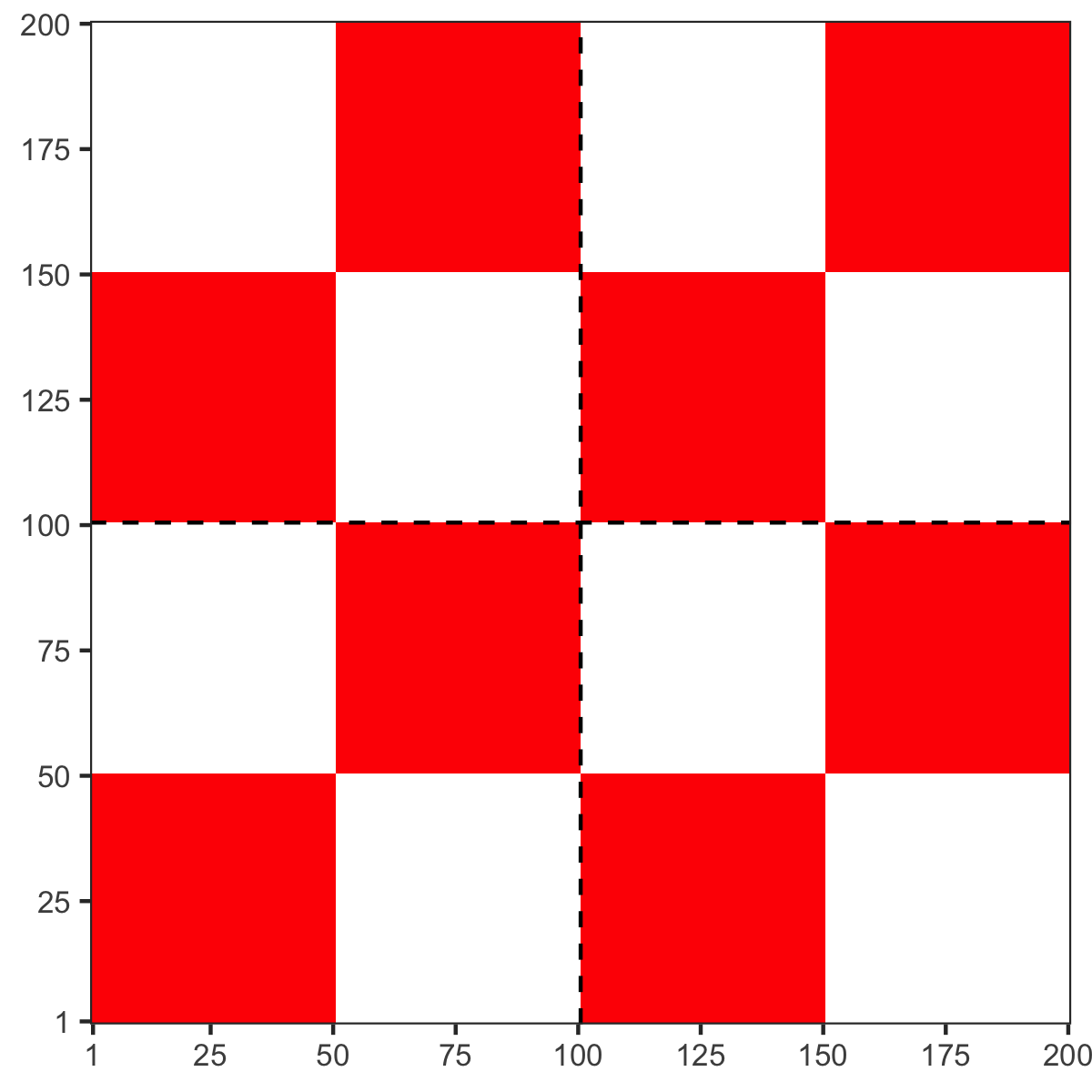}
					\end{subfigure}
					\begin{subfigure}{3cm}
						\includegraphics[width=\textwidth,height=3cm]{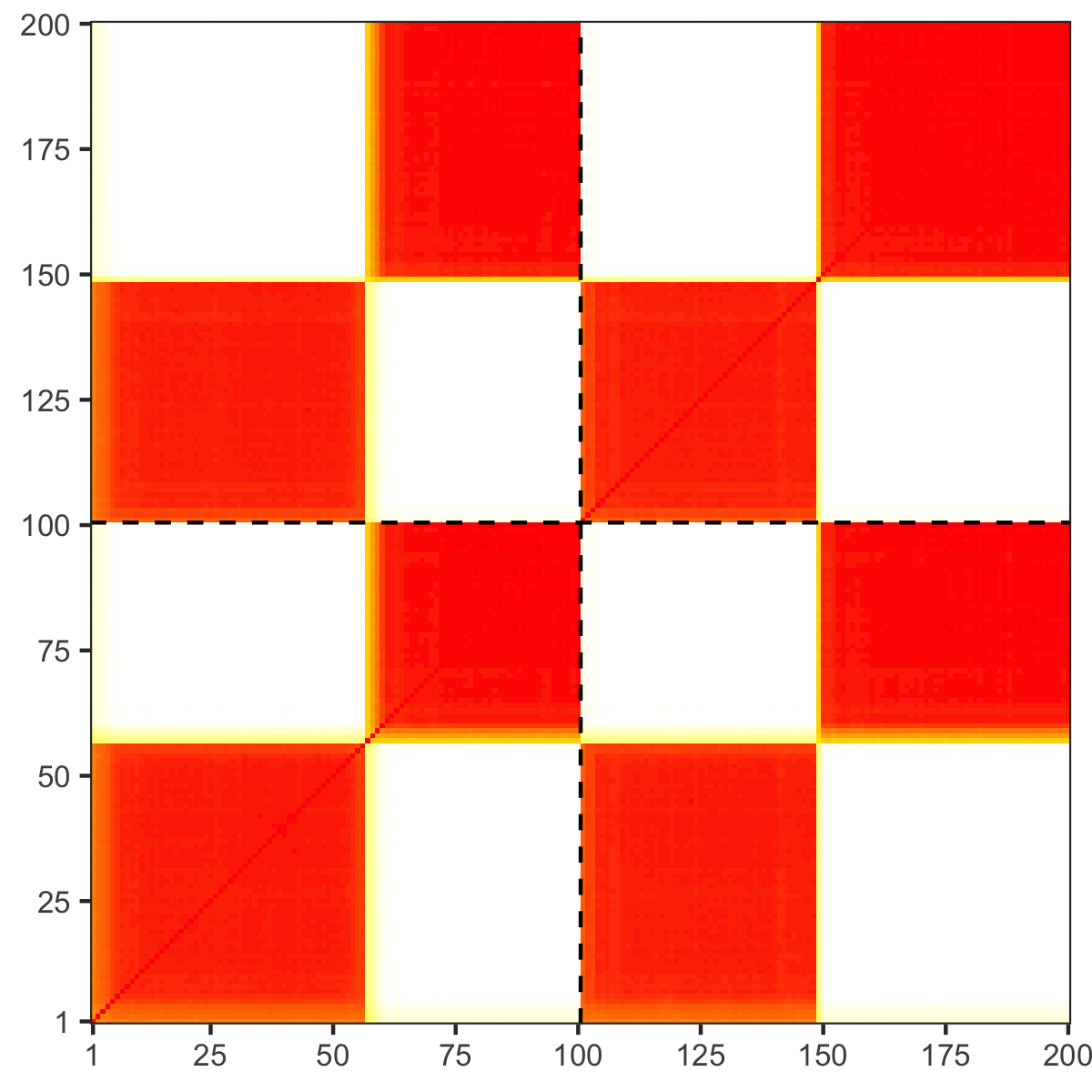}
					\end{subfigure}
					\begin{subfigure}{3cm}
						\includegraphics[width=1.15\textwidth,height=3cm]{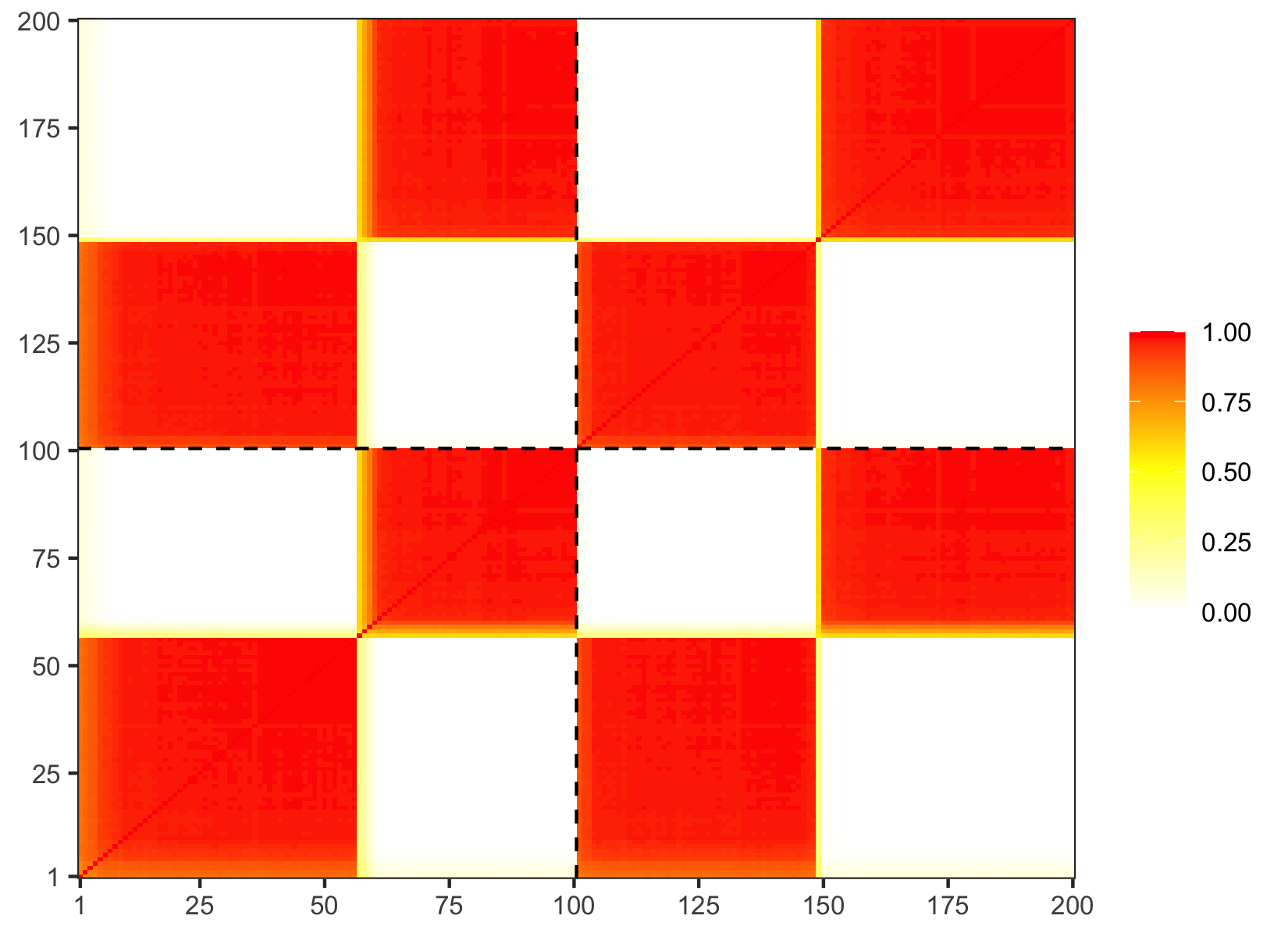}
					\end{subfigure}\\
					\vspace*{0.1cm}
					\rotatebox[origin=c]{90}{\makebox[3cm]{Scenario \RNum{2}}}
					\rotatebox[origin=c]{90}{\makebox[3cm]{ Pop 1 \hspace*{0.15cm} Pop 2}}
					\begin{subfigure}{3cm}
						\includegraphics[width=\textwidth,height=3cm]{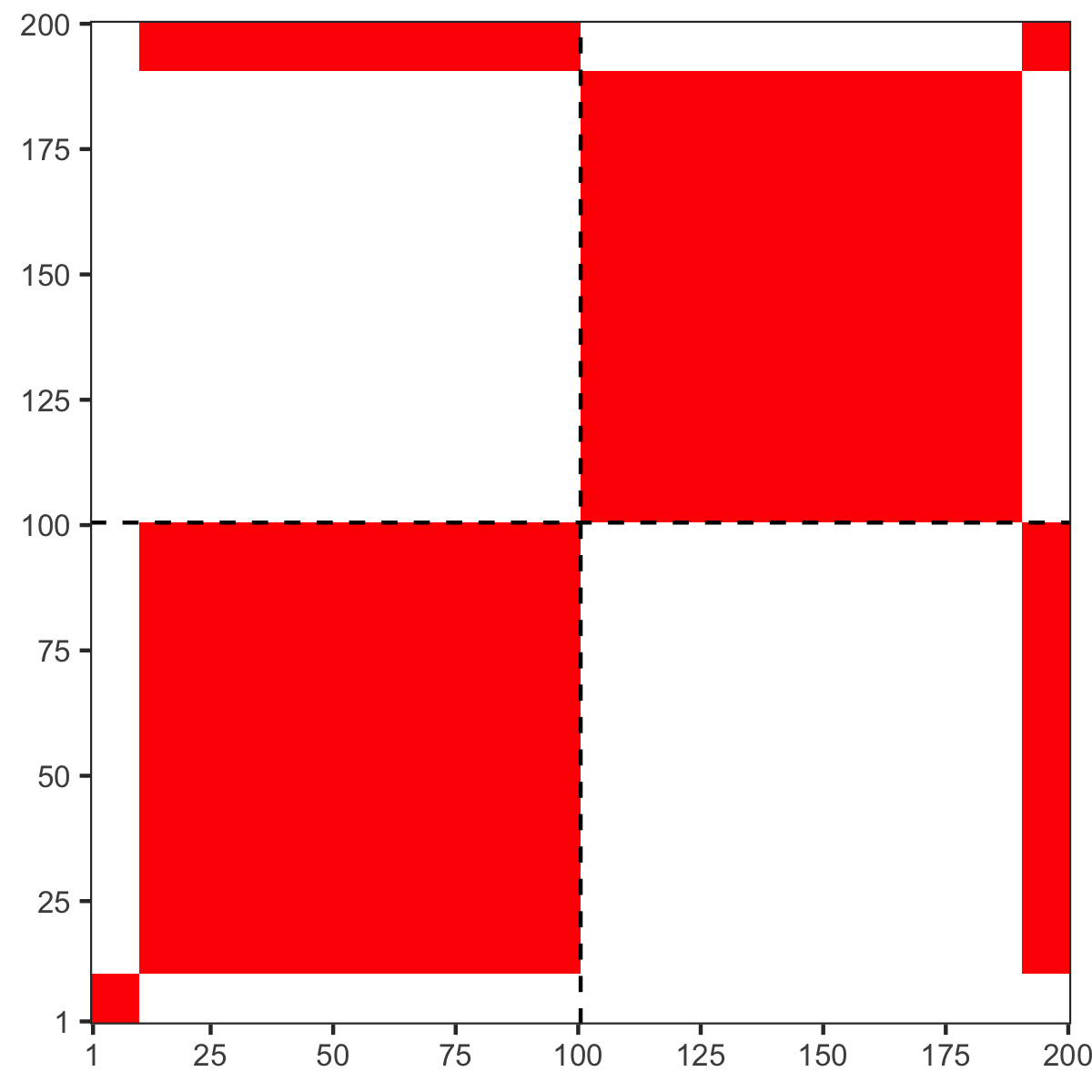}
					\end{subfigure}
					\begin{subfigure}{3cm}
						\includegraphics[width=\textwidth,height=3cm]{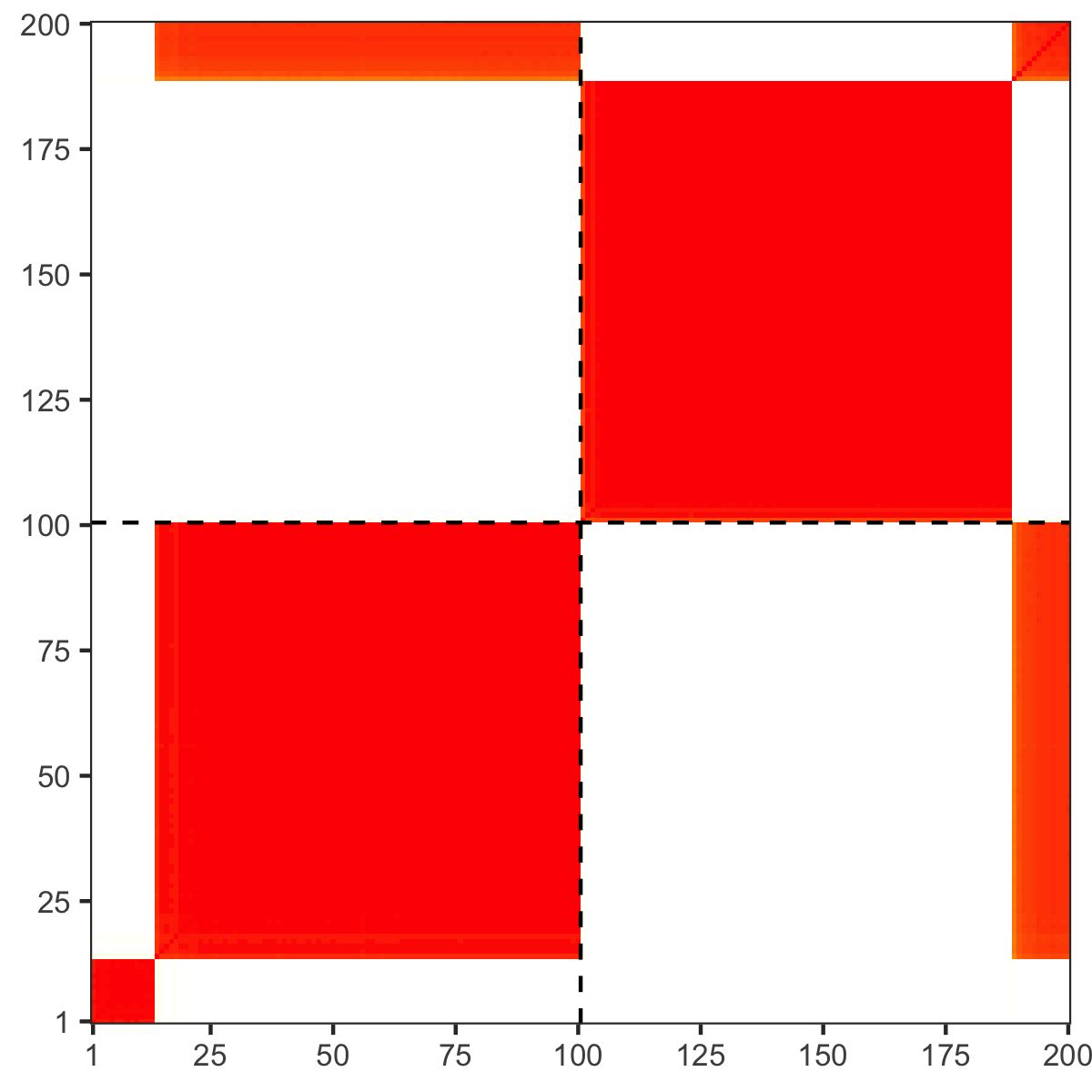}
					\end{subfigure}
					\begin{subfigure}{3cm}
						\includegraphics[width=1.15\textwidth,height=3cm]{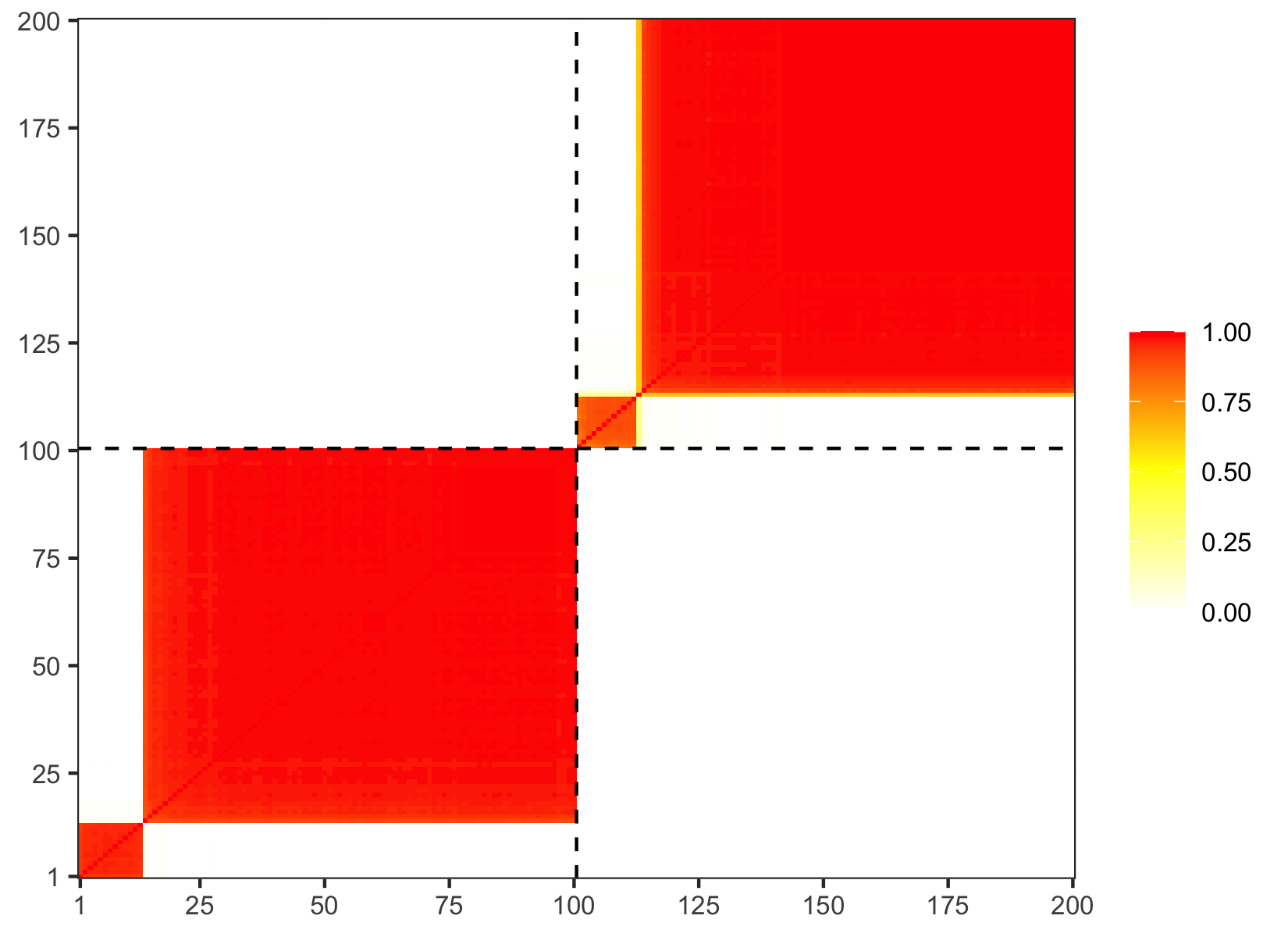}
					\end{subfigure}\\
					\vspace*{0.1cm}
					\rotatebox[origin=c]{90}{\makebox[3cm]{Scenario \RNum{3}}}
					\rotatebox[origin=c]{90}{\makebox[3cm]{ Pop 1 \hspace*{0.15cm} Pop 2}}
					\begin{subfigure}{3cm}
						\includegraphics[width=\textwidth,height=3cm]{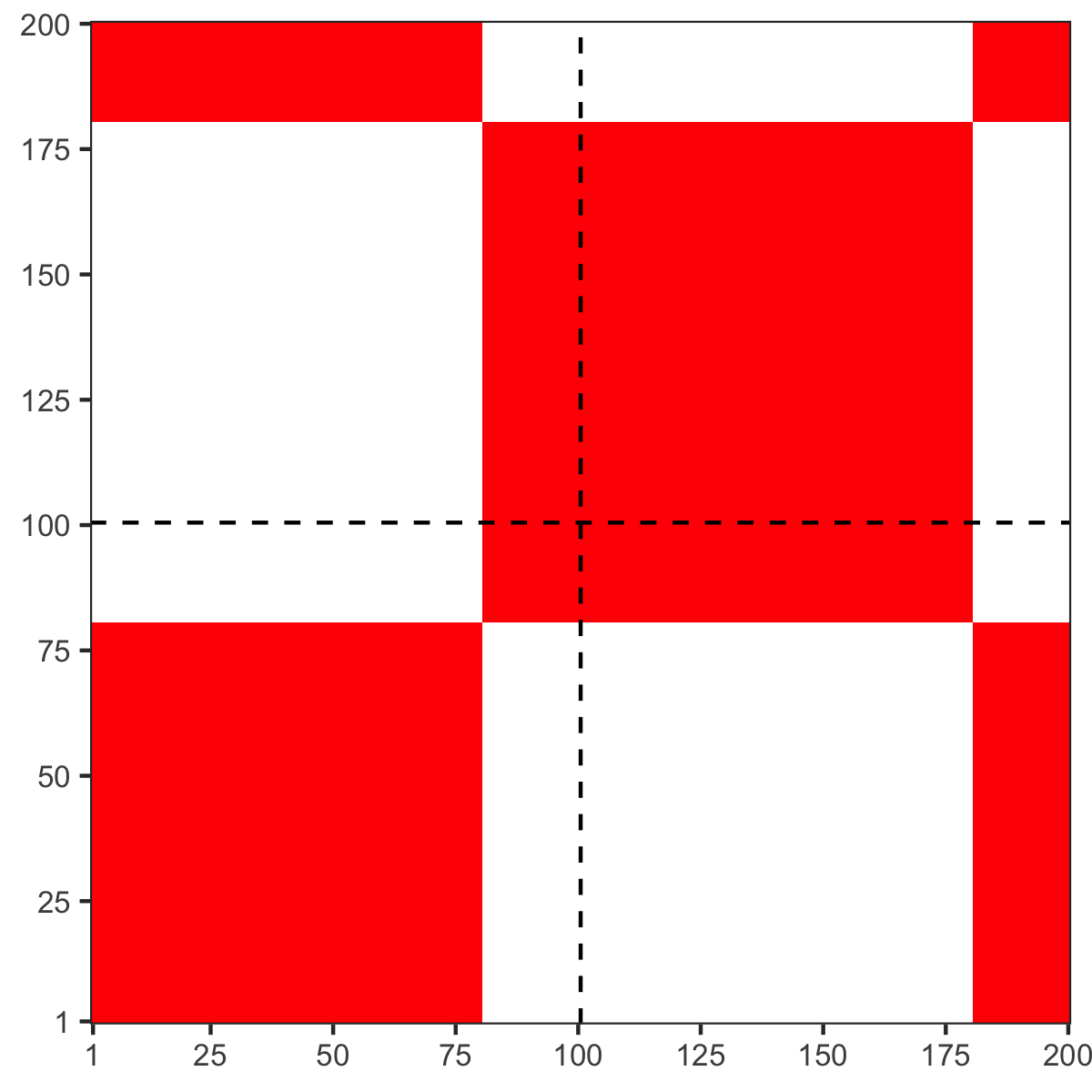}
					\end{subfigure}
					\begin{subfigure}{3cm}
						\includegraphics[width=\textwidth,height=3cm]{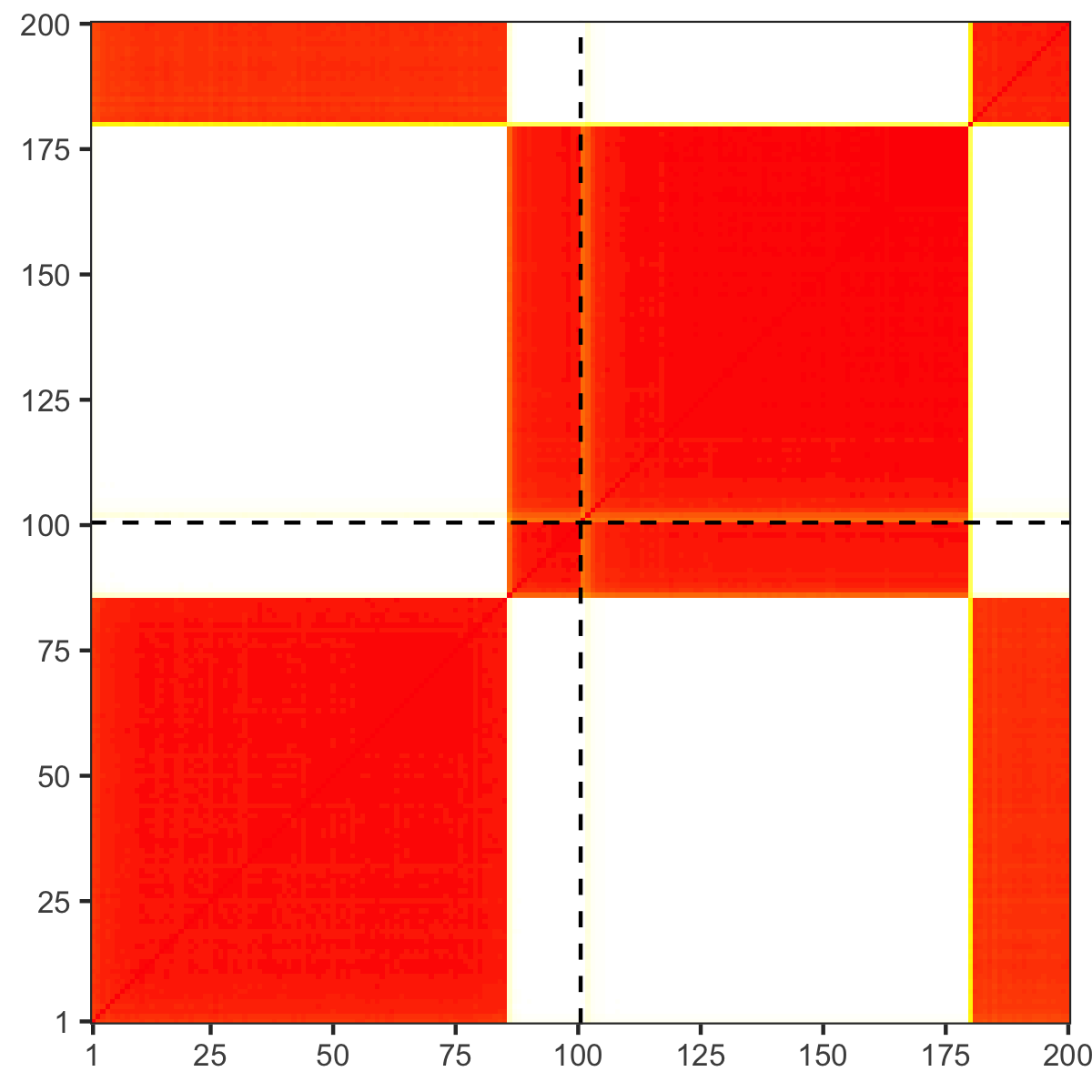}
					\end{subfigure}
					\begin{subfigure}{3cm}
						\includegraphics[width=1.15\textwidth,height=3cm]{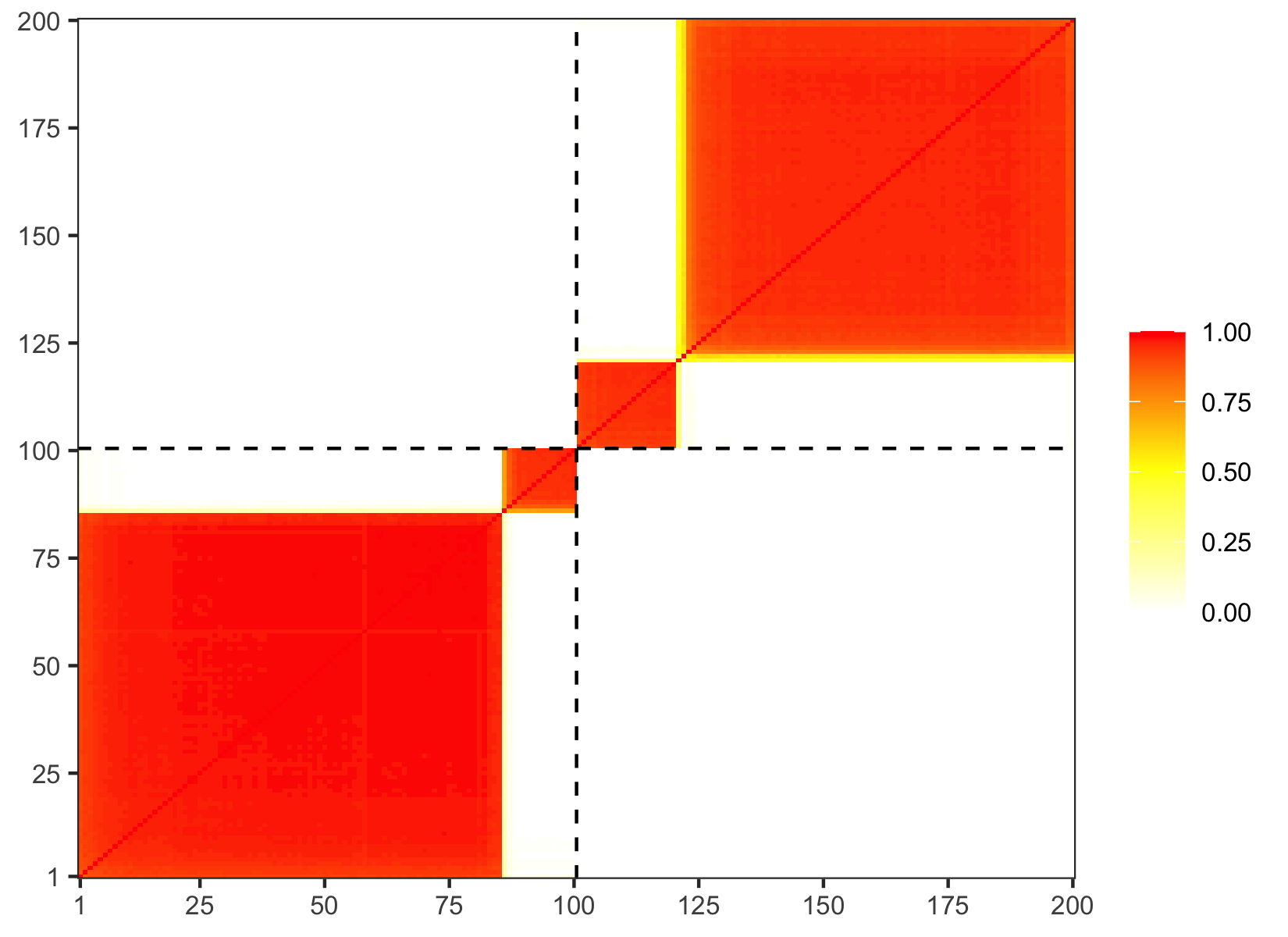}
					\end{subfigure}\\
					\caption{Heatmaps of the true and estimated posterior probability of co-clustering of observations, ordered by population memberships, under the HHDP and the NDP models, for the three different scenarios in Section~\ref{sec:two_populations}.
						\label{fig:Probs}}
				\end{figure}

				Inference is based on $10\,000$ iterations with the first half discarded as burn-in. As for the output, besides obtaining density estimates for the two populations we also determine the point estimate of the clustering of observations that minimizes the variation of information (VI) loss function. See \cite{Meila2007} and \cite{Wade2018a} for detailed discussions on VI and point summaries of probabilistic clustering.
				Additionally, we estimate the probability that observations co-cluster, namely $\P(\zeta_{j,i}=\zeta_{j^\prime,i^\prime} \mid \bm{X})$ through the average over MCMC draws
				\begin{equation*}
					\dfrac{\sum_{b=1}^B \indic\{\zeta_{j,i}^{b}=\zeta_{j^\prime,i^\prime}^{b}\}}{B},
				\end{equation*}
				where $B$ is the number of MCMC iterations.
				These are visualized through heatmaps as in Fig.~\ref{fig:Probs}, 
				with colors ranging from white, if the probability is $0$, to dark red, if the probability is $1$. 
				Our analysis is completed by reporting the estimated distributions of the numbers of mixture components in each scenario.
				
\begin{figure}[!h]
					\makebox[\linewidth]{\hspace*{6cm} NDP \hspace*{6.4cm} HHDP \hspace*{5cm}}	\makebox[\linewidth]{\hspace*{1.6cm} Pop 1 \hspace*{2.2cm} Pop 2 \hspace*{2.8cm} Pop 1 \hspace*{2.2cm} Pop 2 \hspace*{0.5cm}}
					\centering
					\rotatebox[origin=c]{90}{\makebox[3cm]{Scen \RNum{3}}}
					\begin{subfigure}{6.7cm}
						\includegraphics[width=\linewidth, height=3cm]{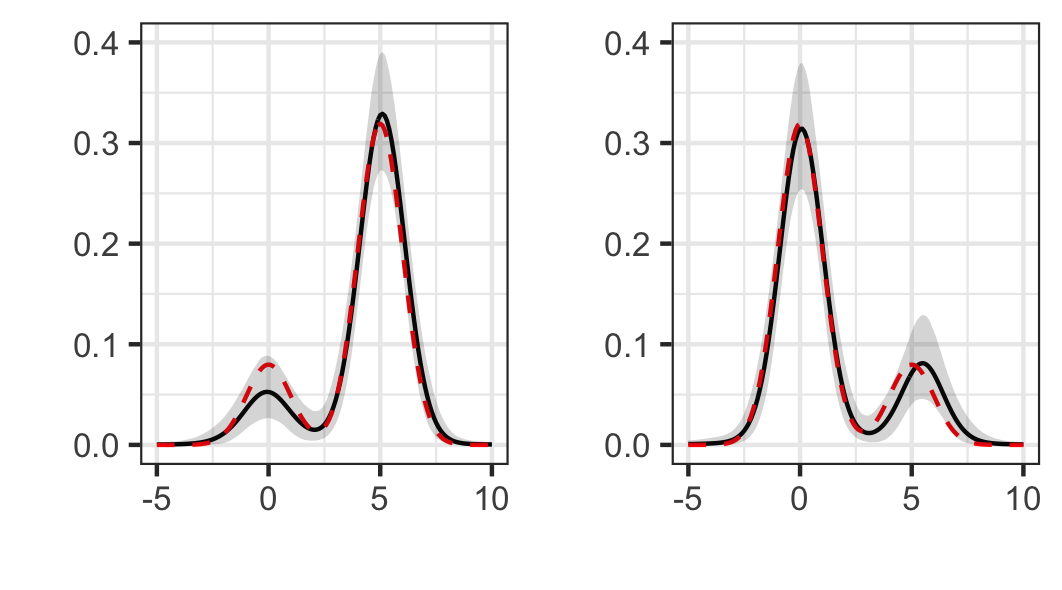}
					\end{subfigure}\hspace*{0.5cm}
					\begin{subfigure}{6.7cm}
						\includegraphics[width=\linewidth, height=3cm]{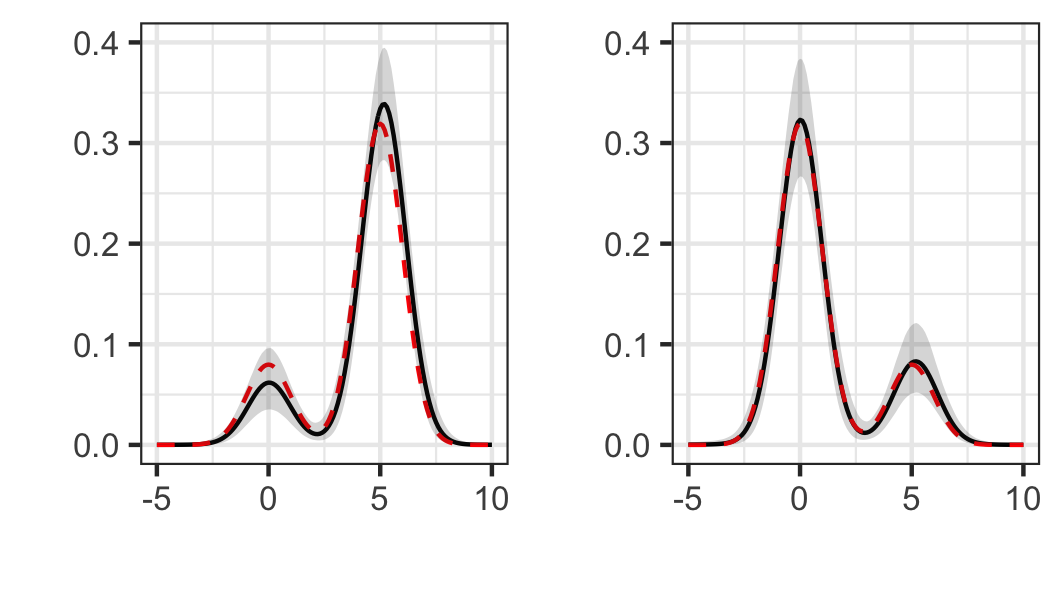}
					\end{subfigure}
					\caption{True (dashed lines), posterior mean (solid lines) densities and 95\% point-wise posterior credible intervals (shaded gray) estimated under the third scenario. \label{fig:Dens}}
				\end{figure}

				As expected, both models yield accurate estimates of the true densities in all scenarios. In \cref{fig:Dens} we report the true and estimated models under the third scenario.
				In terms of clustering, in the first scenario both models correctly cluster together the two populations, thus degenerating to the exchangeable case as they should. However, in the second and third scenarios the NDP makes the two samples $\bm{X}_1$ and $\bm{X}_2$ independent, therefore preventing borrowing of information across the two populations. As the distributions have a shared component, the only way for the NDP to recover correctly the true densities is by missing such a component. Had it been detected, the density estimates of the two populations would have been equal and, thus, far from the truth.
				The point estimate of the observations' clustering 
				in \cref{tab:VI}, the heatmaps of the posterior co-clustering probabilities  in \cref{fig:Probs} and the posterior distributions of the overall number of occupied components in Table \ref{tab:NComp} showcase the theoretical findings, namely 
				that the NDP in the second and third scenarios cannot learn the shared components. Hence, it overestimates the total number of occupied components and does not cluster observations across populations. In contrast, the HHDP model is able to cluster observations across populations, learns the shared components and borrows information also when the model does not degenerate to the exchangeable case. 

				\begin{table}[!h]
					\begin{footnotesize}
						\centering
						\begin{tabular}{c|l|rrrrrrrrrr}
							\multirow{2}{*}{} & \multirow{2}{*}{} & \multicolumn{10}{c}{Overall number of components}\\
							\cline{3-12}
							Scen &	Model		& 1 &	2 				& 3 & 4 & 5 & 6 & 7 & 8 & 9 & $\ge$10\\
							\hline
							\multirow{2}{*}{ \RNum{1}}  & NDP	& 0 & 0.4090 & 0.3615 & 0.1647 & 0.0492 & 0.0136 & 0.0020 & 0 & 0 & 0 \\
							\cline{2-12}
							& HHDP	& 0 & 0.5374 & 0.3743 & 0.0788 & 0.0080 & 0.0016 & 0 & 0 & 0 & 0 \\
							\hline
							\multirow{2}{*}{ \RNum{2}}  & NDP	& 0 & 0 & 0 & 0.2959 & 0.3906 & 0.2151 & 0.0700 & 0.0256 & 0.0024 & 0.0004 \\
							\cline{2-12}
							& HHDP	& 0 & 0 & 0.5742 & 0.3339 & 0.0796 & 0.0116 & 0.0008 & 0 & 0 & 0\\ 
							\hline
							\multirow{2}{*}{ \RNum{3}}  & NDP    & 0 & 0 & 0 & 0.1331 & 0.3055 & 0.2947 & 0.1743 & 0.0608 & 0.0232 & 0.0084 \\ 
							\cline{2-12}
							& HHDP	& 0 & 0.5010 & 0.3966 & 0.0856 & 0.0164 & 0.0004 & 0 & 0 & 0 & 0\\ 
						\end{tabular}
						\caption{Posterior distributions of the number of overall occupied components estimated with the two models under different scenarios.\label{tab:NComp}}
					\end{footnotesize}
				\end{table}
				\begin{table}[!h]
					\begin{footnotesize}
						\centering
						\begin{tabular}{c|rr|rr|rrrr|rrr|rrrr|rr}
							&\multicolumn{4}{c|}{Scenario \RNum{1}} & \multicolumn{7}{c|}{Scenario  \RNum{2}}& \multicolumn{6}{c}{Scenario  \RNum{3}}\\
							\cline{2-18}
							&\multicolumn{2}{c|}{NDP} & \multicolumn{2}{c|}{HHDP} & \multicolumn{4}{c|}{NDP} & \multicolumn{3}{c|}{HHDP} & \multicolumn{4}{c|}{NDP} & \multicolumn{2}{c}{HHDP} \\
							\hline
							Population		& 1 & 2 & 1 & 2 & 1 & 2 & 3 & 4 & 1 & 2 & 3 & 1 & 2 & 3 & 4 & 	1 & 2\\ 
							\hline
							1					 & 56 &    44 & 	56 &    44 & 	87 &    13 &     0 &     0 & 87 &    13 &     0 &85 &    15 &     0 &     0 &	85 &    15\\ 
							2					 & 48 &    52 & 48 &    52 & 	0 &     0 &    88 &    12 & 12 &     0 &    88 &0 &     0 &    80 &    20 &21 &    79\\ 
							\hline
						\end{tabular}
						\caption{Frequencies of observations in the two populations allocated to the point estimate of the clustering that minimizes the VI loss with the two models under different scenarios.\label{tab:VI}}
					\end{footnotesize}
				\end{table}
				
				\subsection{Inference with more than two populations}\label{subsec:syn_data_more}
				
\begin{figure}[!h]
				
				\makebox[\linewidth]{\hspace*{0.7cm} Pop 1 \hspace*{3.3cm} Pop 2 \hspace*{3.3cm} Pop 3 \hspace*{3.3cm} Pop 4 }
					\centering
					\includegraphics[width=\linewidth, height=3cm]{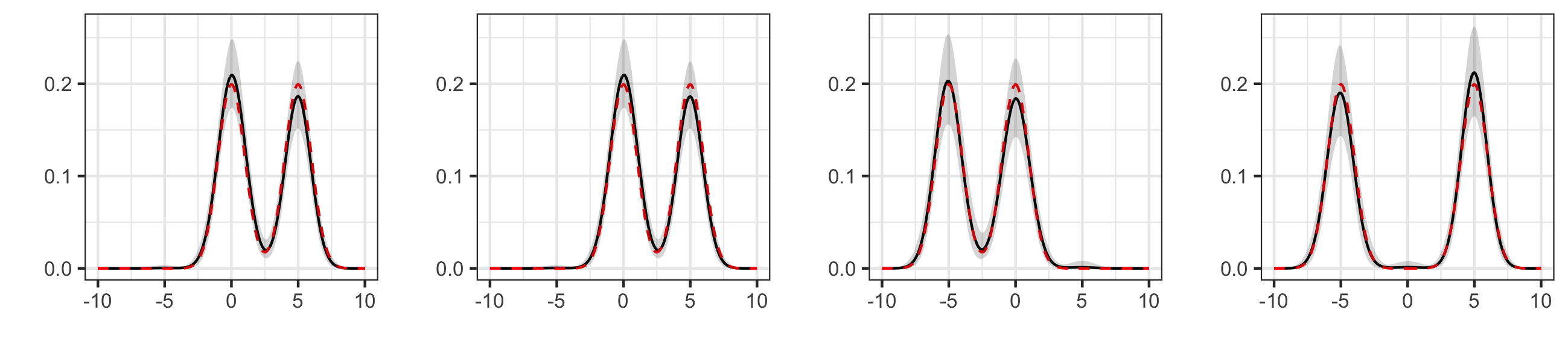}
					\caption{True (dashed lines), posterior mean (solid lines) densities and 95\% point-wise posterior credible intervals (shaded gray) estimated under the fourth scenario. \label{fig:DensMore}}
				\end{figure}

\begin{figure}[!h]
					\centering
					\includegraphics[height=5cm]{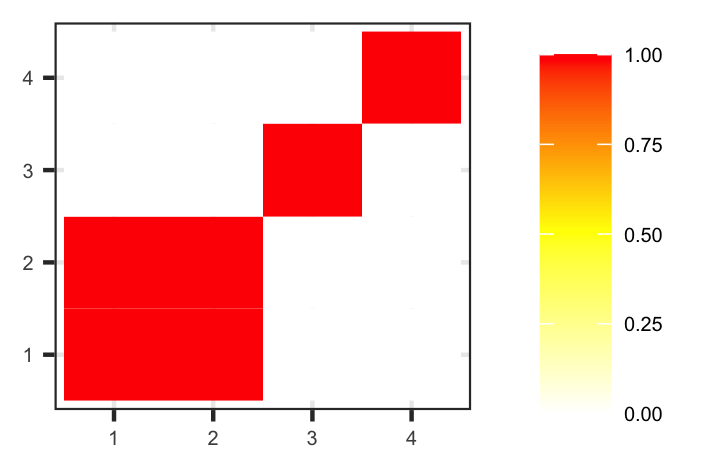}
					\captionof{figure}{Heatmap of the estimated posterior probabilities of co-clustering of the population estimated with the HHDP mixture model under the fourth scenario in Section~\ref{subsec:syn_data_more}. \label{fig:HeatmapMore}}
				\end{figure}

				Here we consider $J=4$ populations and deal with the same scenario discussed in \cite{beraha2021semi}. 
				More precisely, we simulate independently across populations $I_j=100$ (for $j=1,\ldots,4$) observations as follows 
				\begin{align*}
					X_{1,i} \ed X_{2,i} \simiid 0.5 \Norm(0,1) + 0.5 \Norm(5,1) \quad
					X_{3,i} \simiid 0.5 \Norm(0,1) + 0.5 \Norm(-5,1) \quad  X_{4,i} \simiid 0.5 \Norm(-5,1) + 0.5 \Norm(5,1)	
				\end{align*}
				Our prior corresponds to a Gaussian mixture model with the same 
				specification for the HHDP 
				used in the previous Section with $J=2$ population. \cref{fig:DensMore} shows that the HHDP mixture model is able to recover the data generating densities also in this scenario.
				In terms of clustering of populations the point estimate that minimizes the VI loss coincides with the data generating truth.
				\cref{fig:HeatmapMore} reports the heatmaps of the posterior co-clustering probabilities of the four populations that show little uncertainty around the correct point estimate, e.g.\ the estimated probability that populations $1$ and $2$ are correctly clustered together is $0.9858$.

				Finally, the point estimate of the observations' clustering 
				in \cref{tab:VIMore} shows the HHDP model is able to cluster observations across populations, learns the shared components and borrows information also when there are more than two populations. 
				
				\begin{table}[!h]
					\centering
					\begin{tabular}{c|rrr}
						observational cluster & 1 & 2 & 3 \\ 
						\hline
						Pop 1 &    53 &    47 &     0 \\ 
						Pop 2 &    56 &    44 &     0 \\ 
						Pop 3 &    48 &     0 &    52 \\ 
						Pop 4 &     0 &    52 &    48 \\ 
						\hline
					\end{tabular}
					\captionof{table}{Frequencies of observations in the four populations allocated to the point estimate of the clustering that minimizes the VI loss with HHDP under the fourth scenario. \label{tab:VIMore}}
				\end{table}
				
				\subsection{Collaborative perinatal project data}
				\label{subsec:real_data_appl}
				A multi-center application is the focus of this section.
				We consider a data set from the Collaborative Perinatal Project (CPP), a large prospective epidemiologic study conducted from 1959 to 1974.
				Pregnant women were enrolled in 12 hospitals between 1959 and 1966 and were followed over time.
				Among several pre--pregnancy measurements, we focus on the birth weight $X_{j,i}$ for non-smoking woman $i$ in center $j$. We assume the following Gaussian mixture model:
				\begin{align*}
					X_{j,i} \mid \mu_{j,i}, \sigma_{j,i}  &\indsim \Norm(\mu_{j,i}, \sigma_{j,i})  &(i=1,\ldots,I_j, \quad j=1;\ldots,12),\\
					\mu_{j,i}, \sigma_{j,i} \mid G_j &\indsim G_j   &(i=1,\ldots,I_j, \quad j=1;\ldots,12).
				\end{align*}
				The same HHDP prior used for the previous synthetic data is placed the vector of random distributions.
				This model specification is coherent with what is suggested by \cite{Dunson2010d} for the CPP data.
				Indeed, it is known that the pregnancy outcome can vary substantially for women from different ethnicity and socioeconomic groups.
				Therefore, we specify a model allowing to capture differences between the centers since different groups of hospitals can serve different women. \cite{Canale2019} provide further analysis of the CPP data.
				
\begin{figure}[!h]
					\centering
					\includegraphics[width= 0.7\linewidth, height=4cm ]{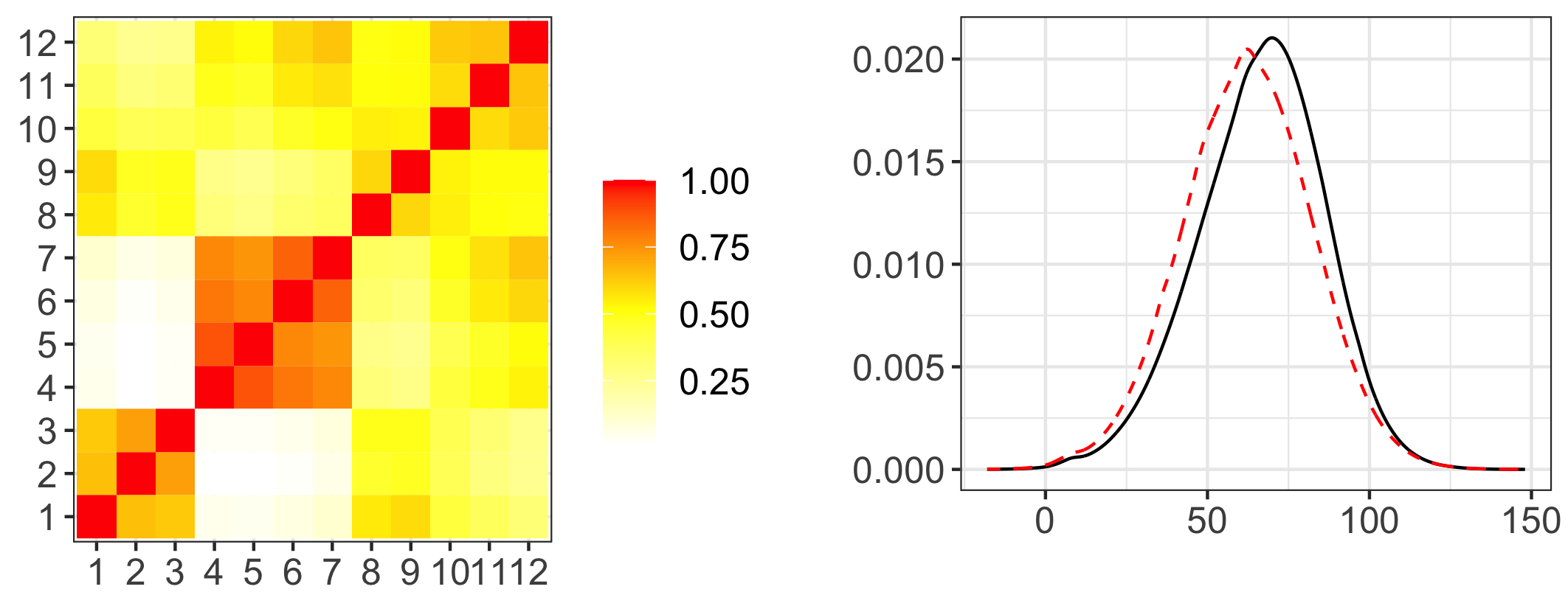}
					\caption{Heatmap of the estimated posterior probability of co-clustering of hospitals and estimated population cluster-specific posterior densities for the CPP data.\label{fig:ProbsCPPDis}}
					\label{fig:HyDPClusDisHeatmap}
				\end{figure}

				The heatmap of the co-clustering posterior probability for the 12 hospitals is shown in \cref{fig:ProbsCPPDis}.
				Such probabilities imply that the clustering point estimate of the hospitals that minimizes the VI loss has two blocks and, in the same figure, the mean posterior densities associated with the two clusters are reported.
				Given the partition of the hospitals, the posterior mean densities are evaluated based on all patients belonging to hospitals in each of the two partition groups.
				The heatmap shows the posterior distribution of the clustering of the hospitals and can be used to perform uncertainty quantification.
				As expected, the lack of well-separated data generating mixtures of Gaussians entails more uncertainty around the point estimate of the clustering of the populations with respect to the numerical experiments.
				However, the heatmap shows that the point estimate of the clustering of distributions is a reliable summary.
				More precisely, the point estimate that minimizes the VI loss entails that the first cluster of hospitals includes the hospitals with (reordered) labels $1,2,3$: these are well-separated from the remaining hospitals according to the posterior probabilities of co-clustering in the heatmap.
				The heatmap shows also that another meaningful point estimate of the clustering of the hospitals is the finer partition $\{\{1,2,3\}, \{4,5,6,7\}, \{8,9,10,11,12\} \}$.
				However, the VI loss suggests a more parsimonious clustering of the hospitals in two blocks, that is $\{\{1,2,3\}, \{4,5,6,7,8,9,10,11,12\} \}$.
				Note that in the second cluster of hospitals (red dashed density in \cref{fig:ProbsCPPDis}) the distribution of the birth weights is slightly shifted on lower values and the two mean densities are similar in the two clusters of populations.
				Coherently the proposed model allows to borrow information across clusters of hospitals for estimating the posterior mean densities of the birth weights.
				Furthermore the model can be used to identify clusters of women shared in the two different clusters of hospitals. 
				Indeed, Table~\ref{tab:SharedNotSharedHyDP_FinDirNDPHDPDataCPP} shows that some clusters of observations are shared across 
				different clusters of hospitals, thus allowing the borrowing of information for estimating the densities of the birth weights in the two groups.

				\begin{table}[!h]
					\centering
					\begin{tabular}{l|rrrrrr}
						number of observational clusters & 0 & 1 & 2 & 3 & 4 & 5 \\ 
						\hline
						only in the second cluster of hospitals & 0.3530 & 0.3670 & 0.2040 & 0.0640 & 0.0100 & 0.0020 \\ 
						only in the first cluster of hospitals & 0.7750 & 0.1850 & 0.0340 & 0.0060 & 0 & 0 \\ 
						shared across clusters of hospitals & 0 & 0.1680 & 0.4800 & 0.2660 & 0.0780 & 0.0080 \\ 
						\hline
					\end{tabular}
					\caption{ Posterior distributions of the number of clusters shared and not shared across the two clusters of hospitals.} 
					\label{tab:SharedNotSharedHyDP_FinDirNDPHDPDataCPP}
				\end{table}

				\section{Discussion}\label{sec:Conclusion}
				As highlighted in the recent literature, NDP mixture models are often not an appropriate tool for clustering simultaneously population distributions and observations. 
				In contrast, the HHDP, overcomes the issues plaguing the NDP, while preserving tractability and clustering flexibility even when the number of populations $J$ is larger than $2$. We have further devised sampling schemes allowing for efficient inference and prediction.  
				This work paves the way for future intriguing research directions that we plan to address in forthcoming work. First, it is natural to move beyond DPs and consider models based on alternative discrete nonparametric priors, such as the Pitman-Yor process and normalized completely random measures, while studying the induced clustering.
				The characterization of the HHDP in terms of the induced random partition suggests a nice connection of our work with recent and exciting advances on time-dependent random partition models such as those proposed, e.g., in  \cite{page2019dependent} and \cite{zanini2019bayesian}. Indeed, these papers define a general framework that can be tailored to HHDP priors for generating time-dependent models suited for analyzing, e.g., longitudinal data thus allowing for the investigation of the joint evolution of observational and distributional clustering through time. 
				The theory we have developed in Sections~3 and 4 provides the necessary tools for successfully carrying out such a program.  Moreover, the general composition scheme, where we have embedded the HHDP, seems a promising and effective approach for addressing other interesting inferential problems, beyond density estimation and clustering. 
				Finally, the general scheme that we have introduced in \eqref{eq:composition_prior} seems an appropriate specification for capturing the inherent complexity and heterogeneity of data that arise when drawing predictions with multivariate species sampling models and when performing inferences in survival and functional data analysis. These will be the object of forthcoming work.
			
\bibliography{HHDP}

\begin{thebibliography}{}

\bibitem[Agrawal et~al., 2013]{Agrawal2013}
Agrawal, P., Tekumalla, L.~S., and Bhattacharya, I. (2013).
\newblock {Nested Hierarchical Dirichlet process for nonparametric entity-topic
  analysis}.
\newblock In {\em Jt. Eur. Conf. Mach. Learn. Knowl. Discov. Databases}, volume
  8189 LNAI, pages 564--579.

\bibitem[Argiento et~al., 2020]{argiento}
Argiento, R., Cremaschi, A., and Vannucci, M. (2020).
\newblock Hierarchical normalized completely random measures to cluster grouped
  data.
\newblock {\em J. Amer. Statist. Assoc.}, 115(529):318--333.

\bibitem[Balocchi et~al., 2021]{balocchi2021clustering}
Balocchi, C., George, E.~I., and Jensen, S.~T. (2021).
\newblock Clustering areal units at multiple levels of resolution to model
  crime incidence in {P}hiladelphia.
\newblock {\em Preprint arXiv: 2112.02059}.

\bibitem[Bassetti et~al., 2020]{bassetti}
Bassetti, F., Casarin, R., and Rossini, L. (2020).
\newblock Hierarchical species sampling models.
\newblock {\em Bayesian Anal.}, 15(3):809--838.

\bibitem[Beraha et~al., 2021]{beraha2021semi}
Beraha, M., Guglielmi, A., and Quintana, F.~A. (2021).
\newblock The semi-hierarchical dirichlet process and its application to
  clustering homogeneous distributions.
\newblock {\em Bayesian Anal.}, 16(4):1187--1219.

\bibitem[Camerlenghi et~al., 2019b]{Camerlenghi2019c}
Camerlenghi, F., Dunson, D.~B., Lijoi, A., Pr{\"{u}}nster, I., and
  Rodr{\'{i}}guez, A. (2019b).
\newblock {Latent nested nonparametric priors}.
\newblock {\em Bayesian Anal.}, 14:1303--1356.
\newblock (With discussion).

\bibitem[Camerlenghi et~al., 2019a]{Camerlenghi2019}
Camerlenghi, F., Lijoi, A., Orbanz, P., and Pr{\"{u}}nster, I. (2019a).
\newblock {Distribution theory for hierarchical processes}.
\newblock {\em Ann. Stat.}, 47(1):67--92.

\bibitem[Camerlenghi et~al., 2018]{Camerlenghi2018e}
Camerlenghi, F., Lijoi, A., and Pr{\"{u}}nster, I. (2018).
\newblock {Bayesian nonparametric inference beyond the Gibbs-type framework}.
\newblock {\em Scand. J. Stat.}, 45(4):1062--1091.

\bibitem[Camerlenghi et~al., 2021]{camerlenghi2021survival}
Camerlenghi, F., Lijoi, A., and Pr{\"u}nster, I. (2021).
\newblock Survival analysis via hierarchically dependent mixture hazards.
\newblock {\em Ann. Stat.}, 49(2):863--884.

\bibitem[Canale et~al., 2019]{Canale2019}
Canale, A., Corradin, R., and Nipoti, B. (2019).
\newblock {Importance conditional sampling for Bayesian nonparametric
  mixtures}.
\newblock {\em Preprint at arXiv: 1906.08147}.

\bibitem[Christensen and Ma, 2020]{christ_ma2020}
Christensen, J. and Ma, L. (2020).
\newblock A {B}ayesian hierarchical model for related densities using {P}\'olya
  trees.
\newblock {\em J. R. Stat. Soc. Ser. B}, 82(1):127--153.

\bibitem[Cifarelli and Regazzini, 1978]{Cifarelli1978c}
Cifarelli, D.~M. and Regazzini, E. (1978).
\newblock {Problemi statistici non parametrici in condizioni di scambiabilita
  parziale e impiego di medie associative}.
\newblock {\em Quaderni Istituto Matematica Finanziaria dell'Universita di
  Torino}.

\bibitem[Denti et~al., 2021]{denti2021common}
Denti, F., Camerlenghi, F., Guindani, M., and Mira, A. (2021).
\newblock A common atom model for the {B}ayesian nonparametric analysis of
  nested data.
\newblock {\em J. Am. Stat. Assoc.}, (in press).

\bibitem[Dunson, 2010]{Dunson2010d}
Dunson, D.~B. (2010).
\newblock {Nonparametric Bayes applications to biostatistics}.
\newblock In {\em Bayesian Nonparametrics}, pages 223--273. Cambridge
  University Press.

\bibitem[Escobar, 1994]{Escobar1994}
Escobar, M.~D. (1994).
\newblock {Estimating normal means with a Dirichlet process prior}.
\newblock {\em J. Am. Stat. Assoc.}, 89(425):268--277.

\bibitem[Escobar and West, 1995]{Escobar1995}
Escobar, M.~D. and West, M. (1995).
\newblock {Bayesian density estimation and inference using mixtures}.
\newblock {\em J. Am. Stat. Assoc.}, 90(430):577--588.

\bibitem[Ewens, 1990]{Ewens1990}
Ewens, W.~J. (1990).
\newblock {Population genetics theory - the past and the future}.
\newblock In {\em Math. Stat. Dev. Evol. Theory}, pages 177--227. Springer,
  Dordrecht.

\bibitem[{\swap{Finetti}{de }}, 1938]{DeFinetti1938}
{\swap{Finetti}{de }}, B. (1938).
\newblock {Sur la condition d'equivalence partielle}.
\newblock {\em Actual. Sci. Ind.}, 739:5--18.

\bibitem[Foti and Williamson, 2015]{Foti2015}
Foti, N.~J. and Williamson, S.~A. (2015).
\newblock {A survey of non-exchangeable priors for Bayesian nonparametric
  models}.
\newblock {\em IEEE Trans. Pattern Anal. Mach. Intell.}, 37(2):359--371.

\bibitem[Ghosal and {van der Vaart}, 2017]{Ghosal2017a}
Ghosal, S. and {van der Vaart}, A. (2017).
\newblock {\em {Fundamentals of nonparametric Bayesian inference}}.
\newblock Cambridge University Press.

\bibitem[Ishwaran and James, 2001]{Ishwaran2001}
Ishwaran, H. and James, L.~F. (2001).
\newblock {Gibbs sampling methods for stick-breaking priors}.
\newblock {\em J. Am. Stat. Assoc.}, 96(453):161--173.

\bibitem[Ishwaran and Zarepour, 2002]{Ishwaran2002a}
Ishwaran, H. and Zarepour, M. (2002).
\newblock {Exact and approximate sum representations for the Dirichlet
  process}.
\newblock {\em Can. J. Stat.}, 30(2):269--283.

\bibitem[James, 2008]{james2008discussion}
James, L. (2008).
\newblock Discussion of {N}ested {D}irichlet {P}rocess paper by {R}odr\'iguez,
  {D}unson and {G}elfand.
\newblock {\em J. Am. Stat. Assoc.}, 483:1131.

\bibitem[Kallenberg, 2005]{Kallenberg2006}
Kallenberg, O. (2005).
\newblock {\em {Probabilistic symmetries and invariance principles}}.
\newblock Springer.

\bibitem[Lo, 1984]{Lo1984a}
Lo, A.~Y. (1984).
\newblock {On a class of Bayesian nonparametric estimates: I. density
  estimates}.
\newblock {\em Ann. Stat.}, 12(1):351--357.

\bibitem[MacEachern, 1999]{MacEachern1999}
MacEachern, S.~N. (1999).
\newblock {Dependent nonparametric processes}.
\newblock In {\em ASA Proc. Sect. Bayesian Stat. Sci.}, pages 50--55.

\bibitem[MacEachern, 2000]{MacEachern2000}
MacEachern, S.~N. (2000).
\newblock {Dependent Dirichlet processes}.
\newblock Technical report, The Ohio State University.

\bibitem[Meilǎ, 2007]{Meila2007}
Meilǎ, M. (2007).
\newblock {Comparing clusterings-an information based distance}.
\newblock {\em J. Multivar. Anal.}, 98(5):873--895.

\bibitem[Muliere and Tardella, 1998]{Muliere1998}
Muliere, P. and Tardella, L. (1998).
\newblock {Approximating distributions of random functionals of
  Ferguson-Dirichlet priors}.
\newblock {\em Can. J. Stat.}, 26(2):283--297.

\bibitem[M\"{u}ller et~al., 2011]{MQR}
M\"{u}ller, P., Quintana, F., and Rosner, G.~L. (2011).
\newblock A product partition model with regression on covariates.
\newblock {\em J. Comput. Graph. Statist.}, 20(1):260--278.

\bibitem[Neal, 2000]{Neal2000}
Neal, R.~M. (2000).
\newblock {Markov chain sampling methods for Dirichlet process mixture models}.
\newblock {\em J. Comput. Graph. Stat.}, 9(2):249--265.

\bibitem[Page and Quintana, 2016]{Page_Quintana_2016}
Page, G.~L. and Quintana, F.~A. (2016).
\newblock Spatial product partition models.
\newblock {\em Bayesian Anal.}, 11(1):265--298.

\bibitem[Page and Quintana, 2018]{Page_Quintana}
Page, G.~L. and Quintana, F.~A. (2018).
\newblock Calibrating covariate informed product partition models.
\newblock {\em Stat. Comput.}, 28(5):1009--1031.

\bibitem[Page et~al., 2022]{page2019dependent}
Page, G.~L., Quintana, F.~A., and Dahl, D.~B. (2022).
\newblock Dependent modeling of temporal sequences of random partitions.
\newblock {\em J. Comput. Graph. Stat.}, (in press).

\bibitem[Pitman, 2006]{Pitman2006}
Pitman, J. (2006).
\newblock {\em {Combinatorial stochastic processes}}.
\newblock Springer.

\bibitem[Quintana et~al., 2022]{ddp2021_Quintana}
Quintana, F.~A., M\"uller, P., Jara, A., and MacEachern, S.~N. (2022).
\newblock The dependent dirichlet process and related models.
\newblock {\em Stat. Sci.}, (in press).

\bibitem[Roberts and Rosenthal, 2009]{Roberts2009}
Roberts, G.~O. and Rosenthal, J.~S. (2009).
\newblock {Examples of adaptive MCMC}.
\newblock {\em J. Comput. Graph. Stat.}, 18(2):349--367.

\bibitem[Rodr{\'{i}}guez et~al., 2008]{Rodriguez2008a}
Rodr{\'{i}}guez, A., Dunson, D.~B., and Gelfand, A.~E. (2008).
\newblock {The nested Dirichlet process}.
\newblock {\em J. Am. Stat. Assoc.}, 103(483):483--1131.

\bibitem[Sethuraman, 1994]{Sethuraman1994a}
Sethuraman, J. (1994).
\newblock {A constructive definition of Dirichlet priors}.
\newblock {\em Stat. Sin.}, 4(2):639--650.

\bibitem[Soriano and Ma, 2019]{soriano_ma2019}
Soriano, J. and Ma, L. (2019).
\newblock Mixture modeling on related samples by {$\psi$}-stick breaking and
  kernel perturbation.
\newblock {\em Bayesian Anal.}, 14(1):161--180.

\bibitem[Teh and Jordan, 2010]{Teh2010b}
Teh, Y.~W. and Jordan, M.~I. (2010).
\newblock {Hierarchical Bayesian nonparametric models with applications}.
\newblock In {\em Bayesian Nonparametrics}, pages 158--207. Cambridge
  University Press.

\bibitem[Teh et~al., 2006]{Teh2006}
Teh, Y.~W., Jordan, M.~I., Beal, M.~J., and Blei, D.~M. (2006).
\newblock {Hierarchical Dirichlet processes}.
\newblock {\em J. Am. Stat. Assoc.}, 101(476):1566--1581.

\bibitem[Wade and Ghahramani, 2018]{Wade2018a}
Wade, S. and Ghahramani, Z. (2018).
\newblock {Bayesian cluster analysis: point estimation and credible balls}.
\newblock {\em Bayesian Anal.}, 13(2):559--626.

\bibitem[Zanini et~al., 2019]{zanini2019bayesian}
Zanini, C. T.~P., M{\"u}ller, P., Ji, Y., and Quintana, F.~A. (2019).
\newblock A {B}ayesian random partition model for sequential refinement and
  coagulation.
\newblock {\em Biometrics}, 75(3):988--999.

\bibitem[Zuanetti et~al., 2018]{Zuanetti2018}
Zuanetti, D.~A., M{\"{u}}ller, P., Zhu, Y., Yang, S., and Ji, Y. (2018).
\newblock {Clustering distributions with the marginalized nested Dirichlet
  process}.
\newblock {\em Biometrics}, 74(2):584--594.

\end{thebibliography}


\begin{thebibliography}{}

\bibitem[Camerlenghi et~al., 2019a]{Camerlenghi2019}
Camerlenghi, F., Lijoi, A., Orbanz, P., and Pr{\"{u}}nster, I. (2019a).
\newblock {Distribution theory for hierarchical processes}.
\newblock {\em Ann. Stat.}, 47(1):67--92.

\bibitem[Camerlenghi et~al., 2018]{Camerlenghi2018e}
Camerlenghi, F., Lijoi, A., and Pr{\"{u}}nster, I. (2018).
\newblock {Bayesian nonparametric inference beyond the Gibbs-type framework}.
\newblock {\em Scand. J. Stat.}, 45(4):1062--1091.

\end{thebibliography}
\end{document}


\begin{center}
		{\Large
			\textbf{\newline{Supplementary materials:\\
					Flexible clustering via hidden hierarchical Dirichlet priors}}
		}
		\newline
		\\
		Antonio Lijoi \textsuperscript{1},
		Igor Pr\"unster \textsuperscript{1},
		Giovanni Rebaudo\textsuperscript{2}
		\\
		\bigskip
		{\mbox{\bf{1}} \small Department of Decision Sciences and BIDSA, Bocconi University, via R\"ontgen 1, 20136 Milan, Italy}
		\\
		{\mbox{\bf{2}} \small Department of Statistics and Data Sciences, University of Texas at Austin, 
		TX 78712-1823, USA}
	\end{center}
	\date{}

\renewcommand{\refname}{{REFERENCES}}
	
\section{Proofs of the main results}

\subsection{Proof of Proposition \ref{eq:MeanVarCovHyDP}}
Note that $G_1^\ast(A) \mid G_0 \sim \Be(\beta G_0(A), \beta (1-G_0(A)))$ and \mbox{$G_0(A) \sim \Be(\beta_0 H(A), \beta_0 (1-H(A)))$}. Hence, 
\[
\E G_0(A)=H(A),\quad \Var[G_0(A)]= \frac{H(A)[1- H(A)]}{\beta_0+1}
\]
and since $G_j \ed G_1^\ast$, 
\begin{align*}
	\E G_j(A) &= 
	\E \E [G_1^\ast(A)\mid G_0] = \E G_0(A) = H(A)\\
	\Var [G_j(A)] 
	&= 
	\E \Var[G_1^\ast(A) \mid G_0 ] + \Var [G_0(A)] = \dfrac{H(A) [1-H(A)](\beta_0+ \beta +1)}{(\beta+1) (\beta_0+1)} .
\end{align*}
Mixed moments are also easy to determine, as $\E G_1^\ast(A) G_2^\ast (A) = \E  \E [G_1^\ast(A) \mid G_0] \E [G_2^\ast(A) \mid G_0]  = \E G_0(A)^2$ and
\begin{align*}
	\E G_j(A) G_{j^\prime}(A)
	&= \E[G_1(A) G_2(A) \mid G_1 = G_2]\, \P (G_1 = G_2) + \E[G_1(A) G_2(A) \mid G_1 \ne G_2]\, \P(G_1 \ne G_2)\\
	& = \dfrac{1}{1+\alpha} \E[G_1^\ast(A)^2] + \dfrac{\alpha}{\alpha +1} \E[G_1^\ast(A) G_2^\ast (A)] \\
	&= 
	\dfrac{1}{1+\alpha} \E[G_1^\ast(A)^2] + \dfrac{\alpha}{\alpha +1} \E [G_0(A)^2].
\end{align*}
One, then, obtains 	
\[
\Cov[G_j(A),G_{j^\prime}(A)] = \E[G_j(A) G_{j^\prime}(A)]  - H(A)^2 = \dfrac{1}{1+\alpha} \Var[G_1^\ast(A)] + \dfrac{\alpha}{\alpha +1} \Var[G_0(A)]
\]
and
\[
\mathrm{Cor}[G_j(A),G_{j^\prime}(A)] = \dfrac{1}{1+\alpha} + \dfrac{\alpha}{\alpha +1} \dfrac{\Var[G_0(A)]}{\Var[G_1^\ast(A)]}
=\dfrac{\beta_0+\beta +1 +\alpha \beta +\alpha}{(1+\alpha)(\beta_0 +\beta +1)}
\]	
so that the conclusion follows. 

\subsection{Proof of Proposition \ref{eq:CovXHyDP}}
Note that $X_{j,i} \ed X^\ast_d$. Thus,
\[
\Cov(X_{j,i},X_{j^\prime,i^\prime})
=
\P(X_{j,i}=X_{j^\prime,i^\prime}) \: \Var(X^\ast_d)
\]
Moreover, if $j=j'$, then
\begin{align*}
\P(X_{j,i^\prime}=X_{j,i}) 
&= \P(X_{j,i^\prime}=X_{j,i} \mid T_{j,i}=T_{j,i^\prime}]\, \P(T_{j,i}=T_{j,i^\prime})+\P(X_{j,i^\prime} = X_{j,i} \mid T_{j,i} \ne T_{j,i^\prime})\: \P(T_{j,i} \ne T_{j,i^\prime})\\
&=\dfrac{1}{\beta+1} + \P(D_{T_{j,i^\prime}} = D_{T_{j,i}} \mid T_{j,i} \ne T_{j,i^\prime}) \,\dfrac{\beta}{\beta+1} 
= \dfrac{\beta+\beta_0+1}{(\beta+1)(\beta_0+1)}
\end{align*}
If $j \ne j^\prime$, then
\begin{align*}
\P(X_{j,i}=X_{j^\prime,i^\prime})
&=
\P(X_{j,i}=X_{j^\prime,i^\prime} \mid G_j = G_{j^\prime})\:
\P(G_j = G_{j^\prime}) 
+ \P(X_{j,i}=X_{j^\prime,i^\prime} \mid G_j \ne G_{j^\prime}) \,\P(G_j \ne G_{j^\prime})\\
&=\P(X_{j,i^\prime}=X_{j,i}) \P(G_j=G_{j^\prime}) + \P(D_{T_{j^\prime,i^\prime}} = D_{T_{j,i}} \mid T_{j,i} \ne T_{j^\prime,i^\prime})\, \P(G_j \ne G_{j^\prime})\\
&=
\dfrac{1}{\beta_0+1} +\dfrac{\beta_0}{(1+\alpha)(1+\beta)(1+\beta_0)}
\end{align*}   
and the conclusion follows.

\subsection{Proof of \cref{th:pEPPFHyDP}}
In order to prove \cref{th:pEPPFHyDP}, we first state the following auxiliary result.
\begin{lemma}\label{le:condpeppf}
The random partition induced by the samples $\{\bm{X}_j: j=1,\ldots,J \}$ drawn from $(G_1,\ldots,G_J) \sim \HyDP\left(\alpha, \beta,\beta_0;H\right)$ given a particular partition of distributions $\Psi^{(J)} = \{B_1,\ldots,B_{R} \}$ is characterized by the pEPPF
\begin{align*}
	\Pi_D^{(n)}\left(\bm{n}_1,\ldots,\bm{n}_J; \alpha, \beta,\beta_0 \mid \Psi^{(J)} = \{B_1,\ldots,B_{R} \}\right) = \Phi_{D,R}^{(n)}\left(\bm{n}_1^\ast,\ldots,\bm{n}^\ast_{R}; \beta, \beta_0 \right),
\end{align*}
where $n^{\ast}_{r,d} = \sum_{j \in B_r} n_{j,d}$ for each $r=1,\ldots,R$, $d=1,\ldots,D$, and  $\Phi_{D,R}^{(n)}\left(\bm{n}_1^\ast,\ldots,\bm{n}^\ast_{R}; \beta, \beta_0 \right)$ is the pEPPF associated to a $R$-dimensional $\HDP(\beta, \beta_0;H)$.
\end{lemma}
Now we can write
\begin{align}
\begin{split}
	&\Pi_D^{(n)}(\bm{n}_1,\ldots,\bm{n}_J; \alpha, \beta, \beta_0 \mid \Psi^{(J)} = \{B_1,\ldots,B_{R} \}) =\\
	&=\E \bigg[\int_{\mathbb{X}^{D}_{\ast}} \prod_{d=1}^D G_1(\mathrm{d}x_d)^{n_{1,d}} \ldots G_J(\mathrm{d}x_d)^{n_{J,d}} \bigm\vert \Psi^{(J)}=\{B_1,\ldots,B_{R} \}\bigg]=\\
	&=\E\bigg[\int_{\mathbb{X}^{D}_{\ast}} \prod_{d=1}^D G_1^{\ast}(\mathrm{d}x_d)^{ n_{1,d}^\ast} \ldots G_{R}^{\ast}(\mathrm{d}x_d)^{ n_{R,d}^\ast} \bigg]=\Phi_{D,R}^{(n)}(\bm{n}_1^\ast,\ldots,\bm{n}^\ast_{R}; \beta,\beta_0 ),
\end{split}
\end{align}
with $\mathbb{X}^{D}_{\ast} = \mathbb{X}^{D} \setminus \{\bm{x} : x_i =x_j \text{ for some }i \ne j \}$ and $( G_{1}^{\ast}, \ldots, G_{R}^{\ast} ) \sim \HDP(\beta,\beta_0; H )$. Moreover, note that the $R$ unique values among $(G_1,\ldots,G_J)$ are not necessarily the first $(G_1^\ast,\ldots,G_{R}^\ast)$ but since $(G_k^\ast)_{k \ge 1}$ are exchangeable the third equality holds.

Therefore, by applying \cref{le:condpeppf}
\begin{align}
\begin{split}
	\Pi_D^{(n)}\big(\bm{n}_1, \ldots, \bm{n}_J \big) &= \sum p\big(\Psi^{(J)}=\{B_1,\ldots,B_{R} \} \big) \Pi_D^{(n)}\big(\bm{n}_1,\ldots,\bm{n}_J; \alpha, \beta,\beta_0 \mid \Psi^{(J)} = \{B_1,\ldots,B_{R} \}\big) =\\
	&= \sum \phi_{R}^{(J)}\big(m_1,\ldots,m_{R};\alpha,\big)  \Phi_{D,R}^{(n)}\big(\bm{n}_1^\ast,\ldots,\bm{n}^\ast_{R}; \beta, \beta_0 \big)
\end{split}
\end{align}

\subsection{Proof of \cref{prop:probDegHyDP}}
In order to derive the posterior probability of degeneracy, we write the marginal likelihood as
\begin{equation*}
p(\bm{X})=\Pi_D^{(n)}(\bm{n}_1,\bm{n}_2) \prod_{d=1}^{D} H(\mathrm{d}{X}^{\ast}_d),
\end{equation*}
where $\{X_1^\ast,\ldots,X_D^\ast\}$ are the $D$ unique values among $\bm{X}$ and $\Pi_D^{(n)}(\bm{n}_1,\bm{n}_2)$ is the pEPPF associated to the proposed model \eqref{eq:peppf_hhdp}, that is
\begin{equation*}
\Pi_D^{(n)}(\bm{n}_1,\bm{n}_2)= \P(G_1=G_2) \Phi_{D,1}^{(n)}(\bm{n}_1+\bm{n}_2) + \P(G_1 \ne G_2) \Phi_{D,2}^{(n)}(\bm{n}_1,\bm{n}_2),
\end{equation*}

Finally, we prove the proposition by applying Bayes theorem
\begin{equation*}
\P(G_1=G_2 \mid \bm{X}) = \dfrac{\P(G_1=G_2) p(\bm{X}\mid G_1=G_2)}{p(\bm{X})} =\dfrac{\Phi^{(n)}_{D,1}(\bm{n}_1+ \bm{n}_2)}{ \Phi^{(n)}_{D,1}(\bm{n}_1+ \bm{n}_2)+\alpha \Phi^{(n)}_{D,2}(\bm{n}_1, \bm{n}_2)},
\end{equation*}
where $\Phi^{(n)}_{D,1}$ and $\Phi^{(n)}_{D,2}$ are the pEPPF and the EPPF of a bivariate and univariate $\HDP(\beta, \beta_0;H)$, respectively.

More precisely, following \cite{Camerlenghi2019,Camerlenghi2018e} we can derive the pEPPF $\Phi^{(n)}_{D,2}$ and the EPPF $\Phi^{(n)}_{D,1}$ of a bivariate and univariate $\HDP(\beta, \beta_0;H)$, respectively. In particular

\begin{equation}\label{eq:EPPFHDP1dim}
\Phi^{(n)}_{D,1}(\bm{n}^\ast)=\dfrac{\beta_0^{D}}{(\beta)_n} \sum_{\bm{\ell}^\ast} \dfrac{\beta^{|\bm{\ell}^\ast|}}{(\beta_0)_{|\bm{\ell}^\ast|}} \prod_{d=1}^{D}(\ell_d^\ast-1)! |\mathcal{s}(n^\ast_{d},\ell^\ast_d)|,
\end{equation}
where $|\mathcal{s}(n,\ell)|$ is the signless Stirling numbers of the first kind and the sum runs over all vectors $\bm{\ell}^\ast = (\ell_1^\ast,\ldots,\ell_D^\ast)$ such that $\ell^\ast_d \in \{1,\ldots,n_{d}^\ast\}$, $|\ell^{\ast}|= \sum_{d=1}^D \ell^\ast_d$ and
\begin{equation}\label{eq:pEPPFHDP2dim}
\Phi^{(n)}_{D,2}(\bm{n}_1, \bm{n}_2) = \dfrac{\beta_0^{D}}{\prod_{j=1}^J (\beta)_{I_j}} \sum_{\bm{\ell}} \dfrac{\beta^{|\bm{\ell}|}}{ (\beta_0)_{|\bm{\ell}|}} \prod_{d=1}^{D}(\ell_{\cdot l}-1)! \prod_{j=1}^2 |\mathcal{s}(n_{j,d},\ell_{j,d})|,
\end{equation}
where $\bm{\ell}=(\bm{\ell}_1,\bm{\ell}_2)$, with each $\bm{\ell}_j =(\ell_{j,1},\ldots,\ell_{j,D}) \in \times_{d=1}^D \{1,\ldots,n_{j,d}\}$ and $|\ell|= \sum_{j=1}^2 \sum_{d=1}^D \ell_{j,d}$.

\section{A marginal Gibbs sampler}\label{subsec:MargGibbs}

The marginal Gibbs sampler that updates $\Delta$, the table dish assignments $T_{j,i}$, and $D_{t}$ can be deduced from the hidden Chinese restaurant franchise presented in \cref{subsec:LCRF}. Let $\bm{S} = \{\Delta, (T_{j,i})_{j,i}, (D_{t})_{t}, ({X}_{j,i})_{j,i} \}$. 
Hence, the algorithm can be summarized as follows
\begin{enumerate}
\item[\texttt{(1)}] Sample the population assignments to the restaurants
\begin{align*} 
	\P(\Delta =1 \mid \bm{X}) = \dfrac{\Phi^{(n)}_{D,1}(\bm{n}_{1}+\bm{n}_2)}{ \Phi^{(n)}_{D,1}(\bm{n}_{1}+\bm{n}_2)+\alpha \, \Phi^{(n)}_{D,2}(\bm{n}_1+\bm{n}_2)},
\end{align*}
where $\Phi^{(n)}_{D,2}$, $\Phi^{(n)}_{D,1}$ are the pEPPF and EPPF of a bivariate and univariate $\HDP(\beta, \beta_0;H)$, respectively.
\item[\texttt{(2)}] Sample the table assignments $T_{j,i}$ and corresponding dishes $D_{T_{j,i}}$ from\\
\begin{align*}
		p(T_{j,i}, D_{T_{j,i}} \mid \bm{S}^{-(T_{j,i},D_{T_{j,i}})} ) \propto \begin{cases} 
			T_{j,i} =t                                   & \frac{q_{r,t,\cdot}^{-(ji)}}{q_{r,\cdot,\cdot}^{-(ji)}+\beta} \,\ p_{D_t}(\{{X}_{j,i}\}) \\ 
			T_{j,i} = t^\text{new}, D_{t^\text{new}} =d                   & \frac{\beta}{q_{r,\cdot,\cdot}^{-(ji)}+\beta}  \frac{\ell_{\cdot,d}^{-(ji)}}{\ell_{\cdot,\cdot}^{-(ji)}+\beta_0} \,\ p_d(\{{X}_{j,i}\})\\
			T_{j,i} = t^\text{new}, D_{t^\text{new}} =d^\text{new} & \frac{\beta}{q_{r,\cdot,\cdot}^{-(ji)}+\beta}  \frac{\beta_0}{\ell_{\cdot,\cdot}^{-(ji)}+\beta_0} \,\ p_{d^\text{new}}(\{{X}_{j,i}\}),
		\end{cases}
	\end{align*}
	where $p_d(\{{X}_{j,i}\})$ is defined by the following equation. For every index set $\mathcal{I}$ 
	\[
	p_d(\{{X}_{j,i}\}_{(j,i) \in \mathcal{I}}) = \frac{\int \prod_{j^\prime i^\prime \in \mathcal{I} \cup \mathcal{I}_d} \mathcal{K}({X}_{j,i}\mid \theta) \mathrm{d}H(\theta) }{\int \prod_{j^\prime i^\prime \in \mathcal{I}_d \setminus \mathcal{I}} \mathcal{K}({X}_{j,i} \mid \theta) \mathrm{d}H(\theta) },
	\]
	where $\mathcal{I}_d = \{(j,i): D_{T_{j,i}}=d \}$. For instance, $p_d(\{{X}_{j,i}\})$ is the marginal conditional probability of $X_{j,i}$ in cluster $d$ given the other observation assigned to cluster $d$.
	\item[\texttt{(3)}] Sample the dish assignments $D_t$ from 
	\begin{align*}
		p(D_{t} \mid \bm{S}^{-t}) 
		\propto \begin{cases}
			d &\frac{\ell_{\cdot,d}^{-t}}{\ell_{\cdot,\cdot}^{-t}+\beta_0} p_d(\{x_{j,i} : T_{j,i}=t \})\\ 
			d^\text{new} & \frac{\beta_0}{\ell_{\cdot,\cdot}^{-t}+\beta_0}  p_{d^\text{new}}( \{x_{j,i}: T_{j,i} =t \}). 
		\end{cases} 
	\end{align*}
\end{enumerate}

\section{Sensitivity analysis for the hyperparameters specification}\label{app:hyperpar}
\begin{figure}[h]
	\makebox[\linewidth]{\hspace*{6.2cm} NDP \hspace*{6.2cm} HHDP \hspace*{5cm}}	\makebox[\linewidth]{\hspace*{2.7cm} Pop 1 \hspace*{2.2cm} Pop 2 \hspace*{3cm} Pop 1 \hspace*{2.2cm} Pop 2 \hspace*{1.3cm}}
	\centering
	\rotatebox[origin=c]{90}{\makebox[3cm]{Parameters 1}}
	\centering
	\begin{subfigure}{6.7cm}
		\includegraphics[width=\linewidth, height=3cm]{NDP_SBSimScen3Ind}
	\end{subfigure}\hspace*{0.5cm}
	\begin{subfigure}{6.7cm}
		\includegraphics[width=\linewidth, height=3cm]{HyDP_FinDirNDPHDPSimScen3Ind}
	\end{subfigure}\\ 
	\rotatebox[origin=c]{90}{\makebox[3cm]{Parameters 0.1}}
	\begin{subfigure}{6.7cm}
		\includegraphics[width=\linewidth, height=3cm]{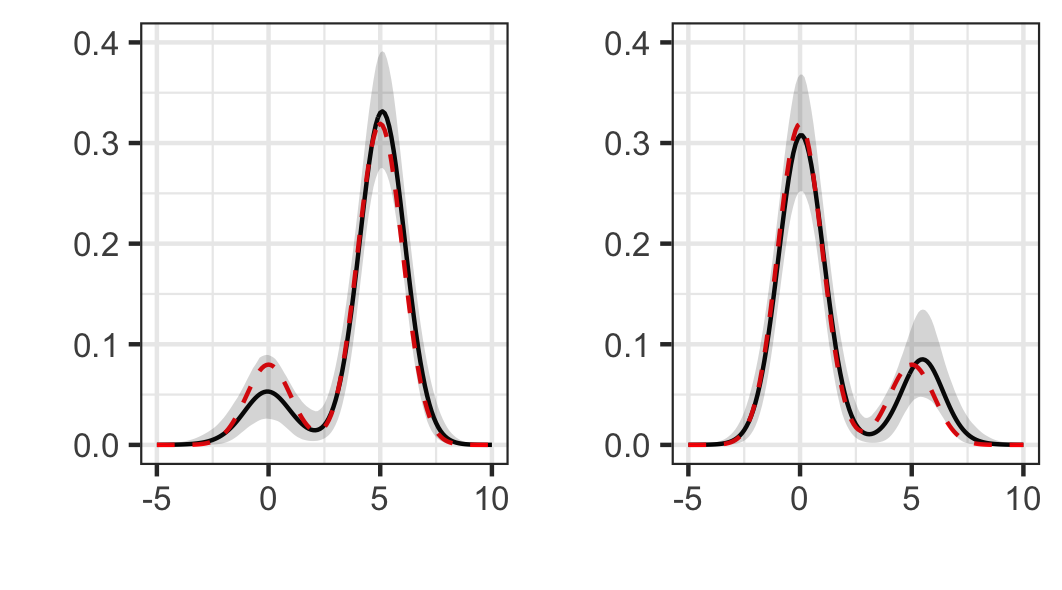}
	\end{subfigure}\hspace*{0.5cm}
	\begin{subfigure}{6.7cm}
		\includegraphics[width=\linewidth, height=3cm]{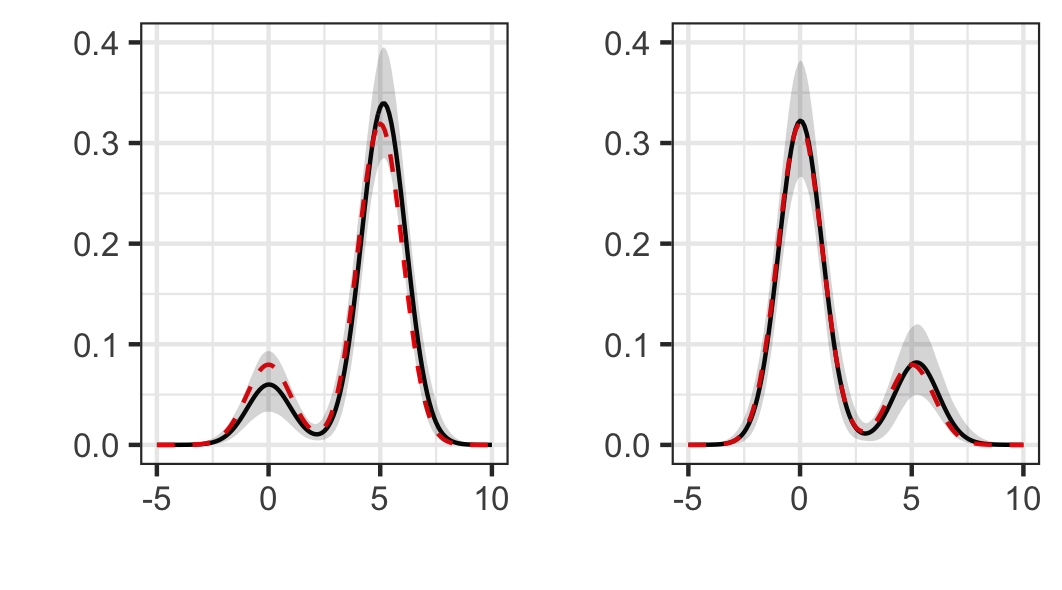}
	\end{subfigure}\\ 
	\rotatebox[origin=c]{90}{\makebox[3cm]{Parameters 3}}
	\begin{subfigure}{6.7cm}
		\includegraphics[width=\linewidth, height=3cm]{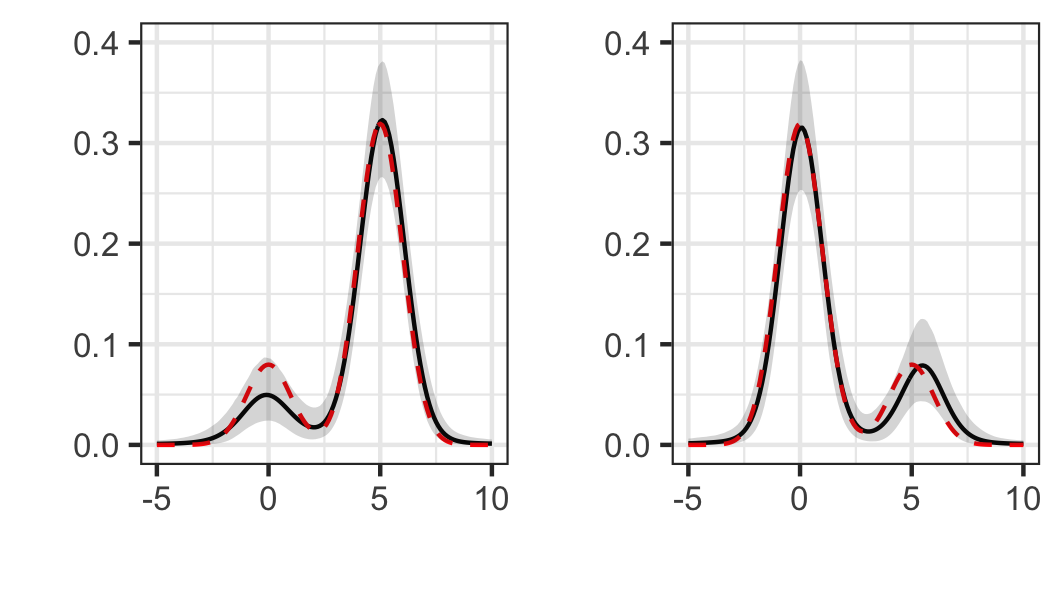}
	\end{subfigure}\hspace*{0.5cm}
	\begin{subfigure}{6.7cm}
		\includegraphics[width=\linewidth, height=3cm]{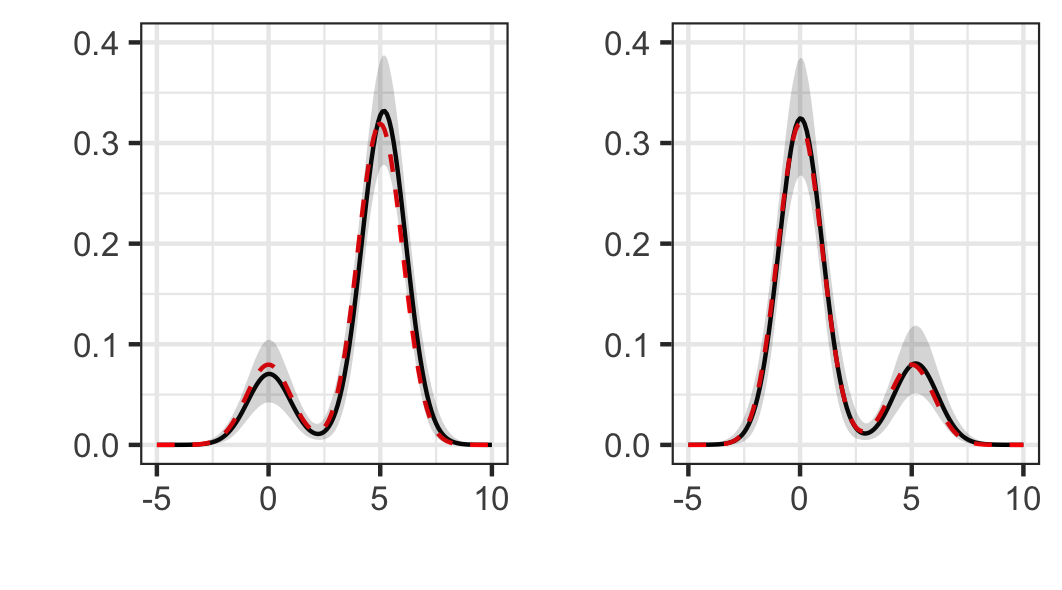}
	\end{subfigure}
	\caption{True (dashed lines), posterior mean (solid lines) densities and 95\% point-wise posterior credible intervals (shaded gray) estimated under different hyperparameters specifications.\label{fig:DensHyper}}
\end{figure}

\begin{figure}[h]
\makebox[\linewidth]{\hspace*{2.6cm} True \hspace*{2cm} HHDP \hspace*{1.9cm} NDP \hspace*{1.6cm}}
					\makebox[\linewidth]{\hspace*{1.8cm} Pop 1 \hspace*{0.1cm} Pop 2\hspace*{0.8cm} Pop 1 \hspace*{0.2cm} Pop 2 \hspace*{0.8cm} Pop 1 \hspace*{0.2cm} Pop 2 \hspace*{1cm}}
	\centering
	\rotatebox[origin=c]{90}{\makebox[3cm]{Parameters 1}}
	\rotatebox[origin=c]{90}{\makebox[3cm]{ Pop 1 \hspace*{0.15cm} Pop 2}}
	\centering
	\begin{subfigure}{3cm}
		\includegraphics[width=\textwidth,height=3cm]{TruClustObsScen3}
	\end{subfigure}
	\begin{subfigure}{3cm}
		\includegraphics[width=\textwidth,height=3cm]{HyDP_FinDirNDPHDPSimScen3IndprobClust}
	\end{subfigure}
	\begin{subfigure}{3cm}
		\includegraphics[width=1.15\textwidth,height=3cm]{NDP_SBSimScen3IndprobClust}
	\end{subfigure}\\
	\vspace*{0.1cm}
	\rotatebox[origin=c]{90}{\makebox[3cm]{Parameters 0.1}}
	\rotatebox[origin=c]{90}{\makebox[3cm]{ Pop 1 \hspace*{0.15cm} Pop 2}}
	\begin{subfigure}{3cm}
		\includegraphics[width=\textwidth,height=3cm]{TruClustObsScen3}
	\end{subfigure}
	\begin{subfigure}{3cm}
		\includegraphics[width=\textwidth,height=3cm]{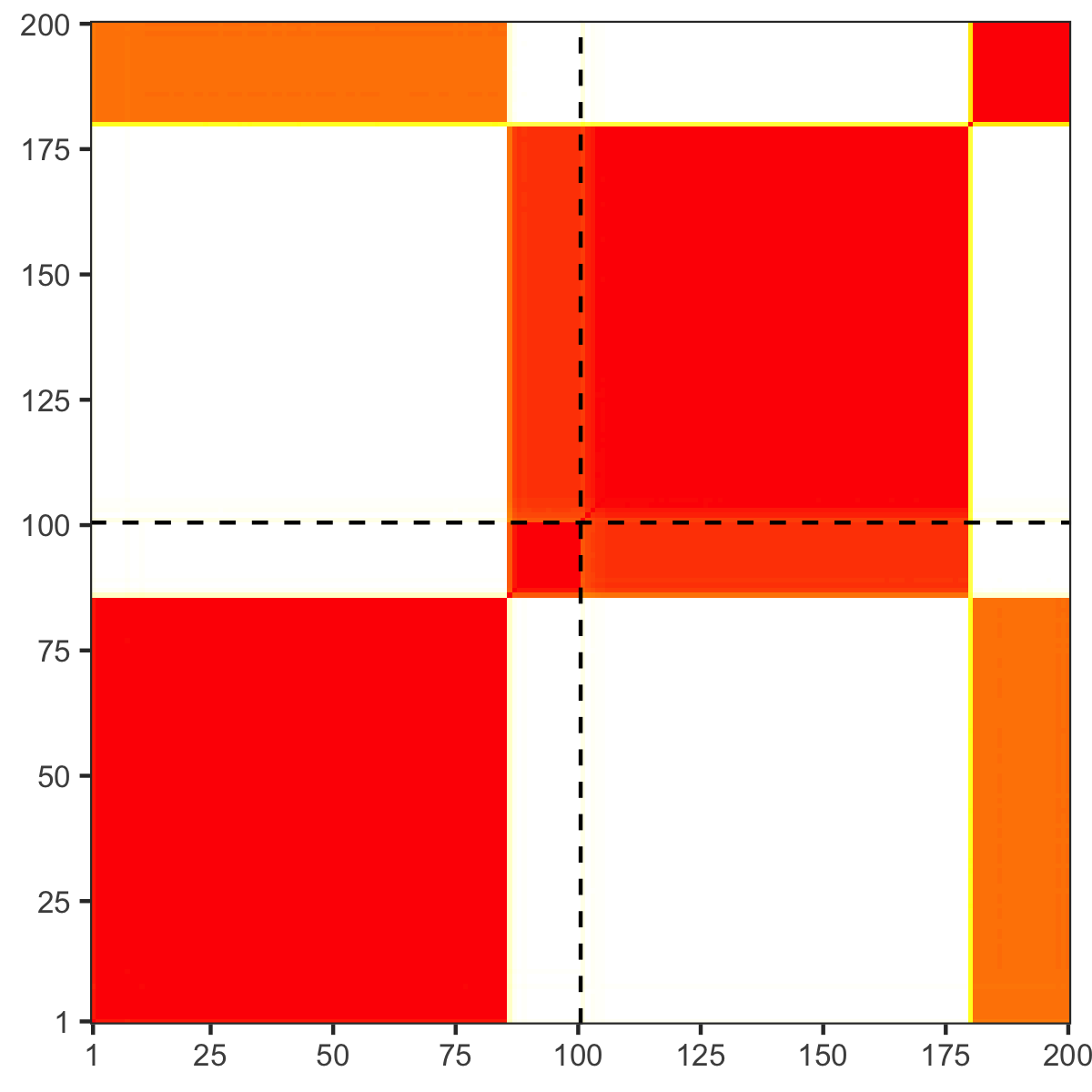}
	\end{subfigure}
	\begin{subfigure}{3cm}
		\includegraphics[width=1.15\textwidth,height=3cm]{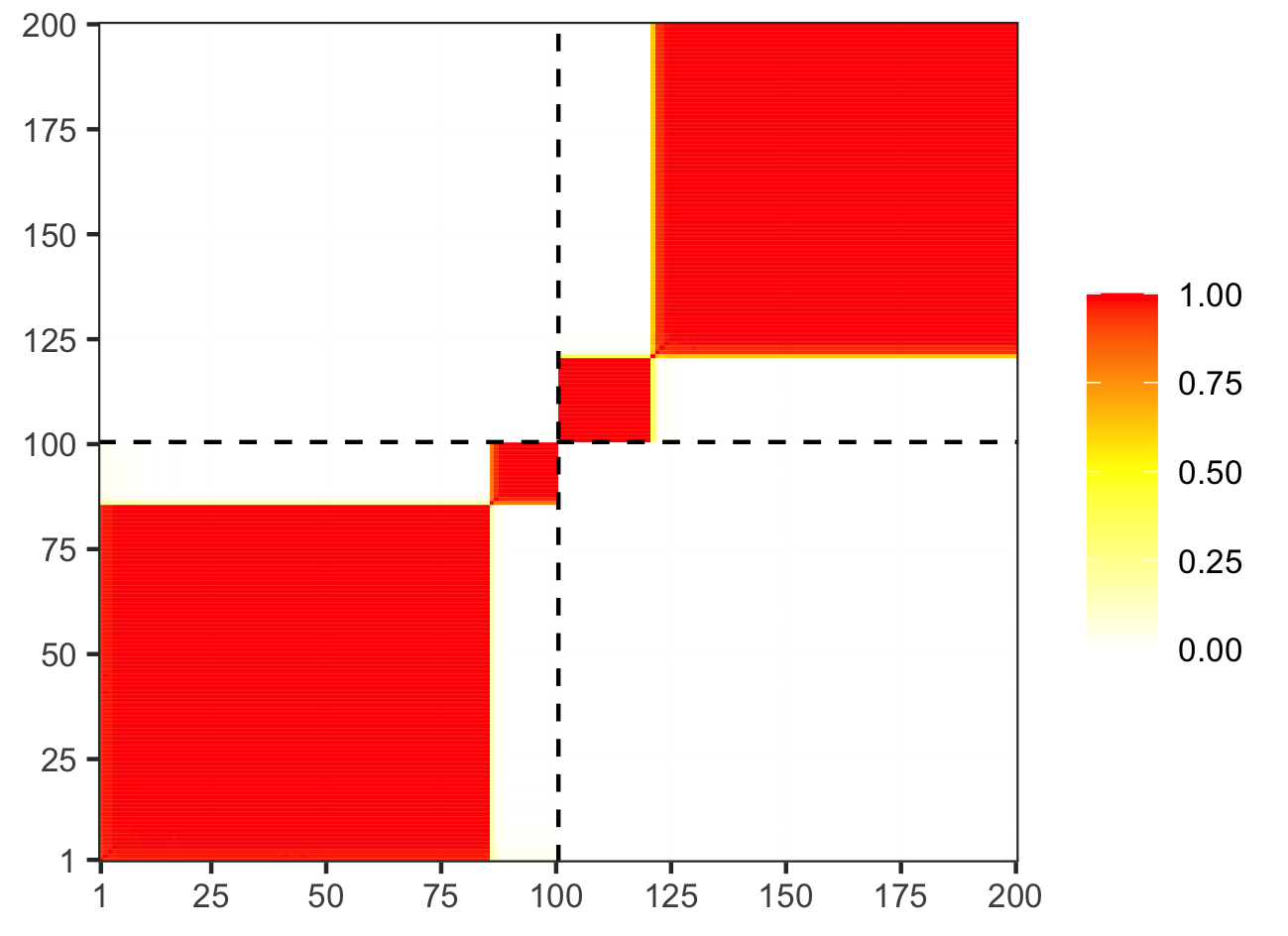}
	\end{subfigure}\\
	\vspace*{0.1cm}
	\rotatebox[origin=c]{90}{\makebox[3cm]{Parameters 3}}
	\rotatebox[origin=c]{90}{\makebox[3cm]{ Pop 1 \hspace*{0.15cm} Pop 2}}
	\begin{subfigure}{3cm}
		\includegraphics[width=\textwidth,height=3cm]{TruClustObsScen3}
	\end{subfigure}
	\begin{subfigure}{3cm}
		\includegraphics[width=\textwidth,height=3cm]{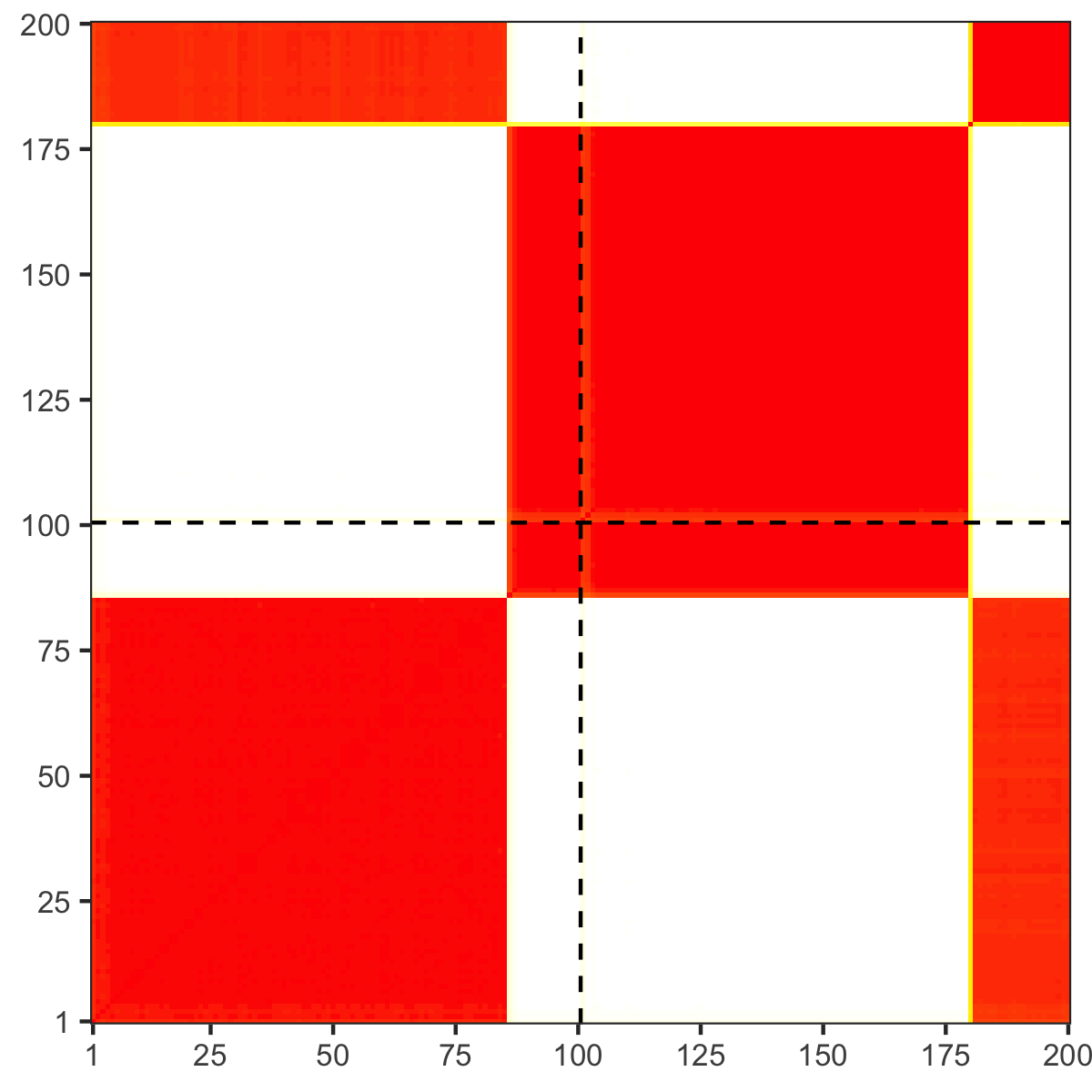}
	\end{subfigure}
	\begin{subfigure}{3cm}
		\includegraphics[width=1.15\textwidth,height=3cm]{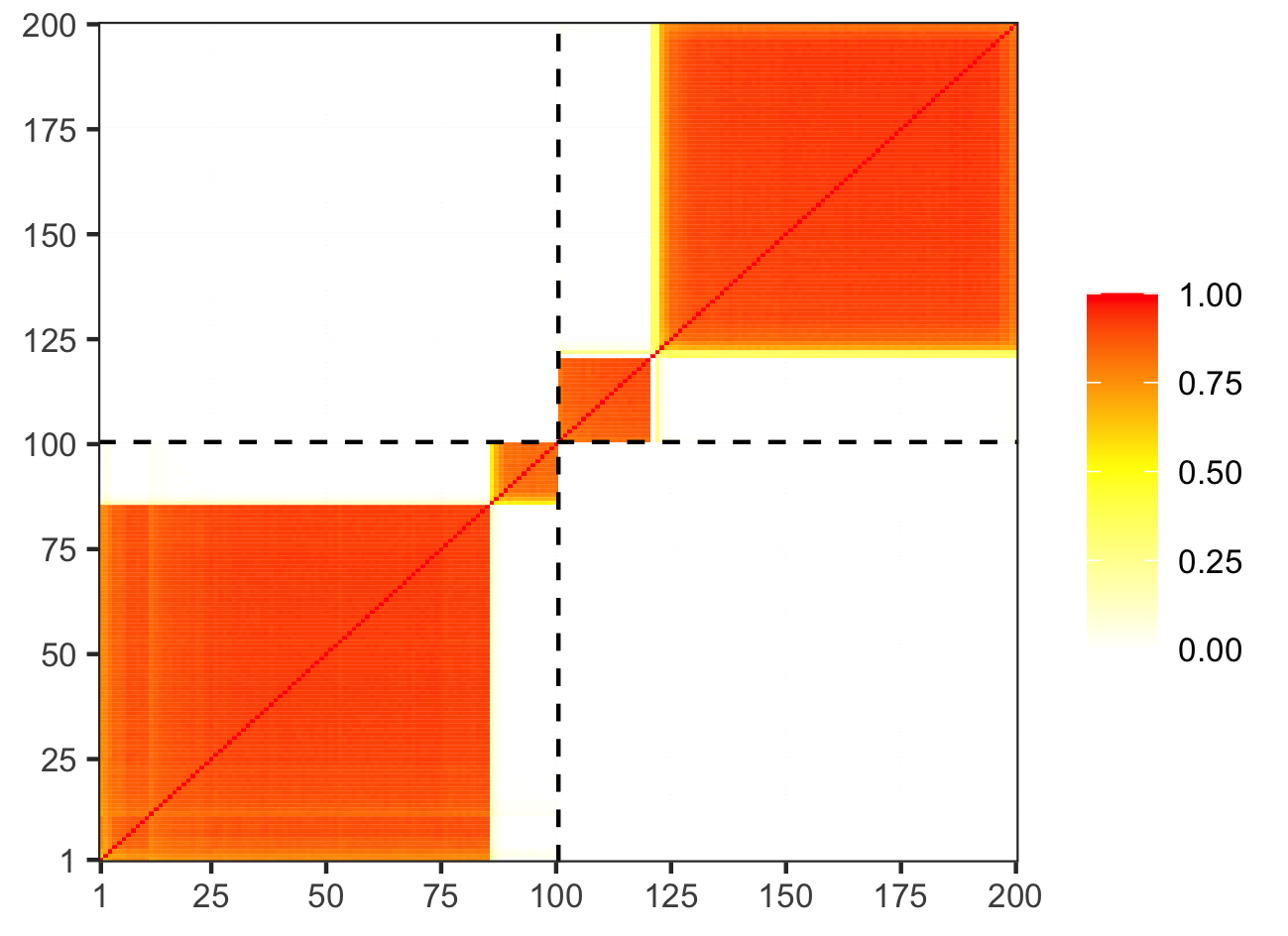}
	\end{subfigure}\\
	\caption{Heat maps of the true and estimated posterior probability of co-clustering of observations, ordered by population memberships, with the two models under different hyperparameters specifications.\label{fig:ProbsHyper}}
\end{figure}

Here we study the robustness with respect to the specification of  hyperparameters in relation to the comparison between the NDP and the HHDP mixture models presented in \cref{sec:Illustration}. The results are reported in terms of density estimates in \cref{fig:DensHyper} and probabilities of co-clustering of the observations in \cref{fig:ProbsHyper} using the finite--dimensional approximations of the $\DP$s with $L=K=50$ and different hyperparameter specifications.
The sensitivity analysis is performed by selecting different values for the concentration parameters. This allows to verify the robustness of the results comparing the two models.
We report the results for the data simulated according to scenario \RNum{3}, in which the two populations share both the Gaussian components, but with different mixture weights.

We perform inference with the model as in \cref{sec:Illustration} with the following specifications for the concentration parameters:
\begin{itemize}
	\item \textbf{Parameters 1:} all the concentration parameters are set equal to $1$, that is\\
	$(G_1,G_2) \sim \NDP(\alpha=1,\beta=1;H)$ and $(G_1,G_2) \sim \HyDP(\alpha=1,\beta=1,\beta_0=1;H)$, respectively.
	\item \textbf{Parameters 0.1:} all the concentration parameters are set equal to $0.1$, that is\\
	$(G_1,G_2) \sim \NDP(\alpha=0.1,\beta=0.1;H)$ and $(G_1,G_2) \sim \HyDP(\alpha=0.1,\beta=0.1,\beta_0=0.1;H)$, respectively.
	\item \textbf{Parameters 3:} all the concentration parameters are set equal to $3$, that is\\
	$(G_1,G_2) \sim \NDP(\alpha=3,\beta=3;H)$ and $(G_1,G_2) \sim \HyDP(\alpha=3,\beta=3,\beta_0=3;H)$, respectively.
\end{itemize}

Importantly the density estimates are essentially the same under the different hyperparameters specifications.
Probabilities of co-clustering change under the different hyperparameter settings coherently with the theory developed in \cref{subsec:PropertiesHyDP}. 
However, in all the scenarios both models do not degenerate to the exchangeable case. This implies that the NDP cannot cluster observations across populations, while the HHDP overcomes this issue.
Therefore, the results of the comparison between the two models presented in \cref{sec:Illustration} are essentially the same.

\section{Choice of the finite-dimensional approximations}\label{app:FinDim}
We now present the inferential results in terms of density estimates in \cref{fig:DensHyper} and probabilities of co-clustering of the observations in \cref{fig:ProbsHyper} for the two specifications in \cref{sec:Illustration}.
We report the results for the data simulated according to scenario \RNum{3} with the following finite-dimensional approximations of the $\DP$s:
\begin{itemize}
	\itemsep0em 
	\item $L=K=50$;
	\item $L=K=30$;
	\item $L=K=70$.
\end{itemize}

\begin{figure}[h]
	\makebox[\linewidth]{\hspace*{6.2cm} NDP \hspace*{6.2cm} HHDP \hspace*{5cm}}	\makebox[\linewidth]{\hspace*{2.7cm} Pop 1 \hspace*{2.2cm} Pop 2 \hspace*{3cm} Pop 1 \hspace*{2.2cm} Pop 2 \hspace*{1.3cm}}
	\centering
		\rotatebox[origin=c]{90}{\makebox[3cm]{ $L=K=5
		0$}}
	\begin{subfigure}{6.7cm}
		\includegraphics[width=\linewidth, height=3cm]{NDP_SBSimScen3Ind}
	\end{subfigure}\hspace*{0.5cm}
	\begin{subfigure}{6.7cm}
		\includegraphics[width=\linewidth, height=3cm]{HyDP_FinDirNDPHDPSimScen3Ind}
	\end{subfigure}\\ 
	\rotatebox[origin=c]{90}{\makebox[3cm]{ $L=K=30$}}
	\begin{subfigure}{6.7cm}
		\includegraphics[width=\linewidth, height=3cm]{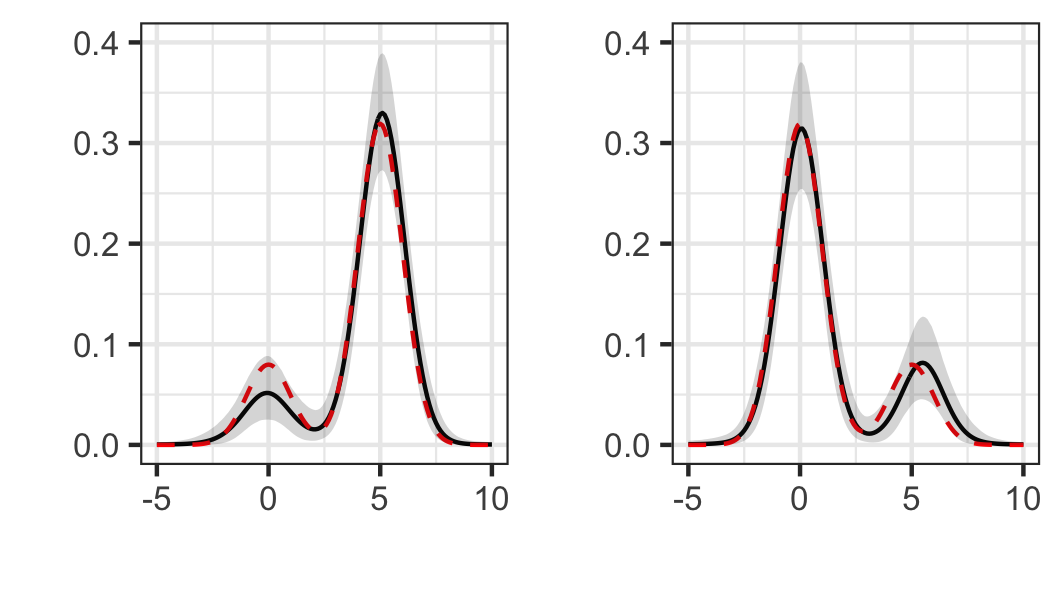}
	\end{subfigure}\hspace*{0.5cm}
	\begin{subfigure}{6.7cm}
		\includegraphics[width=\linewidth, height=3cm]{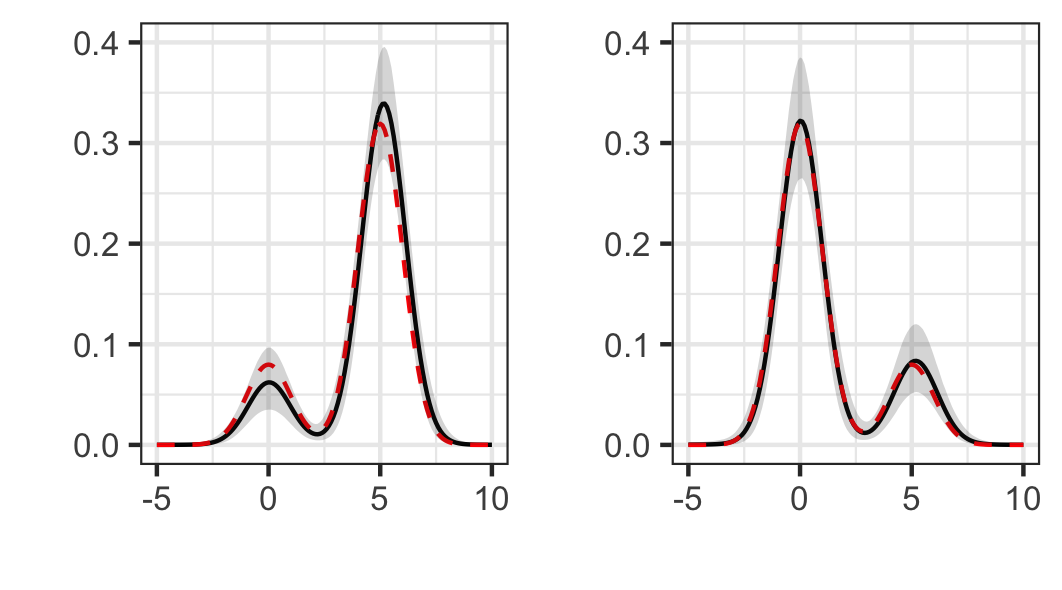}
	\end{subfigure}\\ 
	\rotatebox[origin=c]{90}{\makebox[3cm]{ $L=K=70$}}
	\begin{subfigure}{6.7cm}
		\includegraphics[width=\linewidth, height=3cm]{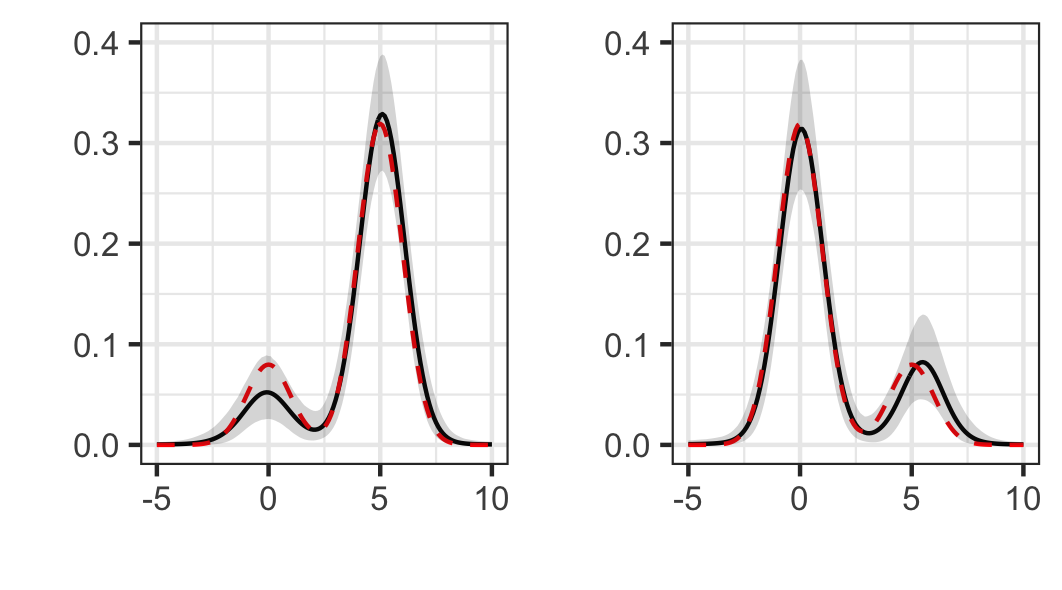}
	\end{subfigure}\hspace*{0.5cm}
	\begin{subfigure}{6.7cm}
		\includegraphics[width=\linewidth, height=3cm]{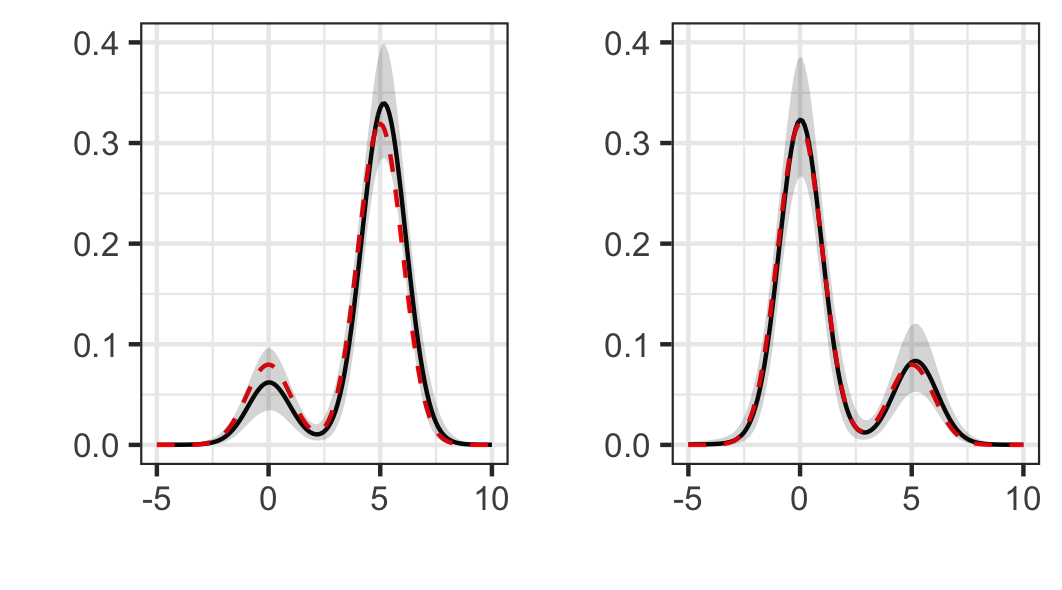}
	\end{subfigure}
	\caption{True (dashed lines), posterior mean (solid lines) densities and 95\% point-wise posterior credible intervals (shaded gray) estimated under different truncation levels.\label{fig:DensTrunc}}
\end{figure}

\begin{figure}[h]
\makebox[\linewidth]{\hspace*{2.6cm} True \hspace*{2cm} HHDP \hspace*{1.9cm} NDP \hspace*{1.6cm}}
					\makebox[\linewidth]{\hspace*{1.8cm} Pop 1 \hspace*{0.1cm} Pop 2\hspace*{0.8cm} Pop 1 \hspace*{0.2cm} Pop 2 \hspace*{0.8cm} Pop 1 \hspace*{0.2cm} Pop 2 \hspace*{1cm}}
	\centering
	\rotatebox[origin=c]{90}{\makebox[3cm]{$L=K=50$}}
	\rotatebox[origin=c]{90}{\makebox[3cm]{ Pop 1 \hspace*{0.15cm} Pop 2}}
	\begin{subfigure}{3cm}
		\includegraphics[width=\textwidth,height=3cm]{TruClustObsScen3}
	\end{subfigure}
	\begin{subfigure}{3cm}
		\includegraphics[width=\textwidth,height=3cm]{HyDP_FinDirNDPHDPSimScen3IndprobClust}
	\end{subfigure}
	\begin{subfigure}{3cm}
		\includegraphics[width=1.15\textwidth,height=3cm]{NDP_SBSimScen3IndprobClust}
	\end{subfigure}\\
	\vspace*{0.1cm}
	\rotatebox[origin=c]{90}{\makebox[3cm]{ $L=K=30$}}
	\rotatebox[origin=c]{90}{\makebox[3cm]{ Pop 1 \hspace*{0.15cm} Pop 2}}
	\begin{subfigure}{3cm}
		\includegraphics[width=\textwidth,height=3cm]{TruClustObsScen3}
	\end{subfigure}
	\begin{subfigure}{3cm}
		\includegraphics[width=\textwidth,height=3cm]{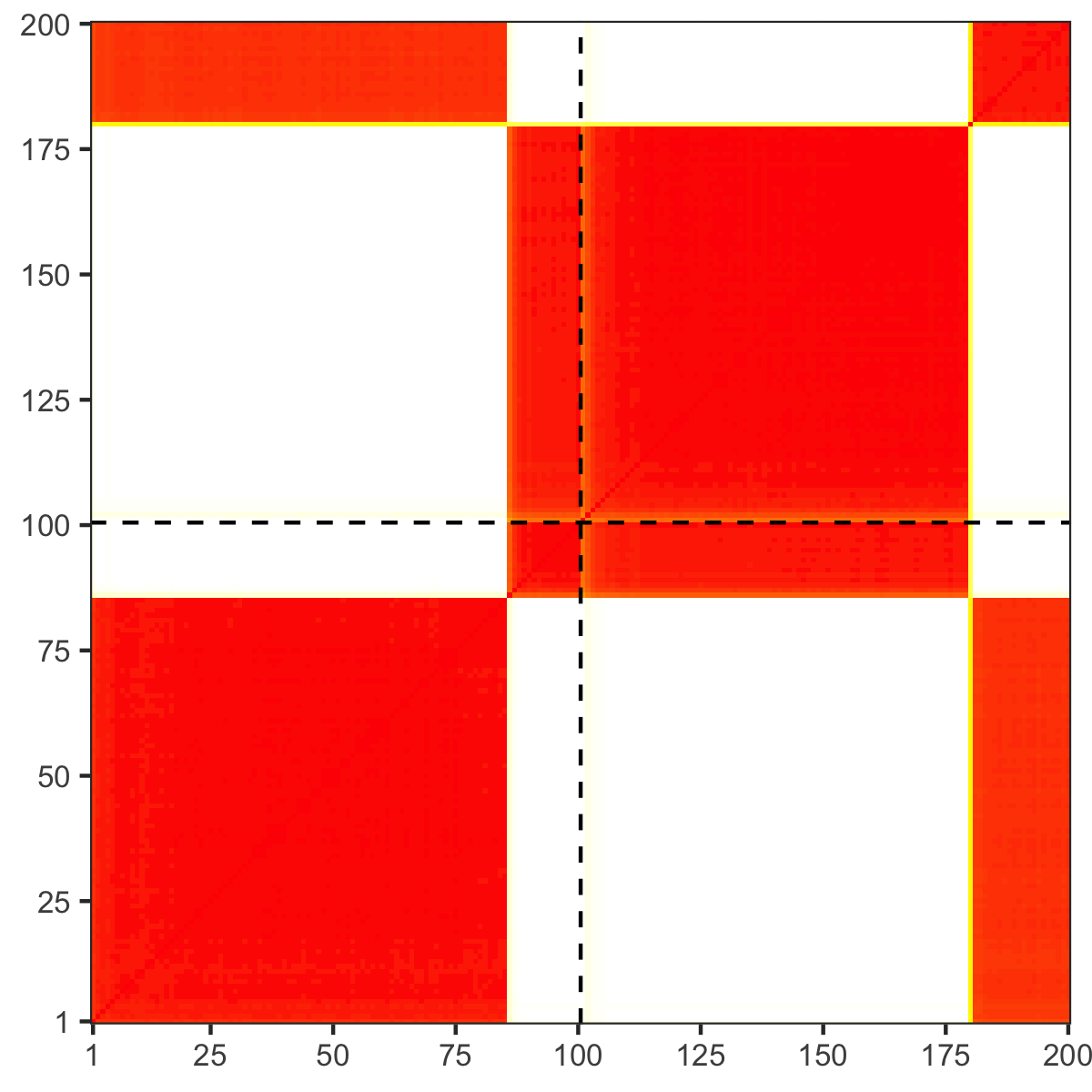}
	\end{subfigure}
	\begin{subfigure}{3cm}
		\includegraphics[width=1.15\textwidth,height=3cm]{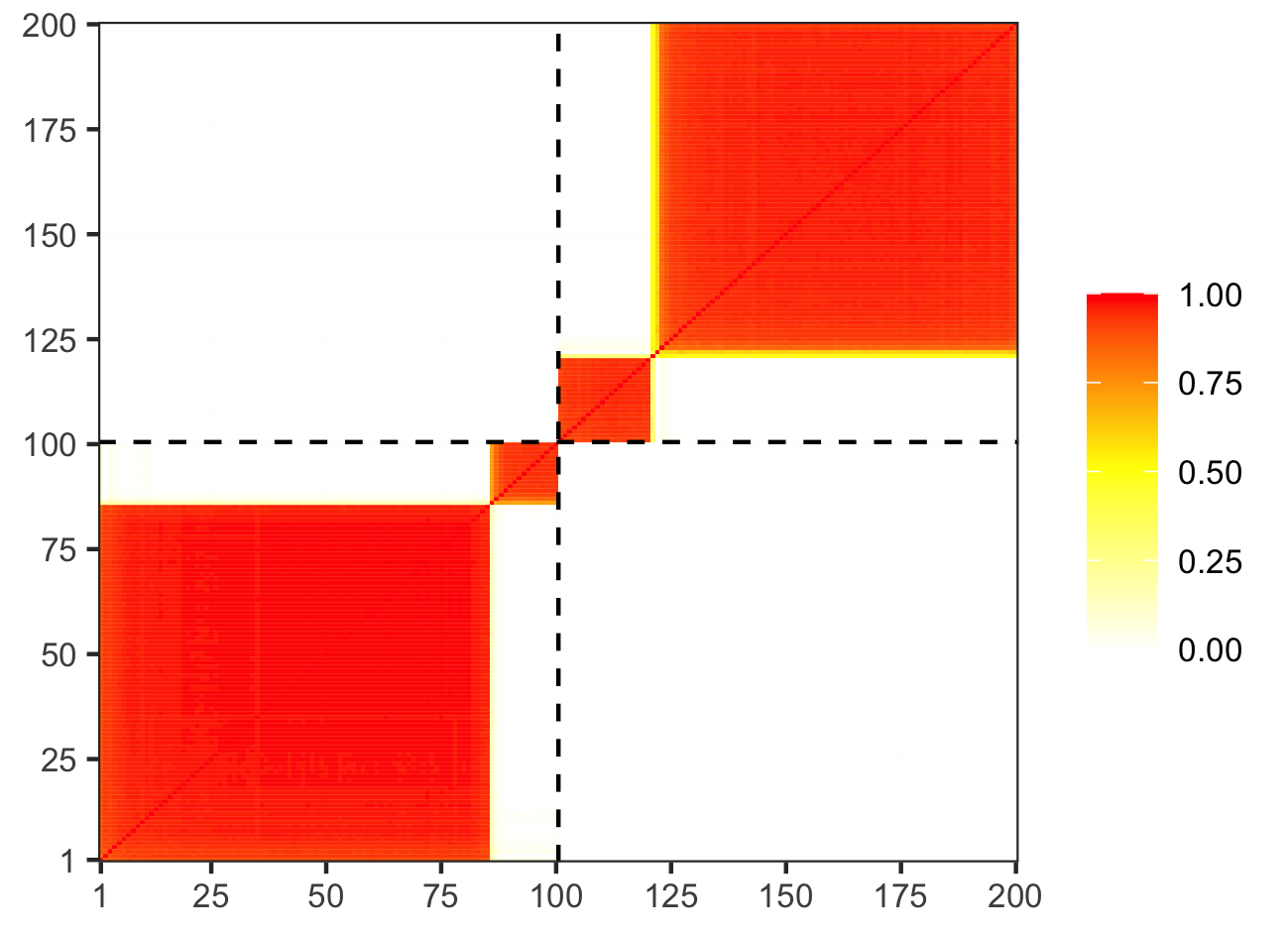}
	\end{subfigure}\\
	\vspace*{0.1cm}
	\rotatebox[origin=c]{90}{\makebox[3cm]{$L=K=70$}}
	\rotatebox[origin=c]{90}{\makebox[3cm]{Pop 1 \hspace*{0.15cm} Pop 2}}
	\begin{subfigure}{3cm}
		\includegraphics[width=\textwidth,height=3cm]{TruClustObsScen3}
	\end{subfigure}
	\begin{subfigure}{3cm}
		\includegraphics[width=\textwidth,height=3cm]{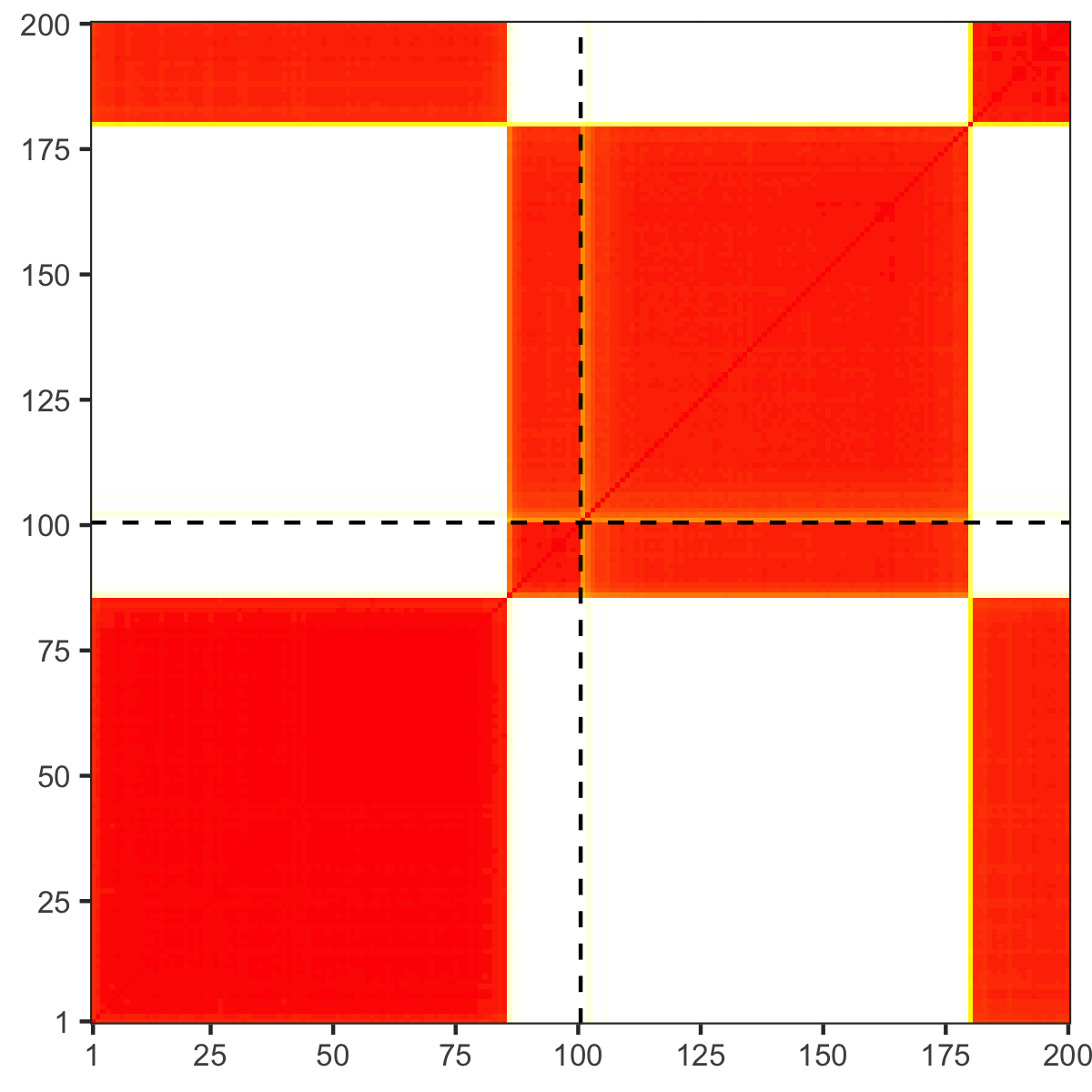}
	\end{subfigure}
	\begin{subfigure}{3cm}
		\includegraphics[width=1.15\textwidth,height=3cm]{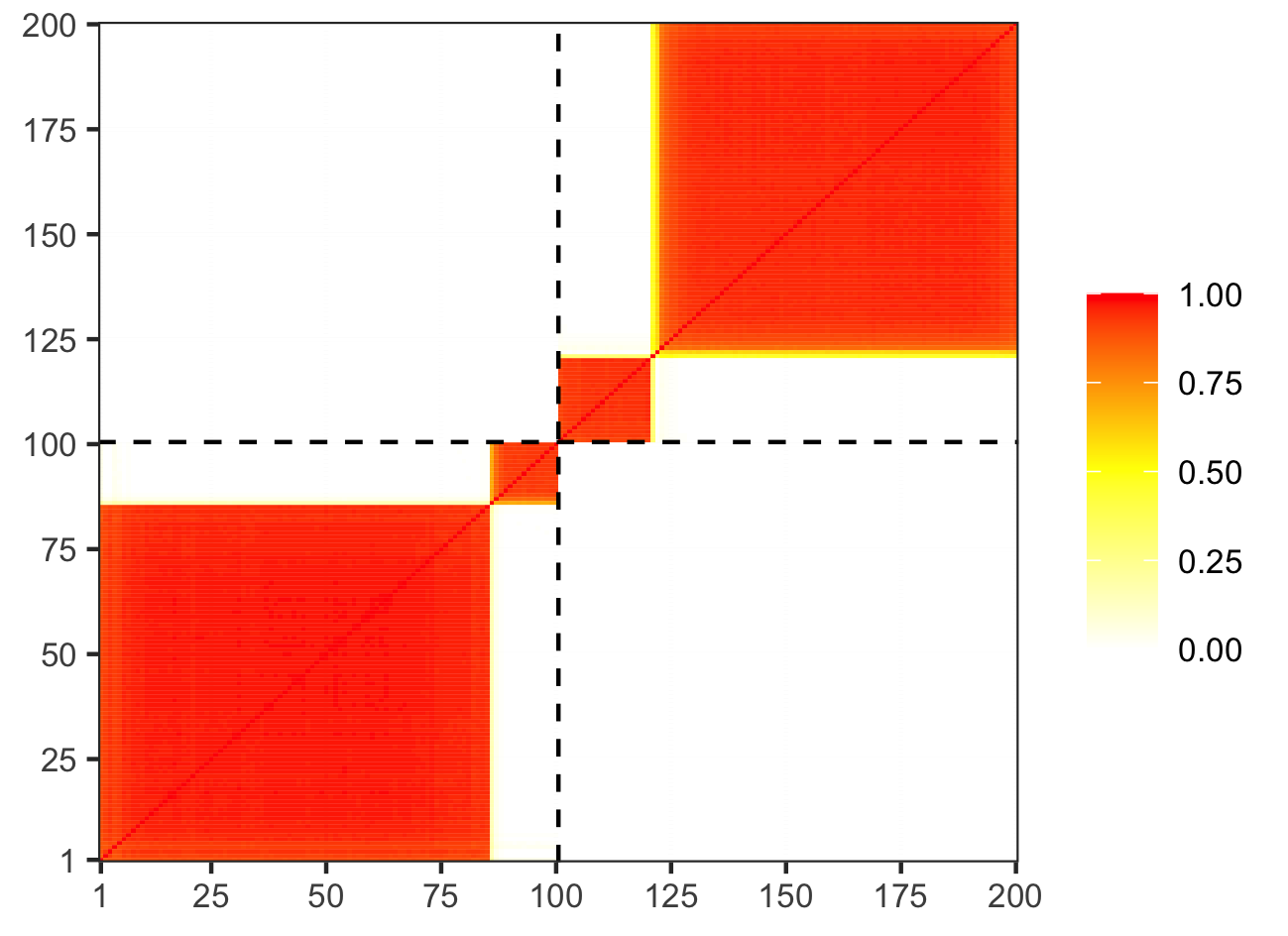}
	\end{subfigure}\\
	\caption{Heat maps of the true and estimated posterior probability of co-clustering of observations, ordered by population memberships, with the two models under different finite-dimensional approximations.\label{fig:ProbsTrunc}}
\end{figure}

Under all the different finite-dimensional approximations the inference is qualitatively the same, corroborating the idea that the finite-dimensional approximations $L=K=50$ proposed for the comparison of the NDP and HHDP in \cref{sec:Illustration} induce a negligible error in our analysis.

\section{Mixing of the MCMC algorithm}\label{app:Mixing}
\begin{figure}[h]
	\centering
	\includegraphics[width=0.7\linewidth]{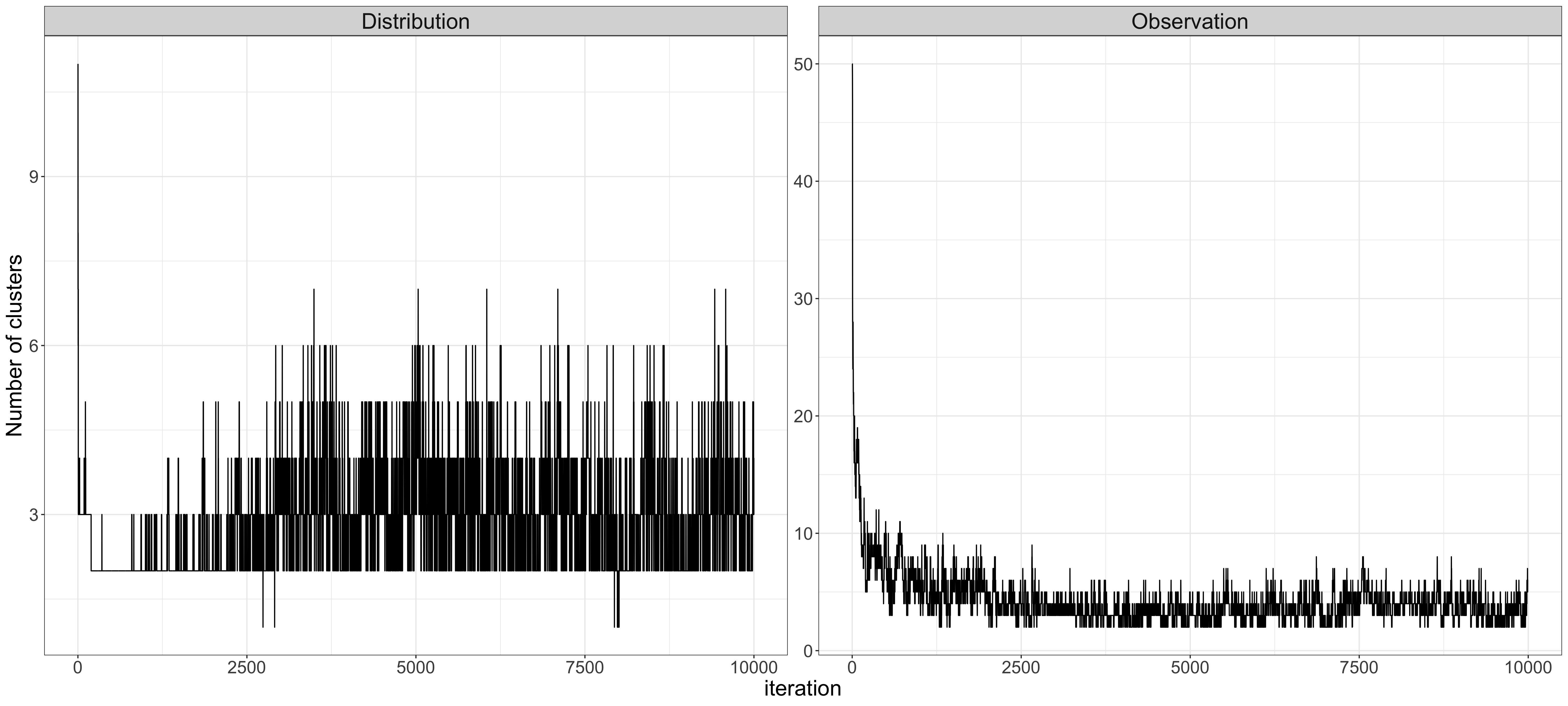}
	\caption{Traceplots of the number of distributional and observational clusters.}
	\label{fig:hydpfindirndphdpdatacppmixing}
\end{figure}

We now investigate the mixing for the number of clusters of both distributions and observations for the collaborative perinatal project application in Section \ref{subsec:real_data_appl} of the main manuscript.
Figure \ref{fig:hydpfindirndphdpdatacppmixing} shows the trace plots of the number of distributional and observational clusters sampled at each iteration (without discarding the burn-in and without performing any thinning). Note that here we started the MCMC with the bad initial guess that both clusterings feature only singletons and still the algorithm performed well.
The traceplots in Figure~\ref{fig:hydpfindirndphdpdatacppmixing} show a good mixing  for the number of clusters of both distributions and observations.  

\bibliography{HHDP_Supp}